\documentclass[11pt,preprint,numberedappendix]{aastex}

\usepackage{graphicx,amsmath,amssymb,epsfig,psfig,arydshln}
\usepackage[T1]{fontenc}
\usepackage{multirow}

\shorttitle{Selection of red QSOs from SDSS and UKIDSS photometry}
\shortauthors{Fynbo et al.}

\begin{document}


\title{Optical/near-infrared selection of red QSOs:
Evidence for steep extinction curves towards galactic centers?
\thanks{Based on
observations collected at the European Organisation for Astronomical Research
in the Southern Hemisphere, Chile, under program 088.A-0098, and on
observations made with the Nordic Optical Telescope, operated
on the island of La Palma jointly by Denmark, Finland, Iceland,
Norway, and Sweden, in the Spanish Observatorio del Roque de los
Muchachos of the Instituto de Astrofisica de Canarias.  
}}

\author{
J. P. U. Fynbo\altaffilmark{2},
J.-K. Krogager\altaffilmark{2},
B. Venemans\altaffilmark{3},
P. Noterdaeme\altaffilmark{4},
M. Vestergaard\altaffilmark{2},
P. M\o ller\altaffilmark{5},
C. Ledoux\altaffilmark{6},
S. Geier\altaffilmark{2,7}
}

\altaffiltext{2}{Dark Cosmology Centre, Niels Bohr Institute, University of
Copenhagen, Juliane Maries Vej 30, DK-2100 Copenhagen O}
\altaffiltext{3}{Max-Planck Institute for Astronomy, K{\"o}nigstuhl 17, 69117 Heidelberg, Germany}
\altaffiltext{4}{CNRS-UPMC, UMR7095, Institut d'Astrophysique de Paris, 98bis bd Arago, 75014, Paris, France}
\altaffiltext{5}{European Southern Observatory, Karl-Schwarzschildstrasse 2, D-85748 Garching bei M\"unchen, Germany}
\altaffiltext{6}{European Southern Observatory, Alonso de C\'ordova 3107,
Vitacura, Casilla 19001, Santiago 19, Chile}
\altaffiltext{7}{Nordic Optical Telescope, 38700 Santa Cruz de La Palma, Spain}

\begin{abstract}
We present the results of a search for red QSOs using a selection based on
optical imaging from SDSS and near-infrared imaging from UKIDSS. Our main goal
with the selection is to search for QSOs reddened by foreground dusty absorber
galaxies. For a sample of 58 candidates (including 20 objects fulfilling our
selection criteria that already have spectra in the SDSS) 46 (79\%) are
confirmed to be QSOs. The QSOs are predominantly dust-reddened except a handul
at redshifts $z\gtrsim3.5$.  However, the dust is most likely located in the
QSO host galaxies (and for two the reddening is primarily caused by Galactic
dust) rather than in intervening absorbers. More than half of the QSOs show
evidence of associated absorption (BAL-absorption). 4 (7\%) of the candidates
turned out to be late-type stars, and another 4 (7\%) are compact galaxies. We 
could not identify the 4 remaining objects. 
In terms of their
optical spectra the QSOs are similar to the QSOs selected in the FIRST-2MASS red Quasar
survey except they are on average  fainter, more distant and only two are
detected in the FIRST survey. As is usually done we estimate the
amount of extinction using the SDSS QSO template reddened by SMC-like dust.  It
is possible to get a good match to the observed (restframe ultraviolet) spectra,
but it is not possible to match the observed near-IR photometry from UKIDSS 
for nearly all the reddened QSOs.
The most likely reasons are that the SDSS QSO
template is too red at optical wavelengths due to contaminating host galaxy
light and that the assumed SMC extinction curve is too shallow.  Three of the
compact galaxies display old stellar populations with ages of several Gyr and
masses of about 10$^{10}$ M$_{\odot}$ (based on spectral-energy-distribution
modeling).  The inferred stellar densities in these galaxies exceed 10$^{10}$
M$_{\odot}$ kpc$^{-2}$, which is among the highest measured for early type
galaxies.  Our survey has demonstrated that selection of QSOs based on near-IR
photometry is an efficient way to select QSOs, including reddened QSOs, with
only small contamination from late-type stars and compact galaxies. This will
be useful with ongoing and future wide-field near-IR surveys such as the VISTA and EUCLID
surveys.
\end{abstract}


\section{Introduction}

QSO absorption line studies have led to major progress in our understanding of
the intergalactic medium (IGM) and of early galaxies \citep{Rauch98,Wolfe05}.
Using high-resolution spectroscopy of Damped Lyman-$\alpha$ Absorbers (DLAs)
detected against the light of background QSOs the chemical enrichment in
galaxies has been probed over most of cosmic history
\citep[e.g.,][]{Prochaska03}.  However, in order to interpret the available
data it is important to understand to which extent the data are biased against
dusty and hence likely more metal-rich sightlines.
This issue has been discussed extensively in the literature for more
than 30 years
\citep[e.g.,][]{Wright81,Pei1991,Pei99,Warren00,Vladilo05,Pontzen09,Erkal2012}.  One approach
by which to gauge the importance of the dust bias is to establish the
absorption statistics from a radio selected sample of QSOs. This approach has
been followed both in the CORALS survey \citep{Ellison01,Ellison04} and the
UCSD survey \citep{Jorgenson06}.  The largest of those studies is the UCSD
survey, which includes the CORALS data. These authors conclude that while
a dust bias is probably a minor effect, a four times larger sample is needed to
absolutely confirm an absence of significant dust bias. The later study of
\citet{Pontzen09} confirm that dust-bias is likely to be a small effect, but
they also find that the cosmic density of metals as measured from DLA surveys
could be underestimated by as much as a factor of 2 due to dust bias.
A factor of 2 is a sufficiently large deficit that it is interesting to pursue this
issue further.

There is positive evidence that dust bias does systematically remove the most
metal rich absorbers form the samples. The first direct
evidence came with the study by \citet{Pei1991} that the spectra of
QSOs with DLAs on average are redder than the spectra of QSOs without
intervening DLAs. This work has been somewhat controversial as some
studies confirmed the finding while other works rejected it 
\citep[e.g.,][]{Murphy04,Frank10,Khare12}. The latest work of
\citet{Khare12} does seem to find robust evidence for significant statistical
excess reddening of QSOs with foreground DLAs. Although it is difficult to
unambiguously establish excess reddening in individual sight-lines to
QSOs as the intrinsic slope cannot be determined a priori there are
also convincing detections of reddening towards individual systems with
foreground DLAs, sub-DLAs or strong metal-line absorbers \citep{Noterdaeme09b,
Noterdaeme10, Kaplan10,Fynbo11,Noterdaeme12,JianGuo2012}. Some of these QSOs
have colors that place them on the edge of or outside of the SDSS QSO
color selection. It is therefore reasonable to expect that there may be
a significant population of metal-rich absorbers that are under-represented
in the current DLA samples.

Motivated by the detection of the reddened QSOs towards metal-rich
absorbers mentioned above we decided to 
carry out a search for QSOs with red colors in order to establish if
some fraction of red QSOs could be red due to dusty foreground absorbers.
The search for red QSOs also has a long history
\citep[e.g.,][]{Webster95,Benn98,Warren00,Gregg02,Glikman04,Hopkins04,Glikman07,
Maddox08,Urrutia09,Banerji12,Glikman12}.
The detection
of red QSOs in most of these works relied on either radio or X-ray 
detections \citep[see][for an extensive discussion]{Warren00}.
However, recently large area surveys in the near-infrared has
made it possible to select QSOs based on near-infrared photometry alone
\citep{Warren00,Warren07,Peth11} and this is the approach 
we have adopted in this work. In Sect.~\ref{selection} we present
our selection criteria. In Sect.~\ref{Observations} and
Sect.~\ref{results} we describe
our observations and analysis of the first sample of candidate red QSOs
selected in the regions of the sky covered both by UKIDSS and SDSS (primarily
stripe 82).

\section{Selection criteria}
\label{selection}

Rather than trying to build an unbiased sample our approach is to make a
tailored search for red QSOs.
Our goal is to make a selection of QSOs that are
redder than the reddened QSO Q\,0918+1636 studied by \citet{Fynbo11}
in the SDSS $g - r$ and
$r - i$ colors. 
In our selection we take advantage of a recent comprehensive
study of the near-IR colors of QSOs found by matching SDSS and UKIDSS catalogs
\citep{Peth11}. Based on this study we select from regions on the sky covered
both by SDSS and UKIDSS (mainly stripe 82)
point sources (classified as such both in UKIDSS and SDSS) with
$J_{AB}-K_{AB}$ $>$ 0.24 (corresponding to $J_{Vega}-K_{Vega}$ $>$ 1.2) and
$i-K_{AB}$ $>$ 0.1 (corresponding to $i-K_{Vega}$ $>$ 2). This
leads to an efficient selection of QSOs and a sufficiently robust rejection of 
stars. In order to select reddened QSOs we next impose $0.8 < g -
r < 1.5$ and $r - i > 0.2$. The limit 
$g - r < 1.5$ comes from the fact that we wish to avoid 
objects that are too red and hence too faint in the optical to allow the
detection of the Lyman-$\alpha$ forest and in particular DLAs.  We demand
$J_{AB} < 19$ in order to have sufficient signal to allow for spectroscopy in
the optical (i.e. again to establish the presence of DLA and metal lines).  We
initially added a color cut of $Y-J > 0$ based on previous experience that
sources with $Y-J < 0$ had problems with the photometry in UKIDSS. However, we
subsequently dropped this criterion as some QSOs do have $Y-J < 0$, in
particular at redshifts between 2 and 3.

\begin{figure}
\resizebox{0.99\hsize}{!}{\includegraphics[angle=90,clip=true]{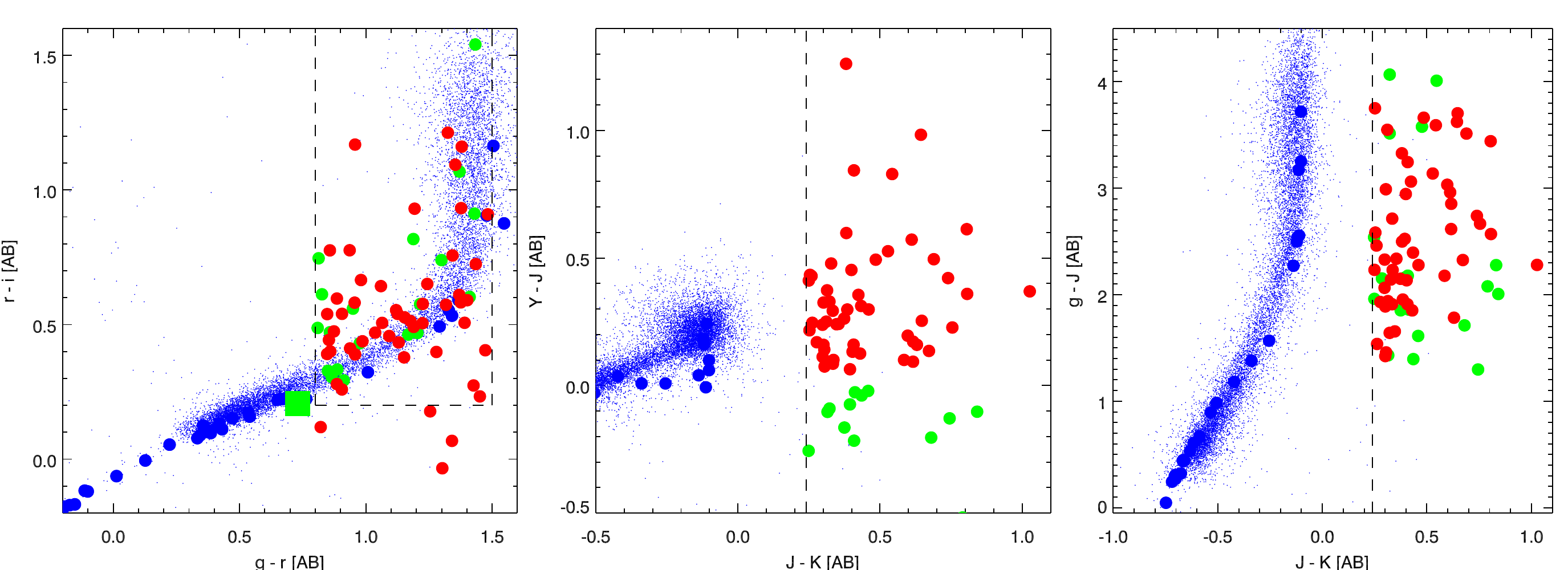}}
\caption[]{
color-color diagrams (in AB magnitudes) illustrating our selection of
candidate reddened quasars. Candidates are shown with big red ($Y-J>0$)
and green ($Y-J<0$) points. From left to right are shown:
$r-i$ vs $g-r$; $Y-J$ vs $J-K$ and $g-J$ vs $J-K$. The regions in color-color space in
which we selected reddened quasars are outlined by dashed lines. The green
square in the panel on the left represents the color of Q\,0918+1636. Also shown
are the colors of a random
sub-sample of all point sources in our catalogues (small blue points) and
a simulated stellar track (big blue points). With our
optical-infrared color selection we are able to robustly 
distinguish reddened quasars from the stellar locus. 
}
\label{cands}
\end{figure}

We found that 54 sources in the overlapping region of UKIDSS and SDSS in stripe 82
fulfill these selection criteria (see Fig.~\ref{cands}). 

By combining the star/galaxy classifiers based on the images in both the
SDSS and the UKIDSS Large Area Survey we are able to very efficiently remove
compact galaxies from our sample of candidates. We tested the 
efficiency of
galaxy removal from our sample with a set of 72464 objects with both Sloan
spectra and UKIDSS photometry. Out of the 39814 objects we selected as point
sources, only 69 (0.2\%) are galaxies based on the SDSS spectrum. Therefore,
our sample should contain less than 1 galaxy. Moreover, only 29 out
of more than 30000 objects spectroscopically classified as stars in the
overlapping region with the UKIDSS equatorial block (which is almost 10 times
larger than stripe 82) data fulfill our selection criteria. Finally, we looked
at the spectroscopically confirmed SDSS QSOs with UKIDSS photometry selected in
an area of about 1500 square degrees and only 36 (out of 21174) satisfied our
criteria. Of these, the majority are either BAL QSOs or red QSOs at 
$z<1$. Interestingly, one of these
QSOs is a red $z=3.62$ QSOs with an intervening DLA - the exact systems we
are looking for (SDSS\ J125306.73+130604.9). Given that stripe 82
has an area of only 250 square degrees we expect that our 54 candidates will
have a different mix of BAL QSOs and dust reddened QSOs. Independent of their
nature this survey will reveal the nature of a group of QSOs that are
missed in most QSO surveys so far.

\section{Observations and Data reduction}
\label{Observations}

In order to confirm or reject the candidate red QSOs we have carried out three
runs of spectroscopic follow-up of the candidates. In order to use observing
nights optimally we selected 7 additional targets at right ascensions around 8
hr with both SDSS and UKIDSS coverage. We also observed a single source with
$J_{AB} > 19$ and 4 sources that do not fulfill the r$_{AB}$ - i$_{AB}$ $>$ 0.2
color criterion. 

In August 2011 and January 2012 we observed 8 candidates at the Nordic Optical
Telescope (NOT) on La Palma equipped with the Andalucia Faint Object Spectrograph and
Camera (Alfosc). At the NOT we used grism 4, which covers the wavelength range
from about 3800 \AA \ to 9000 \AA \ at a resolution of about 350. Redwards of
about 7000 \AA \ the spectra are strongly affected by fringing so the effective
useful wavelength range for faint sources only extends to about 7000 \AA. In
November 2011 we observed another 40 candidates using the New Technology Telescope
(NTT) at La Silla (ESO) equipped with the ESO Faint Object Spectrograph and Camera 2 (EFOSC2). At
the NTT we used grism 6, which covers the wavelength range 3800-7980 \AA \ at a
resolution of about 450. This grism suffers from significant 2nd order
contamination redwards of 6200 \AA. Unfortunately we only realized this after
the observing run so no order sorter filter was used during the observations.
The spectra were taken aligning the slit at the parallactic angle. After the
release of SDSS DR8 we found that 20 of our candidates have been observed by
SDSS and of these 10 were also observed by us. In total we have spectra, either
from our run or from the SDSS, of 58 objects. Of these, 11 are not in our main
sample (i.e., outside stripe 82, have r$_{AB}$ - i$_{AB}$ < 0.2 or $J_{AB} >
19$). For 10 of our main targets from stripe 82 we were unable to
secure a spectrum during the mentioned two observing runs.

The full list of observed targets is provided in Table~\ref{sample}. Observing
conditions during the three runs appeared to be photometric, but with
strongly variable seeing. In particular on the first two nights of the NTT run
the seeing was poor (1.5--2.5 arcsec), whereas the seeing during the last night
of the NTT run and during both NOT runs was below 1 arcsec. During the two
nights with poor seeing we used slits with widths 1.5 and 2.0 arcsec, but we
still had significant slitloss. 
Based on the photometric measurements from SDSS we estimate that the slitloss 
was predominant between 30\% - 50\%. However, the slit loss was for some sources 
significant: the spectral flux was suppressed by more than a factor of two.

\begin{deluxetable}{@{}lllcccclc@{}}
\tablecaption{The sample. 
\label{sample}}
\tablewidth{0pt}
\tablehead{
\colhead{Target} & \colhead{RA(J\,2000)} & \colhead{Dec(J\,2000)} & \colhead{r$_\mathrm{SDSS}$} & \colhead{Telescope} & \colhead{Exptime} \\
         &          &            &   \colhead{(mag)} & & \colhead{(Sec)}    \\  
}
\startdata
CQ2143+0022 & 21 43 17.43 & +00 22 11.0 & 19.63 & SDSS & \\
CQ2144+0045 & 21 44 26.03 & +00 45 17.0 & 17.36 & NOT & 1200 \\ 
CQ2217+0033 & 22 17 14.27 & +00 33 46.5 & 20.16 & NTT & 1200 \\
CQ2227+0022 & 22 27 45.19 & +00 22 16.8 & 20.21 & NTT,SDSS & 1200 \\
CQ2241+0115 & 22 41 19.65 & +01 15 16.3 & 20.61 & NTT & 1800 \\
CQ2241$-$0012 & 22 41 15.86 & $-$00 12 39.8 & 21.12 & NTT,SDSS(DR9) & 2400 \\
CQ2254$-$0001\tablenotemark{1} & 22 54 19.23 & $-$00 01 55.1 & 19.57 & NTT,SDSS(DR9) & 900 \\
CQ2306+0108 & 23 06 39.65 & +01 08 55.2 & 19.36 & NTT,SDSS & 900 \\
CQ2316+0023 & 23 16 59.35 & +00 23 19.4 & 20.26 & NTT & 1200 \\
CQ2324$-$0105 & 23 24 31.32 & $-$00 53 42.6 & 21.18 & NTT & 1200 \\
CQ2342+0043 & 23 42 20.14 & +00 43 43.5 & 20.49 & NTT & 1200 \\
CQ2344$-$0001 & 23 44 58.71 & $-$00 01 38.7 & 19.92 & NTT & 1200 \\
CQ2347$-$0109 & 23 47 07.41 & $-$01 09 03.5 & 19.34 & SDSS & \\
CQ2355+0007 & 23 55 21.48 & +00 07 19.2 & 20.19 & NTT & 1200 \\
CQ2355$-$0041 & 23 55 26.78 & $-$00 41 54.2  & 19.36 & NOT,SDSS & 2000 \\
CQ0009$-$0020 & 00 09 33.18 & $-$00 20 22.3 & 19.42 & NTT & 900 \\
CQ0022+0020 & 00 22 55.07 & +00 20 55.1 & 20.57 & NTT & 2000 \\
CQ0027$-$0019\tablenotemark{1} & 00 27 58.75 & $-$00 19 23.7 & 19.20 & NOT,SDSS & 800 \\
CQ0043+0000 & 00 43 27.46 & +00 00 02.0 & 19.82 & SDSS & \\
CQ0046$-$0011 & 00 46 31.21 & $-$00 11 46.1 & 20.58 & NTT,SDSS(DR9) & 1200 \\
CQ0105+0000 & 01 05 03.39 & +00 00 34.1 & 20.12 & NTT & 1000 \\
CQ0107+0016 & 01 07 07.86 & +00 16 06.1 & 20.15 & NTT & 1000 \\
CQ0127+0114 & 01 27 02.52 & +01 14 12.6 & 17.83 & NOT& 600 \\
CQ0129$-$0059 & 01 29 25.80 & $-$00 59 00.1 & 19.33 & NTT & 900 \\
CQ0130+0013 & 01 30 11.42 & +00 13 14.5 & 19.29 & SDSS & \\
CQ0202+0010 & 02 02 48.74 & +00 10 46.5 & 20.91 & SDSS & \\
CQ0211+0030 & 02 11 53.71 & +00 30 45.9 & 20.23 & NTT & 1800 \\
CQ0212$-$0023 & 02 12 26.85 & $-$00 23 14.2 & 21.25 & NTT & 2400 \\
CQ0220$-$0107 & 02 20 07.65 & $-$01 07 31.0 & 18.54 & NTT,SDSS(DR9) & 900 \\
CQ0222$-$0019\tablenotemark{1} & 02 22 23.13 & $-$00 19 38.7 & 19.18 & NTT,SDSS(DR9) & 1000 \\
CQ0229$-$0029 & 02 29 07.71 & $-$00 29 09.5 & 20.27 & SDSS & \\
CQ0239+0115 & 02 39 24.90 & +01 15 15.6 & 20.56 & NTT & 2400 \\
CQ0242$-$0000 & 02 42 30.65 & $-$00 00 29.6 & 19.50 & NTT & 900 \\
CQ0247$-$0052 & 02 47 17.29 & $-$00 52 05.8 & 20.27 & NTT & 1500 \\
CQ0255+0048 & 02 55 18.58 & +00 48 47.55 & 19.02 & SDSS & \\
CQ0303+0105 & 03 03 40.01 & +01 05 57.3 & 19.73 & NTT,SDSS & 1200 \\
CQ0310+0055 & 03 10 36.84 & +00 55 21.71 & 19.73 & SDSS & \\
CQ0311+0103 & 03 11 42.35 & +01 03 55.2 & 20.12 & NTT,SDSS & 1200 \\
CQ0312+0032 & 03 12 14.24 & +00 32 35.2 & 19.42 & NTT & 900 \\
CQ0312+0035 & 03 12 13.37 & +00 35 54.6 & 19.81 & NTT & 1200 \\
CQ0321$-$0105 & 03 21 18.21 & $-$01 05 39.9 & 18.38 & SDSS & \\
CQ0326+0106 & 03 26 03.80 & +01 06 03.9 & 20.30 & NTT,SDSS & 1500 \\
CQ0327+0006 & 03 27 55.04 & +00 06 15.9 & 19.94 & NTT & 1200 \\
CQ0329$-$0057 & 03 29 19.89 & $-$00 57 15.5 & 19.98 & NTT,SDSS & 1200 \\
CQ0332$-$0013 & 03 32 48.39 & $-$00 13 15.3 & 19.71 & NTT & 1200 \\
CQ0336+0112 & 03 36 59.45 & +01 12 39.2 & 20.91 & NTT & 2400 \\
CQ0338+0004 & 03 38 06.76 & +00 04 33.9 & 20.02 & SDSS & \\
CQ0339+0022\tablenotemark{2} & 03 39 27.11 & +00 22 41.8  & 20.39 & NTT & 1200 \\
CQ0350$-$0031 & 03 50 53.29 & $-$00 31 14.7  & 19.83 & NOT,SDSS & 2000 \\
CQ0354$-$0012 & 03 54 18.16 & $-$00 12 56.7 & 19.14 & NTT & 4000 \\
CQ0354$-$0030 & 03 54 46.04 & $-$00 30 29.5 & 20.36 & NOT & 1800 \\
CQ0822+0004\tablenotemark{3} & 08 22 49.97 & +00 04 32.2  & 19.92 & NTT & 900 \\
CQ0822+0435\tablenotemark{3} & 08 22 02.32 & +04 35 26.0  & 19.50 & NOT & 1800 \\
CQ0826+0728\tablenotemark{3} & 08 26 24.71 & +07 28 20.8  & 20.09 & NOT & 1800 \\
CQ0831+0930\tablenotemark{3} & 08 31 42.36 & +09 30 29.8  & 19.56 & NTT & 900 \\
CQ0831$-$0022\tablenotemark{3} & 08 31 26.43 & $-$00 22 26.7  & 19.57 & NTT & 900 \\
CQ0832+0121\tablenotemark{3} & 08 32 29.34 & +01 21 05.4  & 19.80 & NTT & 1200 \\
CQ0832+0606\tablenotemark{3} & 08 32 29.02 & +06 06 00.6  & 19.73 & NTT & 1200 \\ 
\enddata
\tablenotetext{1}{r$_{AB}$ - i$_{AB}$ < 0.2}
\tablenotetext{2}{J$_{AB}>19$}
\tablenotetext{3}{Not in stripe 82.}
\end{deluxetable}
The spectra were processed using standard techniques for bias and flat-field
corrections and for wavelength calibration. Cosmic-rays were removed using the
software developed by \citet{LACosmic}. 1-dimensional spectra were extracted using
the optimal extraction technique described by \citet{Horne} as implemented in
IRAF\footnote{IRAF is distributed by the National Optical Astronomy
Observatory, which is operated by the Association of Universities for Research
in Astronomy (AURA) under cooperative agreement with the National Science
Foundation.}.  The spectra were
first flux calibrated using observations of the spectrophotometric standard star
Feige110 observed on the same nights as the science spectra. However, as the standard
star observed for the NTT run is very blue (it is a white dwarf) it had strong
2nd order contamination. Our targets, on the other hand, are red by selection and
hence have week 2nd order contamination. This flux calibration made the 
flux-calibrated spectra appear too blue at $\lambda > 6200$ \AA\ 
compared to the existing SDSS photometric measurements for the objects. 
We hence decided to use a red spectrophotometric standard observed on the 3rd of 
June 2011 (LTT7379). Flux-calibrating the NTT spectra using this standard star 
resulted in better agreement with the existing SDSS photometry, but our NTT
spectra still appear sligthly too blue redward of 6200 \AA. Blueward of 6200 \AA\
the responses inferred using the two different standard stars were in agreement
to better than 10 percent.

The spectra were corrected for
Galactic extinction using the extinction maps from \citet{Schlegel98}. Most of the
sources have low Galactic extinction, but 8 have $E(B-V) > 0.1$ and of these
two have $E(B-V) \approx 0.4$.

On February 9 and 10 2012 we observed the field of Q\,0918$+$1636 in the near-IR
using the NOT equipped with the near-IR camera NOTCam in the $Y$, $J$, $H$ and 
$Ks$ bands. The purpose of this observations was to infer if this object,
which basically is the motivation for our QSO search, has near-IR colours 
consistent with our selection criteria.
For photometric calibation we observed the field of RU149
for which $J$, $H$ and $K$ photometry is available in \citet{Hunt98} and 
$Y$-band photometry is available in the UKIDSS survey \citep{Hewett}.

\section{Results}
\label{results}

\subsection{Photometry}

For Q\,0918$+$1636 we derive the following magnitudes on the AB system:
$Y(AB) = 19.46 \pm 0.07$, $J(AB) = 19.32 \pm 0.07$, $H(AB) = 19.29 \pm 0.07$
and $K(AB) = 18.68 \pm 0.10$. In terms of its near-IR colours it falls 
well-within the near-IR color criteria of our sample.

\subsection{Spectroscopy}

In Fig.~\ref{spectra} in Appendix~\ref{appspecs} we show the 1-dimensional
spectra of the 58 candidates observed either by us or SDSS. In these diagrams
we plot both spectra and the photometry from SDSS ($g$, $r$, $i$, $z$) and
UKIDSS ($Y$, $J$, $H$, and $Ks$).  With the exception of four the nature of the
sources can be established robustly.  In Table~\ref{followuptab} we provide our
classification of the candidates in the cases where it could be established. We
also provide the SDSS QSO selection flags as defined in Table~1 of
\citet{Richards02}. We find four of the candidates to be late-type stars, 
four are
compact galaxies, 46 are most likely QSOs and for four candidates we were not
able to establish the nature of the object. 

For each QSO we determine redshifts by visually matching the spectrum with the
template spectrum of \citet{Vandenberk01} and \citet{Telfer02} including
reddening using the SMC, LMC or MW extinction curves as prescribed in
\citet{Pei92}. We here follow the approach of \citet{Urrutia09}.
We assume that the dust is located at the
redshift of the QSO as we find no evidence for absorber galaxies that could 
be responsible for the reddening. The inferred amounts of extinction should only be
considered indicative for a number of reasons. As mentioned, the
flux-calibration of our NTT spectra is affected by 2nd order contamination,
which as discussed will tend to make the NTT spectra appear too blue.
For 10 targets we have either NTT or NOT spectra and an SDSS spectrum.
For those cases the inferred amounts of extinction are typically slightly
higher in the SDSS spectra. In Table~\ref{followuptab} we provide the
redshifts and the amount of extinction $A_\mathrm{V}$ derived in this way.

For the majority of the candidate spectra the QSO template reddened by an SMC-like
extinction curve provide a good match to the data in the observed
optical range. However, the reddened QSO
template provides a poor match to the near-IR photometry from UKIDSS in
most cases. 
We will return to this point in Sect.~\ref{discussQSOs}.

In Appendix~\ref{notes} we provide our notes on each spectrum. 

\subsection{The confirmed QSOs}

The 46 confirmed QSOs are distributed broadly in redshift from $z=0.438$ to
$z=4.01$. There is a curious lack of QSOs between $z=2.6$ and $z=3.4$, which is
unfortunate as this is the redshift range we are primarily interested in for
the detecting of dusty intervening DLAs. The QSOs at $z>3.5$ appear relatively
normal, i.e. similar to the QSO template, and they have entered our selection
due to their strong Lyman-$\alpha$ forest blanketting. Two confirmed QSO are
relatively normal QSOs that are primarily reddened by substantial Galactic
extinction. 

At $z<3.5$ the majority of the remaining confirmed QSOs deviate from the
QSO template, in addition to their redder spectra, either by showing BAL
features or a combination of broad and narrow emission lines (e.g., 
CQ2342+0043 where \ion{C}{3} is broad and \ion{C}{4} is narrow). This fact
suggests that the source of the reddening is likely to be at the QSO 
redshift rather than in intervening galaxies. There is also no evidence for
either strong metal-line absorbers nor intervening galaxies close to the 
line-of-sight in the imaging data.

\begin{figure}
\plotone{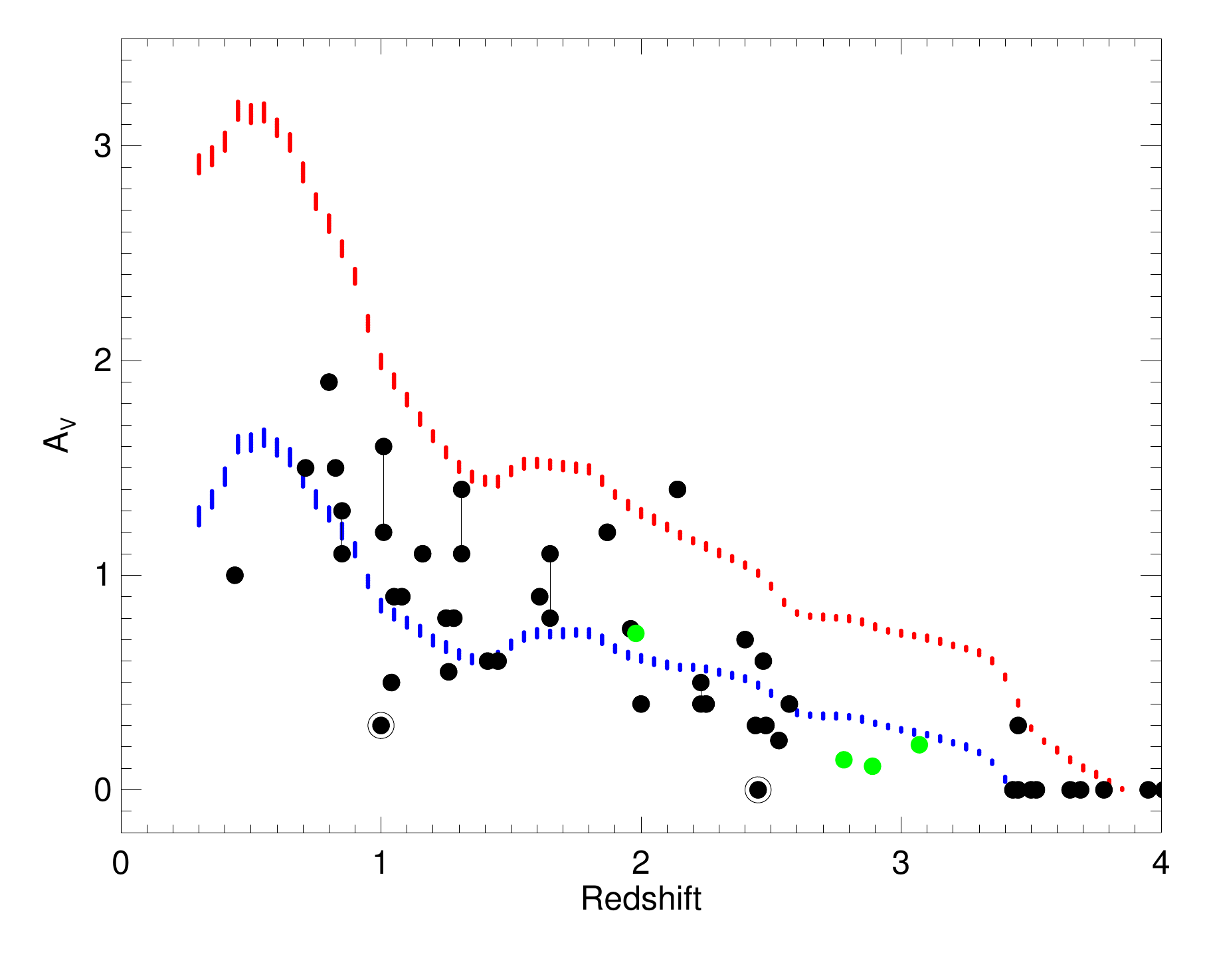}
\caption[]{$A_\mathrm{V}$ vs. redshift for our confirmed QSOs. The red and blue
points correspond to our color cuts in $g-r$ assuming the template QSO 
spectrum reddened by an SMC-like extinction curve.
The greens points correspond to Q0918+1636, Q1135$-$0010, Q1604+2203 and
Q1237+0647 that are all reddened by foreground absorbers \citep{Noterdaeme09b,
Noterdaeme10, Fynbo11, Noterdaeme12}. Note that for these
QSOs the amount of absorption is derived assuming that the dust is at the
redshift of the intervening absorbers, but the data points are plotted at the 
redshifts of the QSOs. Many of the data points fall below the blue
curve, most likely due to the fact that the strong emission lines are absorbed
away, but possibly also due to the fact that our flux calibration suffers from
2nd order contamination. Also, the flux calibration of our NTT spectra results
in too blue spectra due to 2nd order contamination and this will also bias our
measurements towards too low values of $A_\mathrm{V}$. For the objects observed
both by us and by SDSS we show the $A_\mathrm{V}$ deduced from both spectra
connected by a vertical line. The two encircled points represents the two
sources with strong Galactic absorption.  } \label{Avplot} \end{figure}

In Fig.~\ref{Avplot} we show how the derived $A_\mathrm{V}$ distribute with
redshift for the confirmed QSOs in our sample. As mentioned, these
$A_\mathrm{V}$ measurements are mainly indicative, but other surveys have
determined $A_\mathrm{V}$ in a similar manner \citep[e.g.,][]{Urrutia09} and
hence the values are still interesting for comparison. With red and blue
symbols we show the upper and lower limits on $A_\mathrm{V}$, respectively,
calculated using our color cuts in $g-r$ and by using the QSO template reddened
by the SMC extinction curve. Many of the points fall below the blue curve,
presumably due to an extra reddening caused by the fact that the emission lines
are absorbed, but also due to the problem with flux calibration of our NTT
spectra mentioned above.


\begin{deluxetable}{@{}lllllcclc@{}}
\tablecaption{The result of the spectroscopic follow-up. We here provide the QSO
selection flag from SDSS (if any) our classification (QSO, star, or galaxy),
the inferred redshift (when two values are given it is our measurement and the one
based on the SDSS spectrum), and extinction.
\label{followuptab}}
\tablewidth{0pt}
\tablehead{
\colhead{Target} & \colhead{SDSS\_flag} & Type & Redshift & $A_\mathrm{V}$\\
}
\startdata
CQ2143+0022 & QSO\_CAP & QSO & 1.26 & 0.55 \\
CQ2144+0045 & none & M-dwarf & & \\
CQ2217+0033 & none & M-dwarf & & \\
CQ2227+0022 & none & QSO & 2.23,2.24 & 0.4,0.8 \\
CQ2241+0115 & QSO\_HIZ & unknown &  &  \\
CQ2241$-$0012 & QSO\_HIZ & QSO &  & \\ 
CQ2254$-$0001 & QSO\_HIZ & QSO & 3.69,3.71 & 0.0 \\
CQ2306+0108 & STAR\_CARBON, QSO\_HIZ & QSO & 3.65,3.64 & 0.15,0.0 \\ 
CQ2316+0023 & none & QSO & 2.1? & \\
CQ2324$-$0105 & none & QSO & 2.25? &  \\
CQ2342+0043 & QSO\_REJECT & QSO & 1.65,1.65 & 0.8,1.1 \\
CQ2344$-$0001 & none & QSO & 1.04 & 0.5 \\
CQ2347$-$0109 & QSO\_SKIRT & QSO & 1.08 & 0.80 \\
CQ2355$-$0041 & QSO\_HIZ & QSO & 1.01,1.01 & 1.0,1.6 \\ 
CQ2355+0007 & none & unknown & & \\
CQ0009$-$0020 & none & Galaxy & 0.387 & \\ 
CQ0022+0020 & none & QSO & 0.80? & 1.9 \\
CQ0027$-$0019 & QSO\_CAP, QSO\_HIZ & QSO & 3.52,3.52 & 0.0,0.0 \\ 
CQ0043+0000 & QSO\_FIRST\_CAP & STAR? & & \\
CQ0046$-$0011 & none & QSO & 2.44,2.467 & 0.3 \\
CQ0105+0000 & none & Galaxy & 0.278 & \\ 
CQ0107+0016 & none & QSO & 2.47 & 0.6  \\
CQ0127+0114 & QSO\_FIRST\_CAP & QSO & 1.16,1.14 & 1.1,1.1 \\
CQ0129$-$0059 & QSO\_FIRST\_CAP, QSO\_HIZ & QSO & 0.71 & 1.5  \\ 
CQ0130+0013 & QSO\_REJECT & QSO & 1.05 & 0.8 \\
CQ0202+0010 & QSO\_REJECT, QSO\_HIZ & QSO & 1.61 & 0.9 \\
CQ0211+0030 & ROSAT\_D, ROSAT\_E & QSO & 3.45 & 0.30 \\
CQ0212$-$0023 & QSO\_REJECT & QSO & 1.87? & 1.2 \\
CQ0220$-$0107 & none & QSO & 3.43,3.467 & 0.0  \\
CQ0222$-$0019 & QSO\_HIZ & QSO & 3.95,3.947 & 0.0 \\
CQ0229$-$0029 & QSO\_FAINT, QSO\_HIZ & QSO & 2.14,1.97 & 1.40 \\
CQ0239+0115 & QSO\_REJECT & QSO? & 0.867 & 1.70 \\ 
CQ0242$-$0000 & STAR\_CARBON, QSO\_HIZ & QSO & 2.48 & 0.30 \\
CQ0247$-$0052 & none & QSO & 0.825 & 1.5 \\
CQ0255+0048 & QSO\_HIZ & QSO & 4.01 & 0.0 \\
CQ0303+0105 & QSO\_HIZ & QSO & 3.45 & 0.0  \\
CQ0310+0055 & QSO\_FAINT, QSO\_HIZ & QSO & 3.78 & 0.0 \\
CQ0311+0103 & QSO\_FAINT, QSO\_HIZ & QSO & unknown, 3.27 & 0.30? \\
CQ0312+0032 & none & QSO & 1.25 & 0.8 \\ 
CQ0312+0035 & none & QSO & 1.28 & 0.8 \\
CQ0321$-$0105 & QSO\_HIZ & QSO & 2.40 & 0.7 \\
CQ0326+0106 & SERENDIP\_DISTANT & QSO & 0.85,0.85 & 1.0,1.3 \\
CQ0327+0006 & none & QSO & 3.50 & 0.15 \\
CQ0329$-$0057 &SERENDIP\_MANUAL  & QSO & 1.31,1.31 & 1.1,1.4 \\
CQ0332$-$0013 & none & QSO & 0.438 & 1.0 \\
CQ0336+0112 & none & M-dwarf & & \\
CQ0338+0004 & QSO\_CAP & QSO & 1.45 & 0.6 \\
CQ0339+0022 & none & QSO & 1.41 & 0.5 \\
CQ0350$-$0031 & QSO\_CAP & QSO & 2.00,2.00 & 0.4,0.4 \\
CQ0354$-$0012 & none & QSO & 2.45 & 0.0 \\
CQ0354$-$0030 & none & QSO & 1.00 & 0.3 \\
CQ0822+0004 & QSO\_HIZ & Galaxy & 0.378 & \\
CQ0822+0435 & none & M-dwarf &      & \\
CQ0826+0728 & none & QSO     & 1.77 & 0.6 \\
CQ0831+0930 & none & QSO & 1.96 & 0.75 \\
CQ0831$-$0022 & none & QSO & 2.53 & 0.23 \\
CQ0832+0121 & none & Galaxy & 0.166 & \\
CQ0832+0606 & none & QSO & 2.57 & 0.4 \\
\enddata
\end{deluxetable}

\begin{figure} \plotone{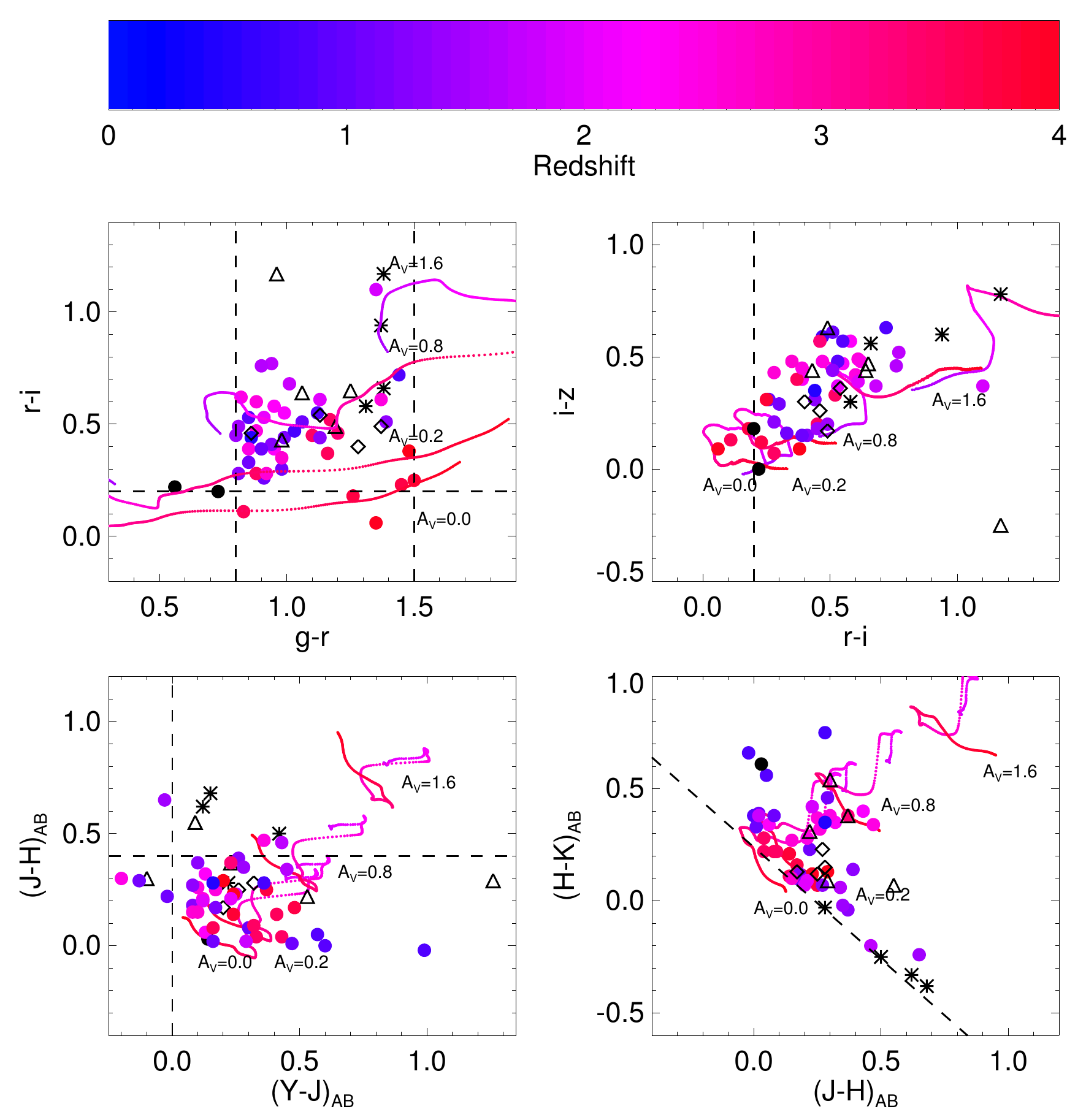} 
\caption[]{
Color-color diagrams based on the SDSS and UKIDSS photometry. Filled black circles:
Q\,0918+1636 and Q\,1135$-$0010. In the lower two panels we only include
Q\,0918+1636 as we do not have near-IR photometry for Q\,1135$-$0010.  Filled
colored points: QSOs with measured redshifts. Asterisk: M-dwarfs. Rhombes:
Galaxies.  Triangles: objects of unknown nature or unknown redshift. The dashed
lines mark the boundaries of our selection criteria. The four colored 
tracks marked by
small colored points outline the expected colors of the QSO template in the
redshift range from 1.5 to 4.0 and reddened
by $A_\mathrm{V} = 0, 0.2, 0.8$ and 1.6 mag assuming that the dust is at the QSO
redshift.
}
\label{colplot}
\end{figure}

\begin{figure}
\plotone{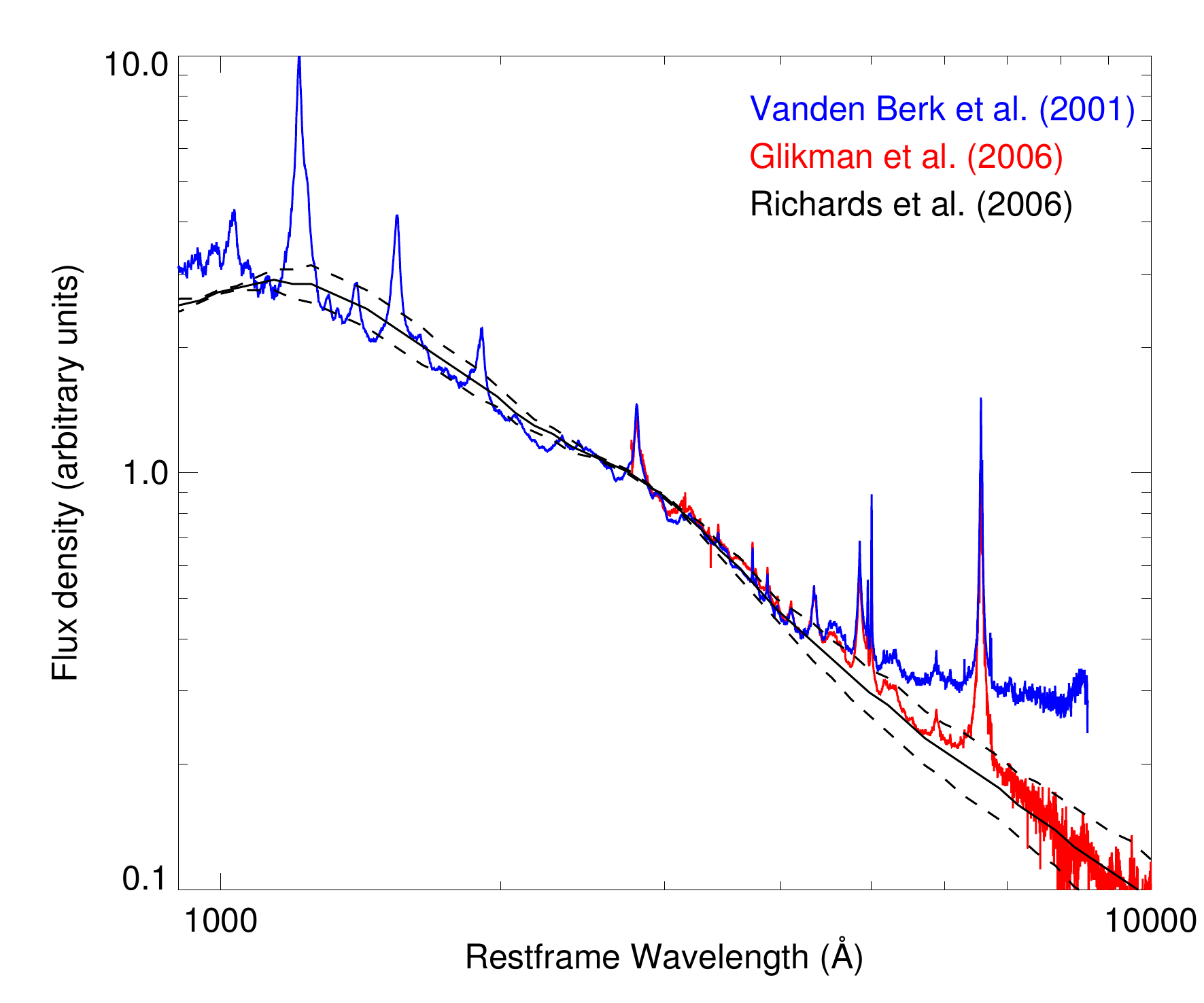}
\caption[]{
Here we compare the QSO templates of \citet{Vandenberk01}, \citet{Glikman06}
and \citet{Richards06}. For the \citet{Richards06} templates we also provide
the $\pm 1\sigma$ curves relative to the average template. As seen, 
the \citet{Vandenberk01} template has 
excess emission in the restframe optical range presumably due to 
host galaxy contamination.
}
\label{compocompare}
\end{figure}

\begin{figure}
\plotone{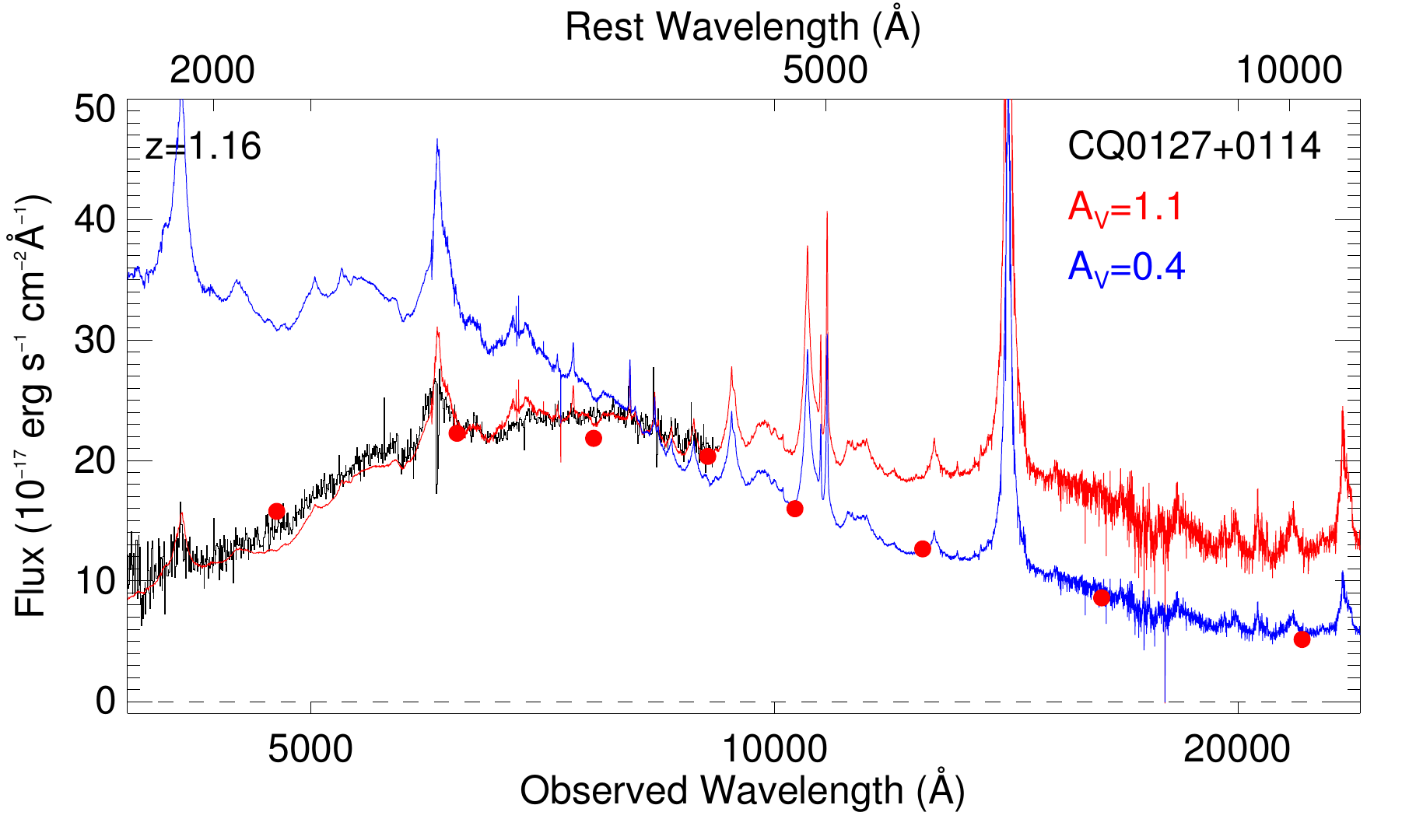}
\caption[]{
Here we use the merged QSO template described in the text to explore if the shape
of the spectral-energy-distributions can be matched using this template (without
host galaxy contamination)
and the SMC extinction curve in the
specific example of CQ0127+0114. The red solid curve is our best match to the 
optical data whereas the solid blue curve is our best fit to the near-IR data.
As seen, it is possible to match the optical and near-IR regions separately,
with different amounts of extinction, but it is not possible to match the
full spectral range. An extinction curve with a steep UV rise is needed to 
explain the full spectral energy distribution.
}
\label{compoMplot}
\end{figure}

\subsection{Stars}

Based on \citet[][their Table~9]{Hewett} M-dwarfs are expected to have
$J-K<1.2$ corresponding to $(J-K)_\mathrm{AB}<0.24$. In the compilation of
\citet{Hewett} there is one exception, namely an M8.5 dwarf with $J-K = 1.25$.
In the lower right panel of Fig.~\ref{colplot} the four
stars in our survey all lie very close to the cut-off in $J-K$. The colors and
spectra of the two stars that are M-dwarfs indicate that they are relatively
early type M-dwarfs (from M0 to M6) for which $J-K < 1.0$. They have presumably
entered our sample either due to photometric errors or due to infrared excess
emission, i.e. from circumstellar disks.

\subsection{Galaxies}
Three of the four galaxies we have identified are old and very compact
stellar systems (CQ0105+0000, CQ0822+0004, and CQ0832+0121). The
fourth object identified as a galaxy, CQ0009$-$0020, displays strong
forbidden Oxygen and Neon emission lines. It also
seems to have a relatively old underlying stellar continuum with a clear
4000-\AA \ break. The galaxy fraction among our candidates is somewhat higher
than expected, but is still sufficiently low that galaxy contamination is 
not a major obstacle for QSO searches based on this method.

\section{Discussion and Conclusions}

\subsection{Implications for QSO searches}
\label{discussQSOs}

79\% of our candidate dust-reddened QSOs are confirmed QSOs.
The contamination from stars and compact galaxies are each about $\sim$7\% 
and the
remaining 7\% remains unclassified. The confirmed QSOs have a very broad
redshift range from 0.4 to 4.0. Most of the QSOs appear significantly
dust-reddened with an amount of extinction broadly consistent with our
color-selection criterion on $g-r$. In most cases the
spectral shape of the QSOs in the observed optical range can be well-matched by
the template QSO spectrum reddened by an SMC-like extinction curve assuming
that the dust is at the QSO redshift. However, that same model fails for most
of the reddened QSOs in the near-IR. We consider the two most likely reasons
for this failure in the following. First, there is a problem with the QSO
template which, as also discussed by \citet{Vandenberk01}, most likely contains
significant contamination from host galaxy light redward of 5000 \AA.
Second, it is plausible that the SMC extinction curve is too shallow.
There is growing evidence that the extinction curve towards the Galactic center
is significantly steeper in the optical and near-IR than the standard MW
extinction curve parametrisation \citep[e.g.,][]{Sumi2004,
Nishiyama2008,Nishiyama2009,Gosling2009}. Given that we obviously also probe
regions towards galactic centers it is plausible that these environments have
dust with similar extinction properties, i.e. very steep extinction curves, as
towards the Galactic center. Steep extinction curves have also been 
identified against GRB afterglow light \citep{Tayyaba12}. 

To investigate the role of the QSO template
we follow \citet{JianGuo2012} and build a template by merging the
\citet{Vandenberk01} template at $\lambda_\mathrm{rest} < 3000$ \AA \
with the \citet{Glikman06} near-IR template at $\lambda_\mathrm{rest}
> 3000$ \AA. In Fig.~\ref{compocompare} we compare this template
with the \citet{Vandenberk01} template and with the QSO templates 
from \citet{Richards06}. As can be seen, the templates agree well in
the reframe UV, but at $\lambda_\mathrm{rest} > 4000$ \AA \ the 
\citet{Vandenberk01} template clearly has excess light presumably due
to host contamination.
In Fig.~\ref{compoMplot} we re-plot the spectrum of CQ0127+0114 (here the 
SDSS spectrum) and attempt to match its spectral-energy-distribution using 
the merged QSO template described above. As can be seen here, we can match 
the optical and near-IR sections separately with different amounts of 
extinction, but not the full spectral range.
We plan to investigate the second point, i.e. the question if
an extinction curve similar to that observed towards the Galactic centre can be
inferred from our spectra, further in a future work.

More than half of the confirmed QSOs show signs of 
associated absorption (BAL, low-BALs, P-Cygni profiles) or unusual emission
line profiles, i.e. broad \ion{C}{3} emission and narrow \ion{C}{4} 
emission (e.g., CQ2342+0043, CQ0826+0728). 

It is interesting to compare our confirmed QSOs with the QSOs identified
in the FIRST-2MASS red Quasar survey of \citet{Urrutia09}. Their survey
is targeting point sources detected by both FIRST and 2MASS, whereas
only two of our targets are FIRST sources and the majority of our
targets are too faint to be detected by 2MASS. For these reasons,
the QSOs in our sample extend out to higher redshifts than the QSOs
in the FIRST-2MASS survey. 
The FIRST-2MASS QSO color distribution is relatively wider.
As an example, $g-r$ ranges from 0 to 2.5 for the FIRST-2MASS QSOs, 
compared to our (selected) range of colors, $0.8<g-r<1.5$. Our study confirms
their finding that QSOs with BAL-features are very common among
dust-reddened QSOs and extend the result to QSOs at higher redshifts.
A study of point sources selected only by $J-K>1.2$ and a $K$-band
limit of about 19 would be very interesting as it would give a broad
census of the AGN population at $0.5<z<3.5$. The many AGN 
with $J-K<1.2$ are predominantly at low redshift (where the
host galaxy contributes to the near-IR photometry) or at 
$z\approx4$, where the \ion{Mg}{2} emission enters the 
$J$-band. Here it is also interesting that \citet{Banerji12} 
have done a study of $J-K>2.5$ 
point sources from UKIDSS and these are all heavily reddened
QSOs.

The incompleteness of classic color/morphology selection techniques
of QSO samples was also addressed by the purely magnitude
limited VVDS QSO sample \citep{Gavignaud2006, Bongiorno2007} 
where it was shown that faint QSO samples may miss up to
35\% of the population due partly to the color selection and
partly to the requirement that the candidates must be point-sources.
We have shown here that at least part of this missed
population can be understood as dust-reddened QSOs. 
 
About half of our confirmed QSOs were not flagged as QSOs based 
on the SDSS photometry and our survey therefore provides some insight
into the completeness of the SDSS QSO catalog. In total, there are 
10925 spectroscopically confirmed quasars in Stripe82 in DR8, the 
latest data release. However, only for a subset of these is UKIDSS 
photometry available. Of the spectroscopically confirmed quasars 
in Stripe82 with UKIDSS photometry 2246 have $J_{AB} < 19.0$. Of these
only 927 are point sources in both UKIDSS and SDSS. Only 14 of the
927 fullfil our additional color criteria. We have in this survey
detected 45 QSOs with $J_{AB} < 19.0$ in stripe 82 and of these only 4 were
already spectroscopically confirmed by SDSS. Hence, it is clear that
SDSS is incomplete for red QSOs, but these constitute a tiny
fraction of all QSOs with $J_{AB} < 19.0$.

Finally, we note that a very similar study to ours, i.e. a QSO 
search based on SDSS and UKIDSS photometry, has recently been 
submitted for publication \citep{Wu2012}. Their study has a 
slightly different aim, namely a targetted search for QSOs at
$z=2.2$ - 3.5.

\subsection{Implications for QSO host galaxies}
Our survey has found that there exists a subsample of
dust-reddened QSOs which is missed by classic QSO selection
methods. This result is consistent with those of \citet{Gavignaud2006}
and \citet{Urrutia09}.

We have further found that the reddening is in general caused
by the QSO host galaxy rather than intervening absorbers, and
that the reddened galaxies only make up a comparatively small
fraction of the underlying QSO host galaxy population. The
implication is that either QSO host galaxies in general contain
very little dust, or the QSO itself is able to create a dust-free
sightline along the emission cone.

From Fig.~\ref{Avplot} it is seen that there is an apparent drop
in the number of reddened
hosts at redshifts larger than $z=2.6$ (note that the green points
in Fig.~\ref{Avplot} do not belong to the sample). At redshifts larger than
$z=3.5$ our color selection is not well tuned to find reddened
QSOs, but in the range $z=2.6-3.5$ our selection is tuned for
$A_\mathrm{V}$ in the range 0.2 to 0.8 and we find only a single
reddened host at $z=3.45$. At redshifts around 1-2 our sample
contains 16-17 reddened hosts per unit redshift so the sudden drop
at $z=2.6$ to 1.1 per unit redshift appears to be
significant. This could possibly indicate that QSO host galaxies at $z>2.6$
contain less dust than those below $z=2.6$. 
A similar result has
been reported for high redshift galaxies selected by their
Ly$\alpha$ emission line \citep{Nilsson2009} where it was found
that there is a significant evolution in the dust properties
of Ly$\alpha$ emitters from $z=3$ to $z=2$. In particular it has been
found that there is a very sharp transition in the ULIRG
fraction of Ly$\alpha$ emitters at a redshift of z$\approx 2.5$
\citep{NilssonMoller}. The drop we see in the dust
properties of QSO hosts at $z=2.6$ appears to be equally sudden
and it is possible that the two observations are
related to a common evolutionary event. However, we obviously need a larger
sample to confirm this tentative result.

\subsection{Implications for the study of compact Galaxies}
Three of the galaxies we have detected are old stellar systems (CQ0105+0000,
CQ0822+0004, and CQ0832+0121). The spectrum of CQ0822+0004 has Balmer
absorption lines and hence most likely has an age of about a Gyr, whereas the
other two have older stellar populations. We use the 
LePhare software \citep{Arnouts99,Ilbert06} to fit to the SDSS + UKIDSS
photometry fixing the redshift to the measured values. Based thereon we infer stellar
masses and stellar ages of several Gyr (the exact value depending on the 
assumed metallicity) and masses of the order 10$^{10}$
M$_{\odot}$. Interestingly, they are all point sources - even in the $K$-band,
where the seeing is about 0.6 arcsec in the UKIDSS images. This implies that
the galaxies must have stellar densities in excess of 10$^{10}$ M$_{\odot}$
kpc$^{-2}$, which is similar to the compact Distant Red Galaxies 
found at redshifts around 2 \citep[e.g.,][]{VanDokkum98} and among the
highest measured at any redshift \citep{Franx08}. However, the
stellar masses of the compact galaxies found at high redshifts are
significantly larger than for the three galaxies found here. \citet{Taylor10}
find that massive, compact galaxies are much rarer in the local Universe than
at $z\approx2$.

\subsection{Dusty DLAs}
We did not find any obvious cases of QSOs being reddened by foreground 
dusty absorbers and this confirms that such absorbers are rare. We
are currently working on refining our selection criteria with the goal
of selecting fewer strongly reddened $z<2$ QSOs and fewer $z>3.5$ unreddened
QSOs. 


As also discussed by \citet{Wu2012} the outlook for impending advance 
for this kind of QSO surveys based on ongoing near-IR surveys like
the VISTA suveys \citep[e.g.,][]{Henry2012,Jarvis2012,Viking} or the future
EUCLID survey \citep{Euclid} is very good.

\acknowledgements
We thank our anonymous referee for a constructive and helpful report and 
T. Zafar, K. Denny, J. Hjorth, B. Milvang-Jensen, A. de Ugarte Postigo and Richard 
McMahon for helpful discussions. JPUF
acknowledges support from the ERC-StG grant EGGS-278202. The Dark cosmology
centre is funded by the DNRF.  Funding for the SDSS and SDSS-II has been
provided by the Alfred P. Sloan Foundation, the Participating Institutions, the
National Science Foundation, the U.S. Department of Energy, the National
Aeronautics and Space Administration, the Japanese Monbukagakusho, the Max
Planck Society, and the Higher Education Funding Council for England. The SDSS
Web Site is http://www.sdss.org/.  The SDSS is managed by the Astrophysical
Research Consortium for the Participating Institutions. The Participating
Institutions are the American Museum of Natural History, Astrophysical
Institute Potsdam, University of Basel, University of Cambridge, Case Western
Reserve University, University of Chicago, Drexel University, Fermilab, the
Institute for Advanced Study, the Japan Participation Group, Johns Hopkins
University, the Joint Institute for Nuclear Astrophysics, the Kavli Institute
for Particle Astrophysics and Cosmology, the Korean Scientist Group, the
Chinese Academy of Sciences (LAMOST), Los Alamos National Laboratory, the
Max-Planck-Institute for Astronomy (MPIA), the Max-Planck-Institute for
Astrophysics (MPA), New Mexico State University, Ohio State University,
University of Pittsburgh, University of Portsmouth, Princeton University, the
United States Naval Observatory, and the University of Washington. We
acknowledge the use of UKIDSS data.  The United Kingdom Infrared Telescope is
operated by the Joint Astronomy Centre on behalf of the Science and Technology
Facilities Council of the U.K.

\appendix 
\section{The spectra}
\label{appspecs}
\begin{figure}
\plotone{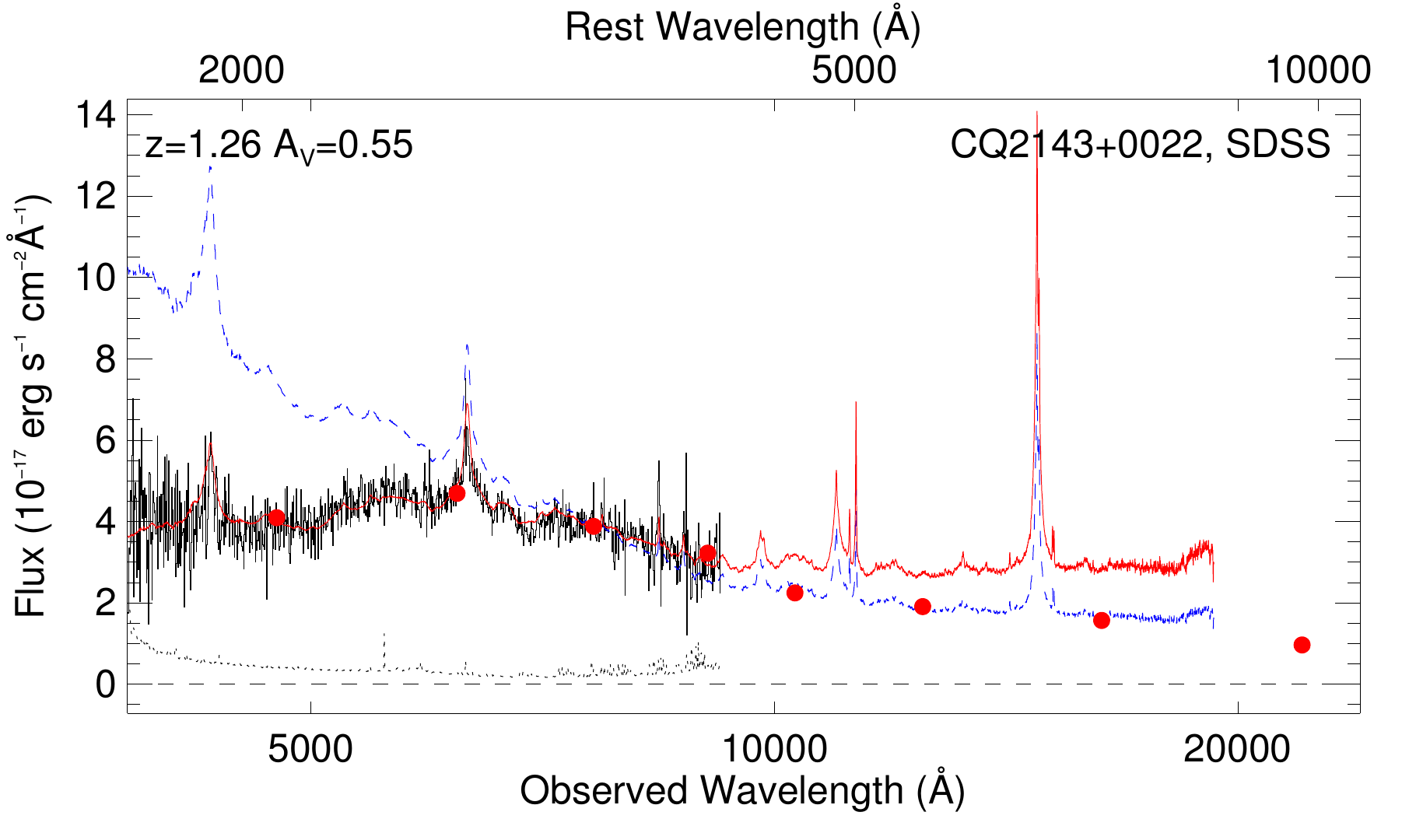}
\caption{Shown are 1-dimensional spectra for the 58 candidate QSOs
observed in our survey or by the SDSS. For each candidate the observed spectrum is plotted
(solid) and the error spectrum (dotted line) is shown.
In the upper left corner the estimated redshift and restframe $V$-band
extinction is provided. With a blue dashed line we show the composite QSO spectrum 
from \citet{Vandenberk01} and \citet{Telfer02} redshifted to the estimated 
redshift and with 
a solid red line the redshifted composite spectrum reddened by the indicated
amount of extinction. Overplotted with filled circles is the SDSS and
UKIDSS photometric
points. The NOT and NTT spectra have been scaled to match the $r$-band 
photometric point from SDSS. Unless otherwise noted we have assumed an SMC-like
extinction curve. Note that the spectra have not been corrected for
telluric absorption. The SDSS spectra have been binned down by a factor of
4 for clarity.
\label{spectra}}
\end{figure}
\begin{figure}
\plotone{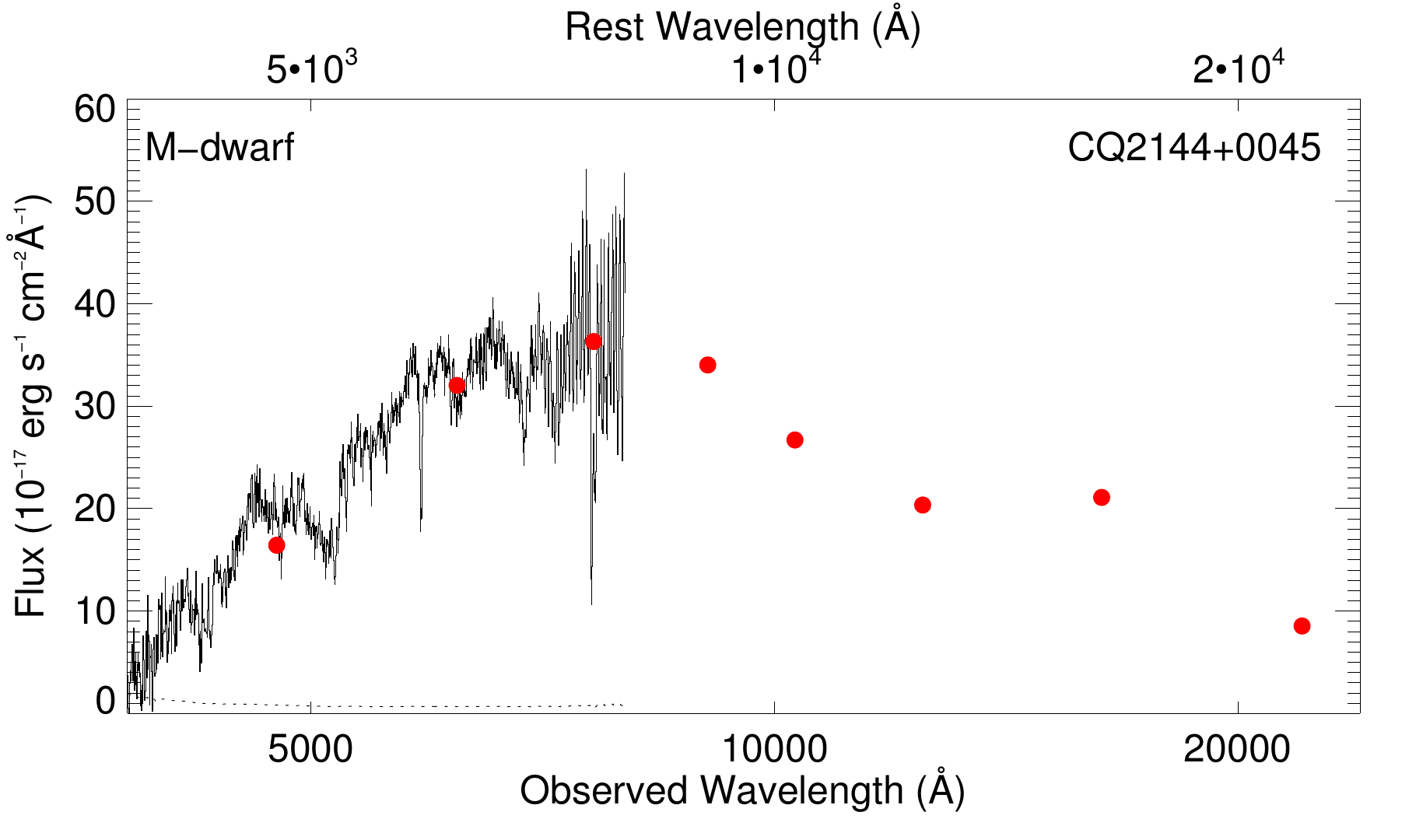}
\end{figure}
\begin{figure}
\plotone{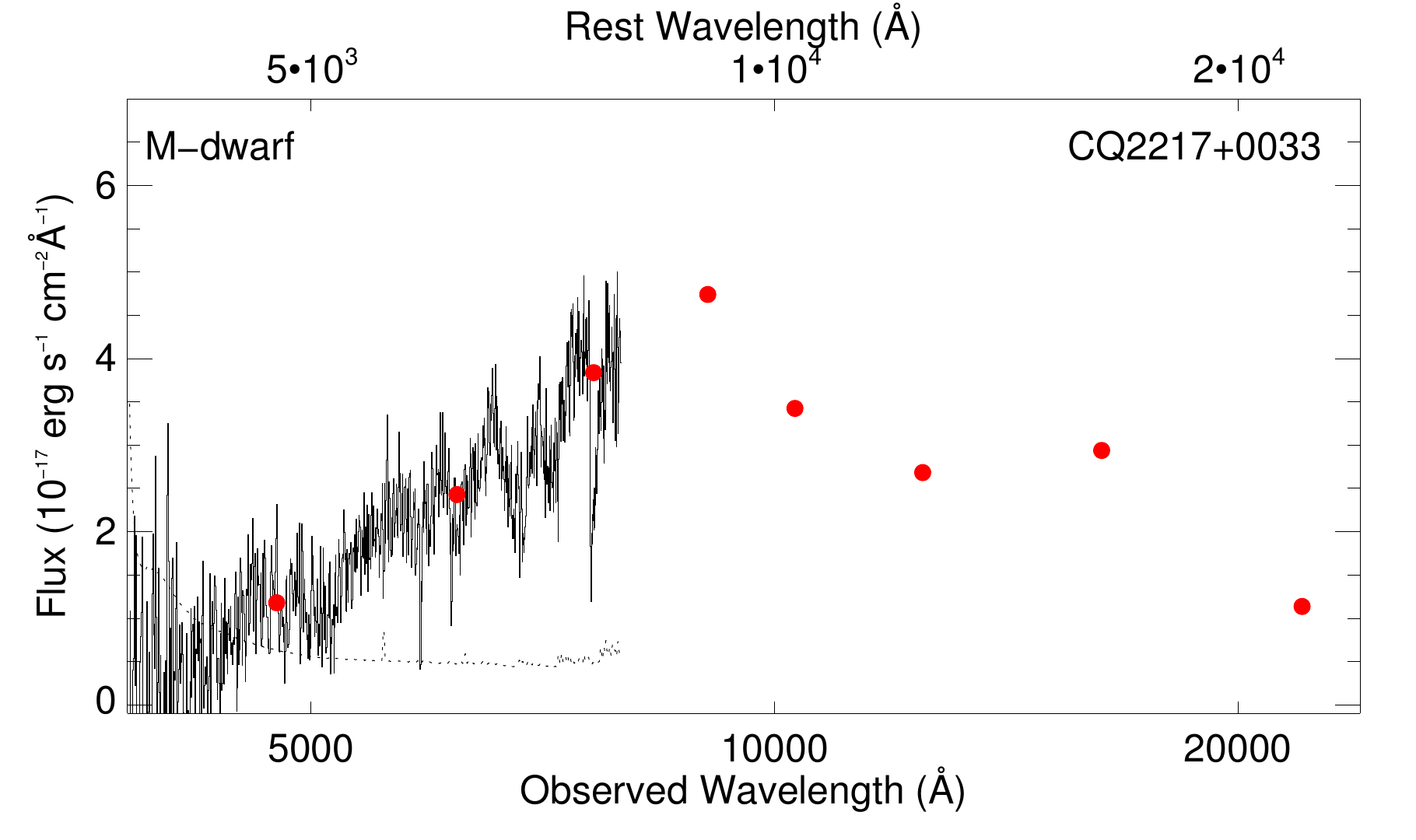}
\end{figure}
\begin{figure}
\plotone{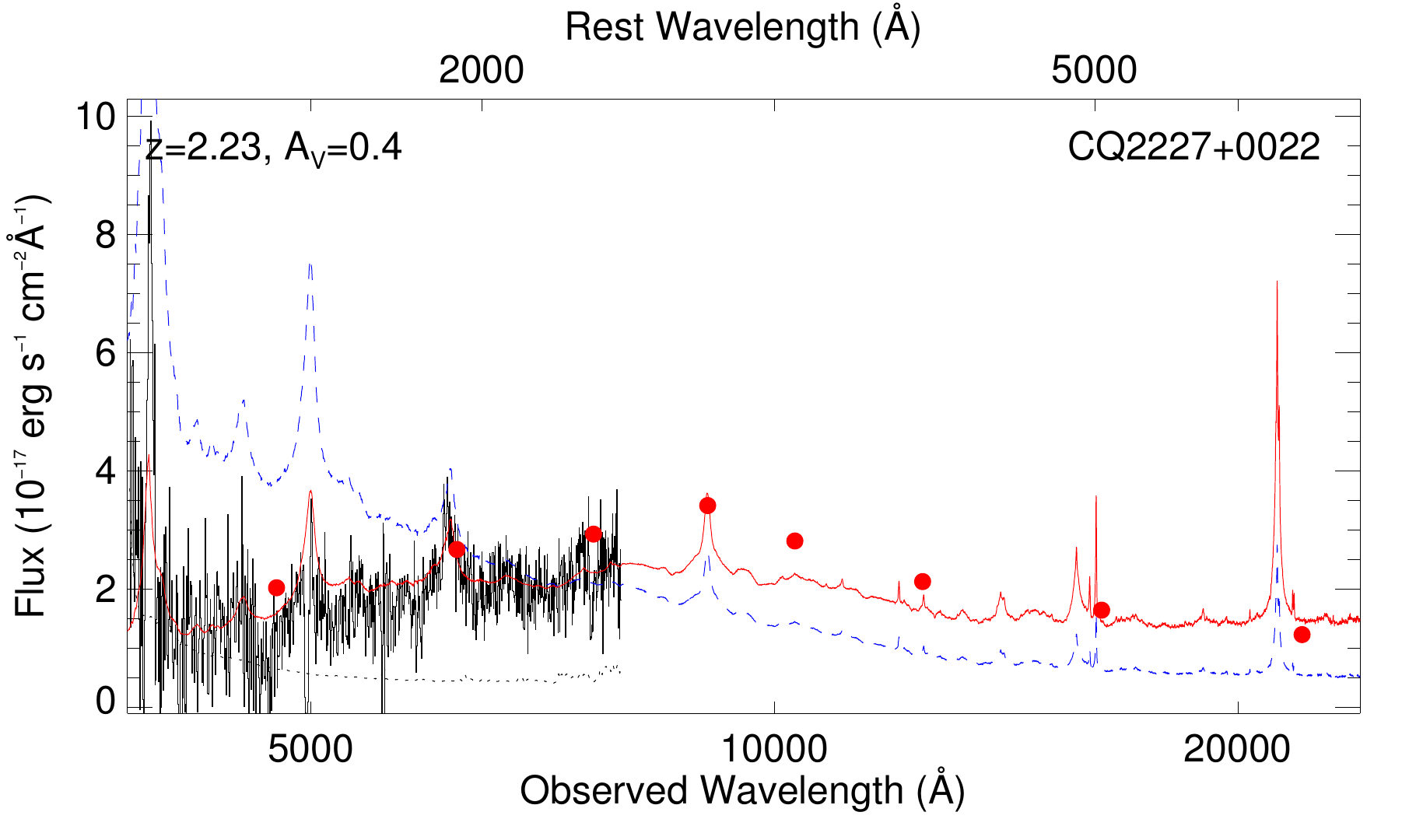}
\end{figure}
\begin{figure}
\plotone{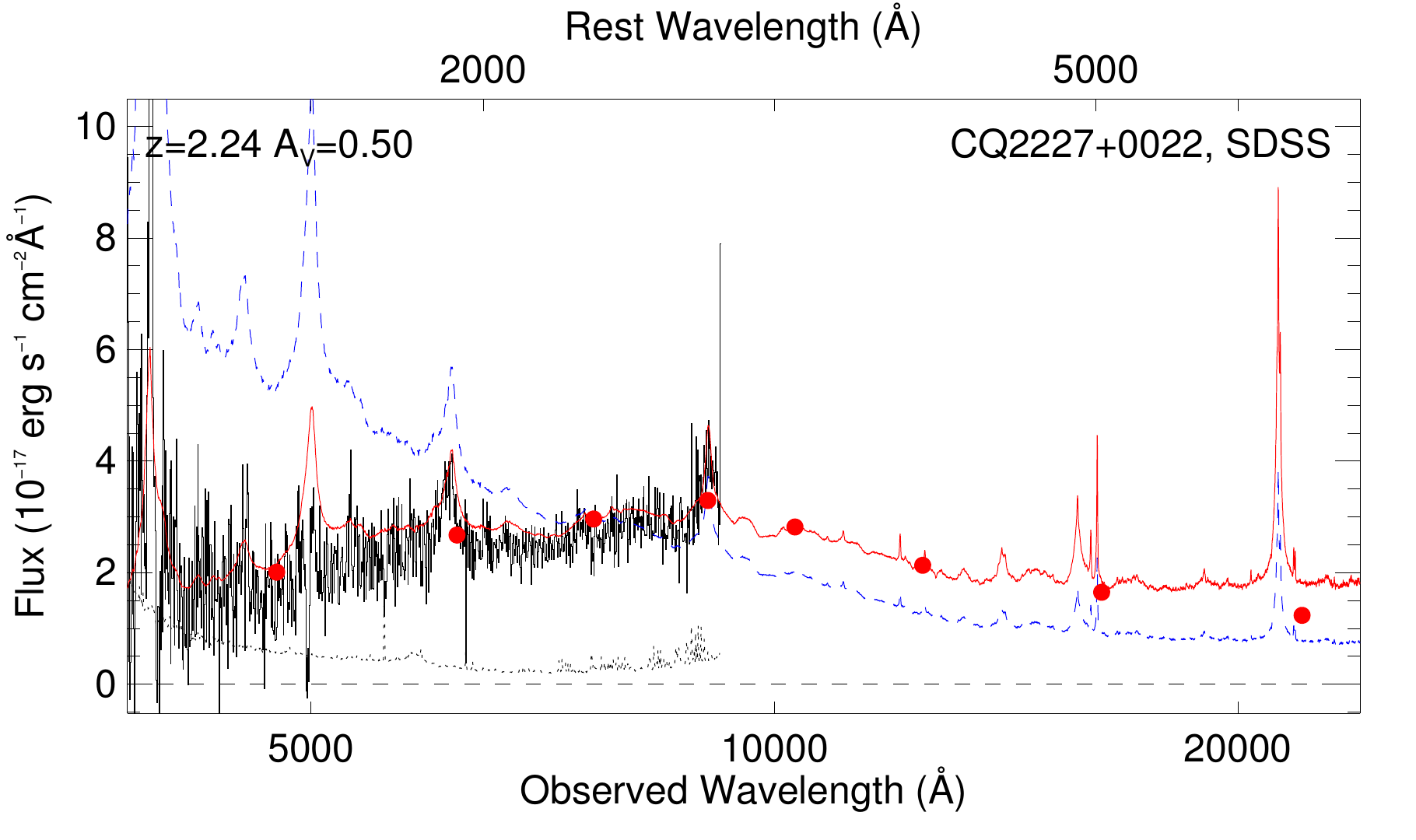}
\end{figure}
\begin{figure}
\plotone{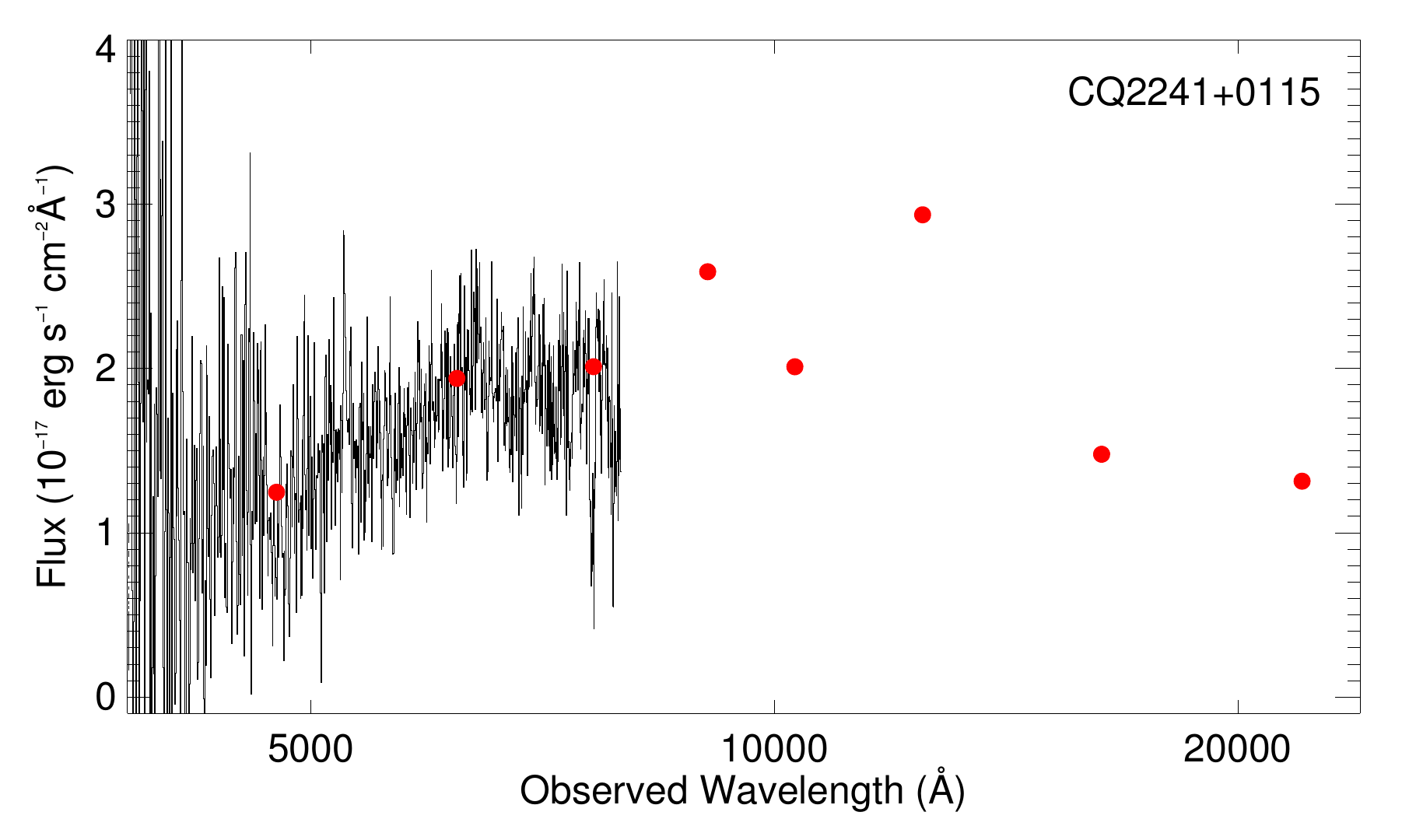}
\end{figure}
\begin{figure}
\plotone{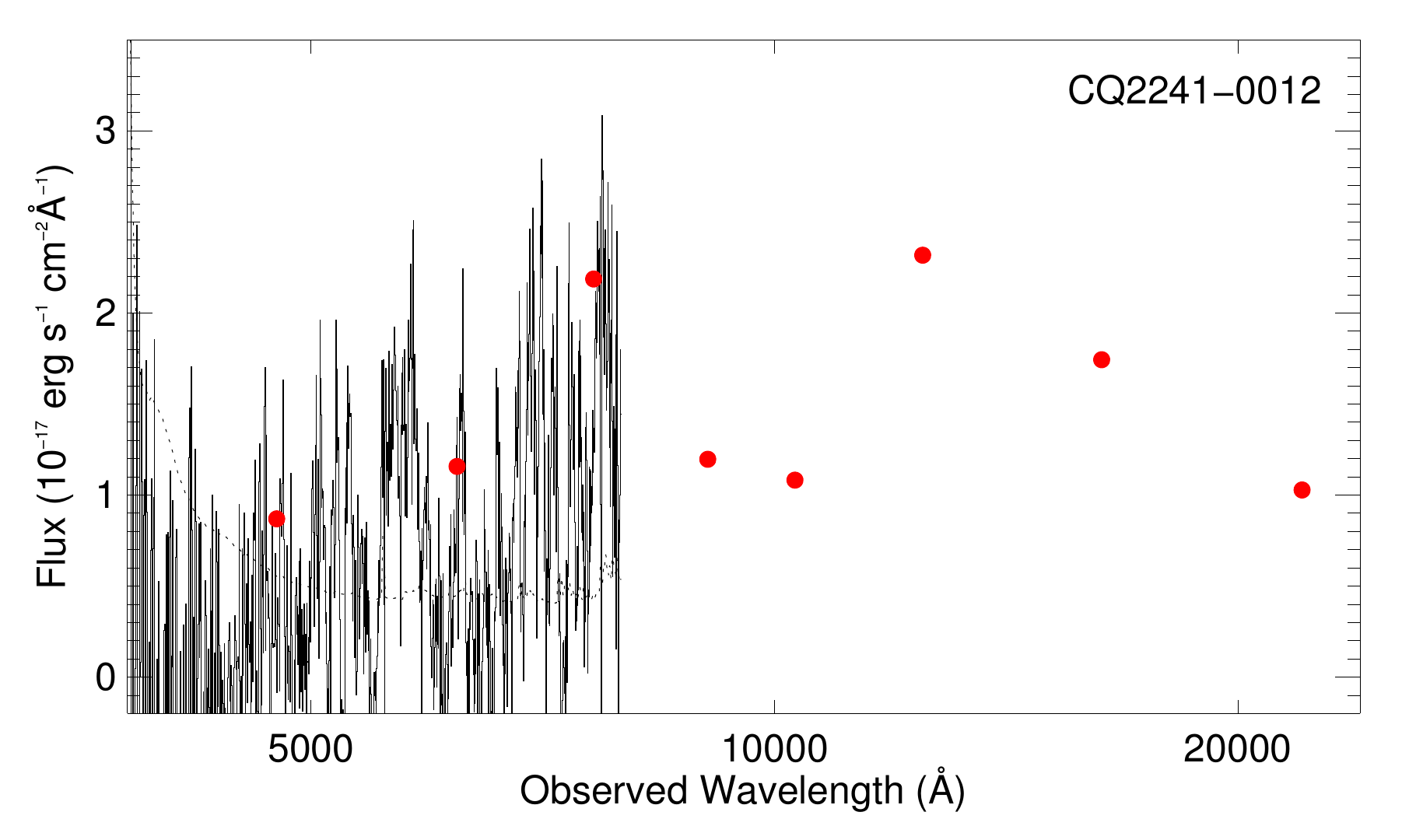}
\end{figure}
\begin{figure}
\plotone{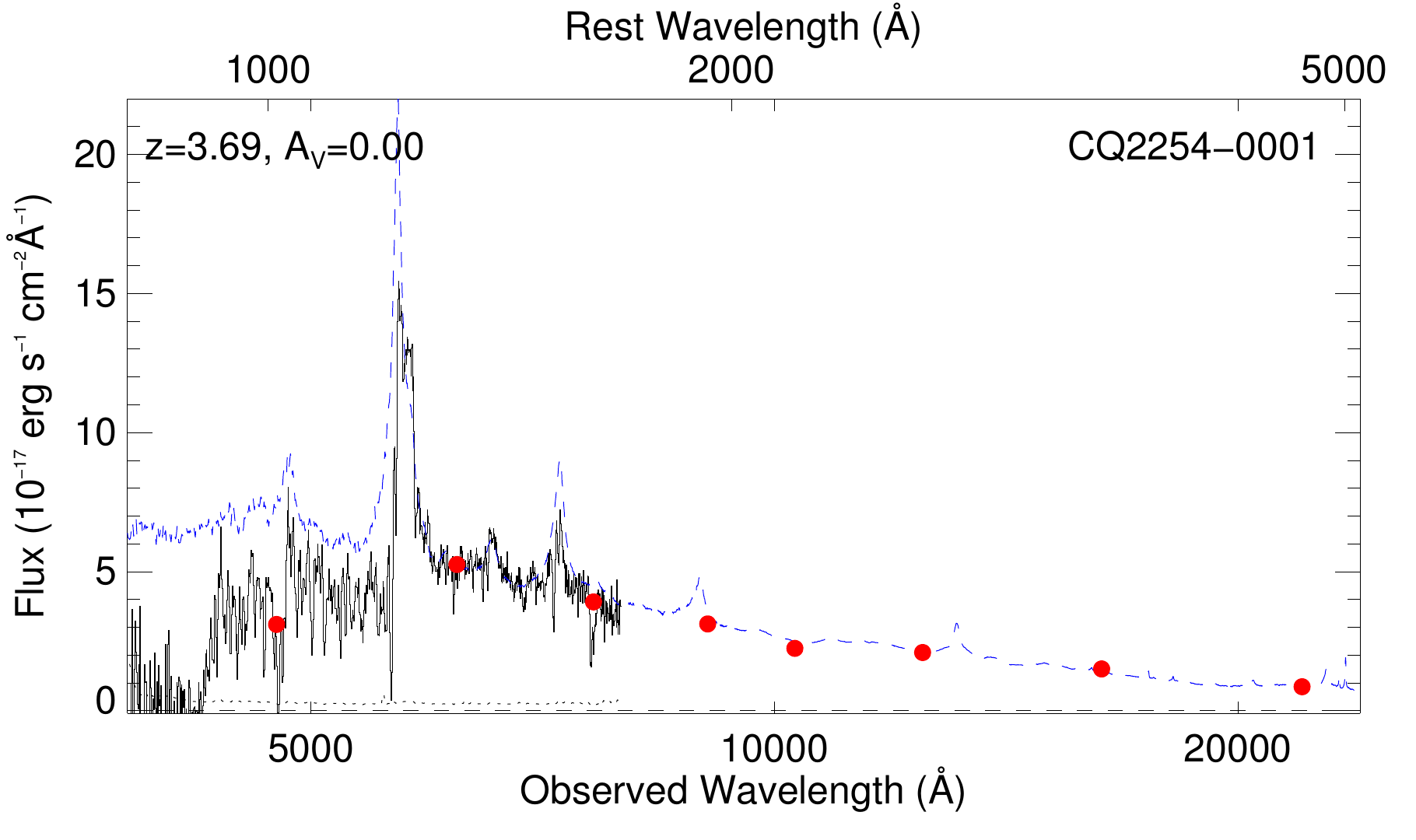}
\end{figure}
\begin{figure}
\plotone{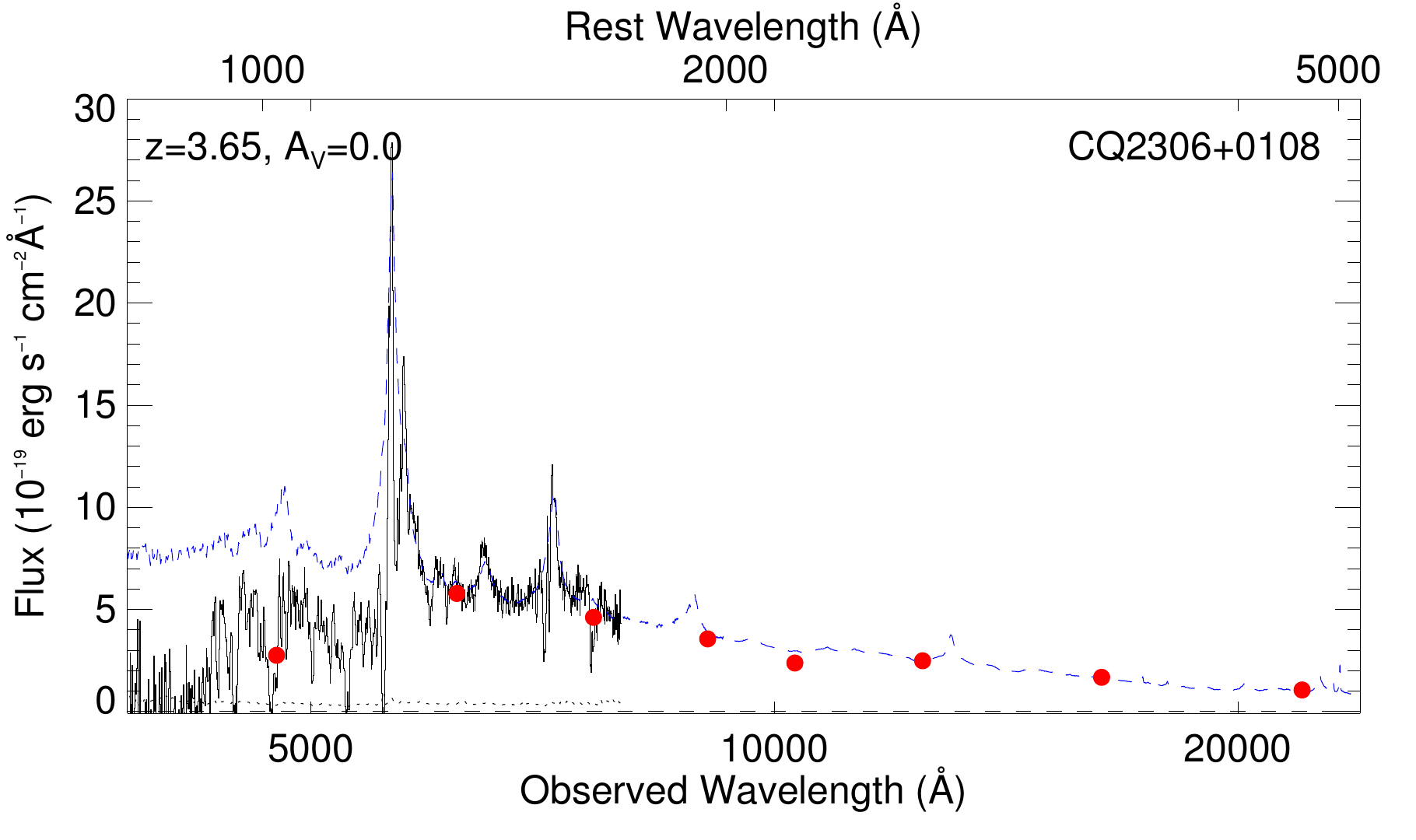}
\end{figure}
\begin{figure}
\plotone{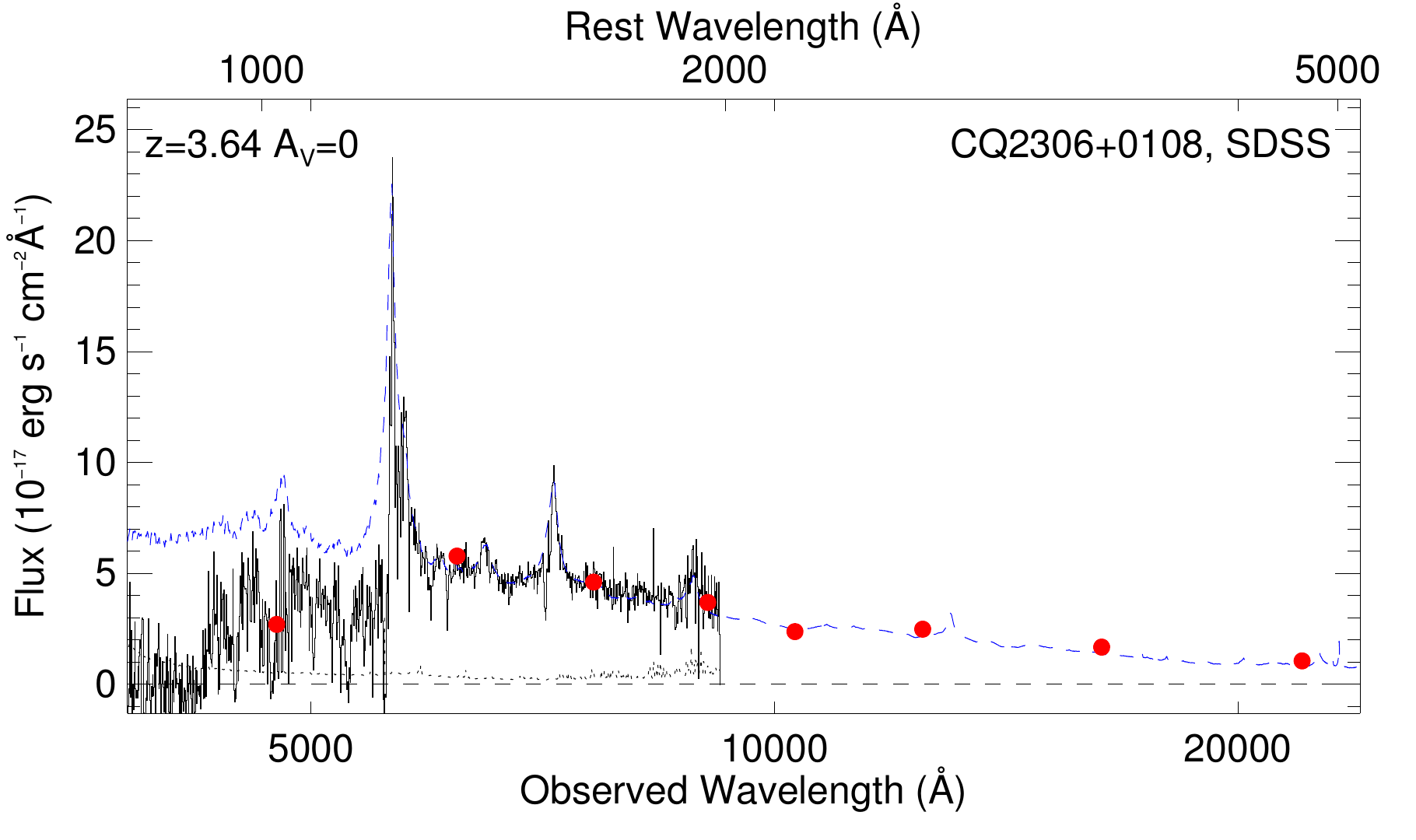}
\end{figure}
\begin{figure}
\plotone{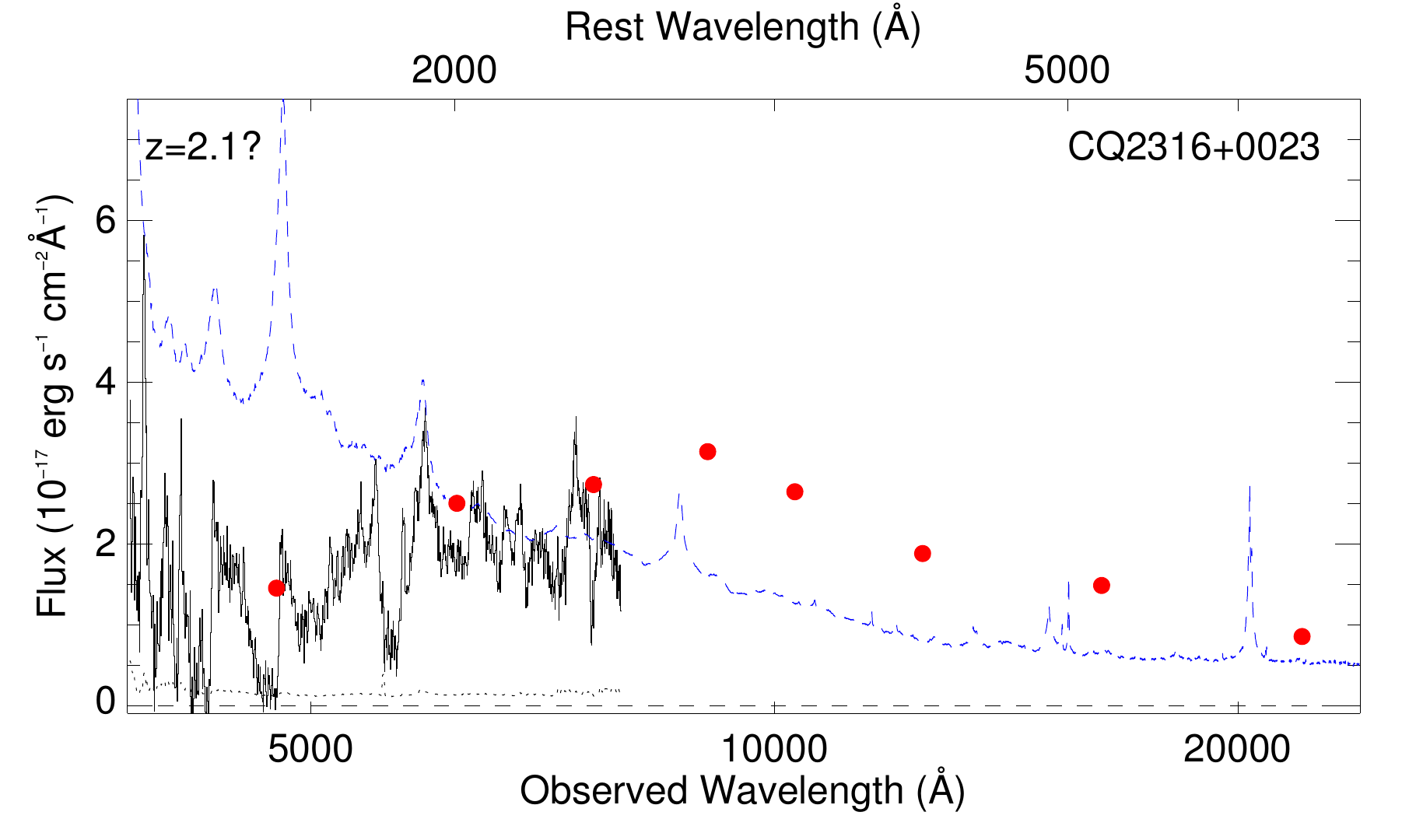}
\end{figure}
\begin{figure}
\plotone{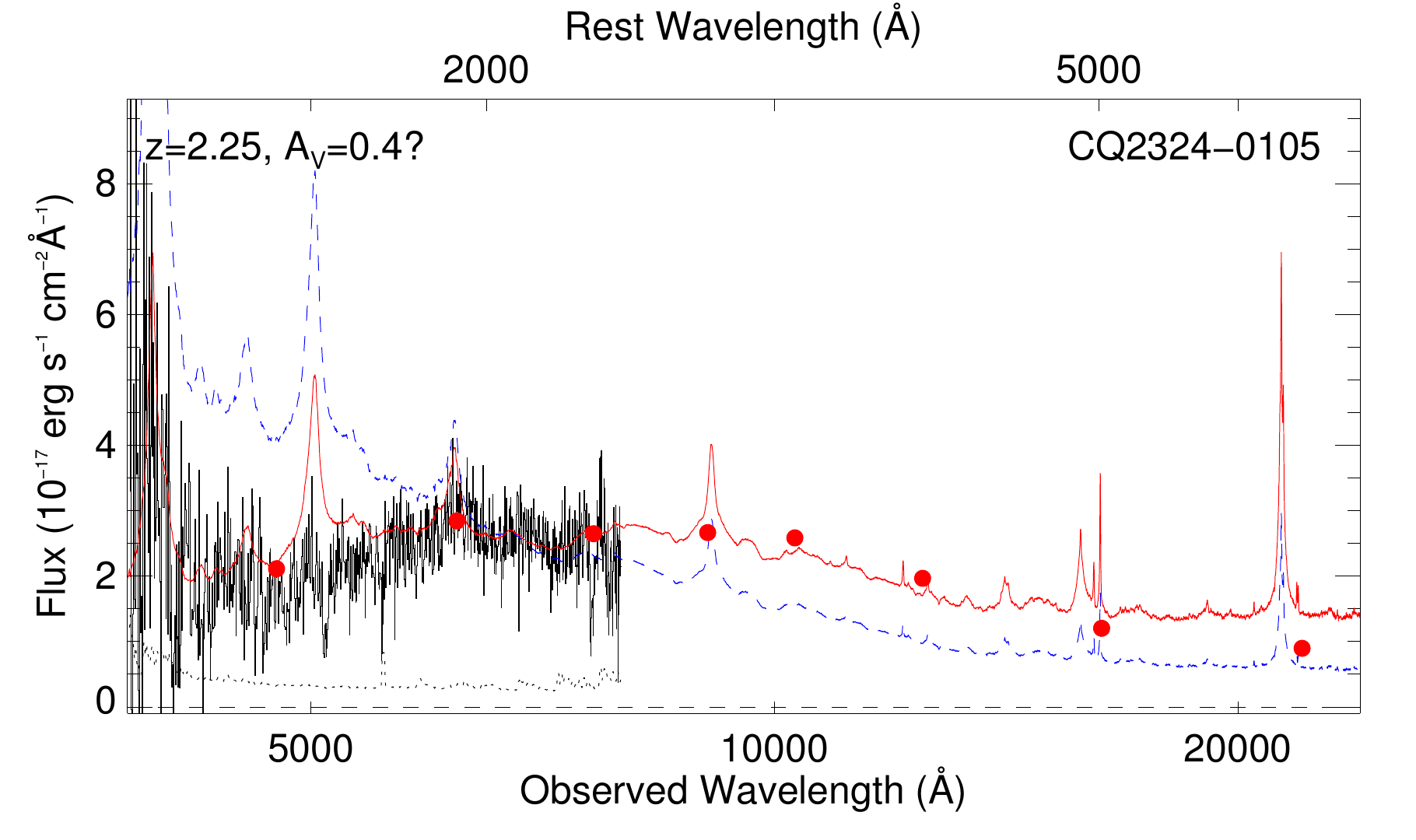}
\end{figure}
\begin{figure}
\plotone{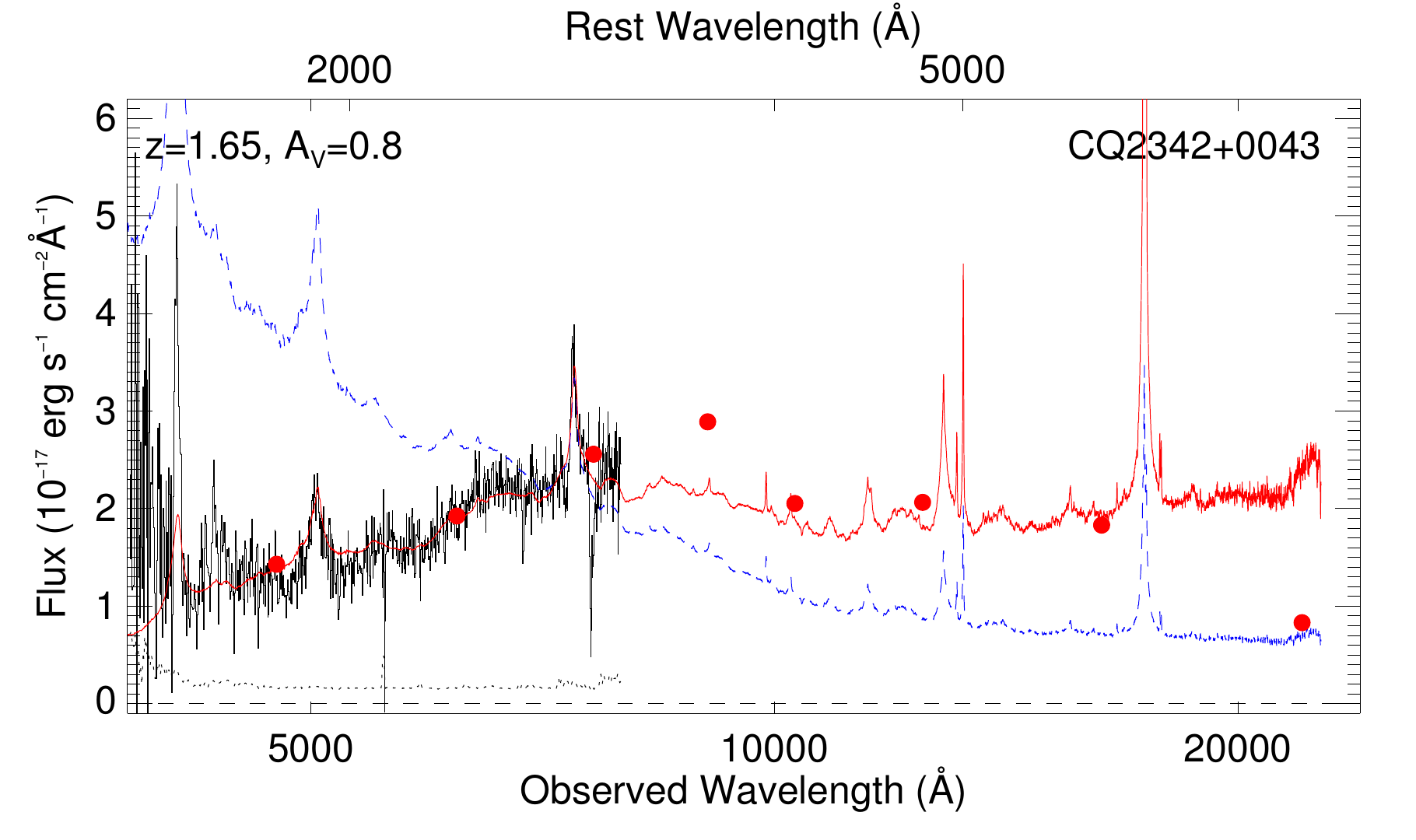}
\end{figure}
\begin{figure}
\plotone{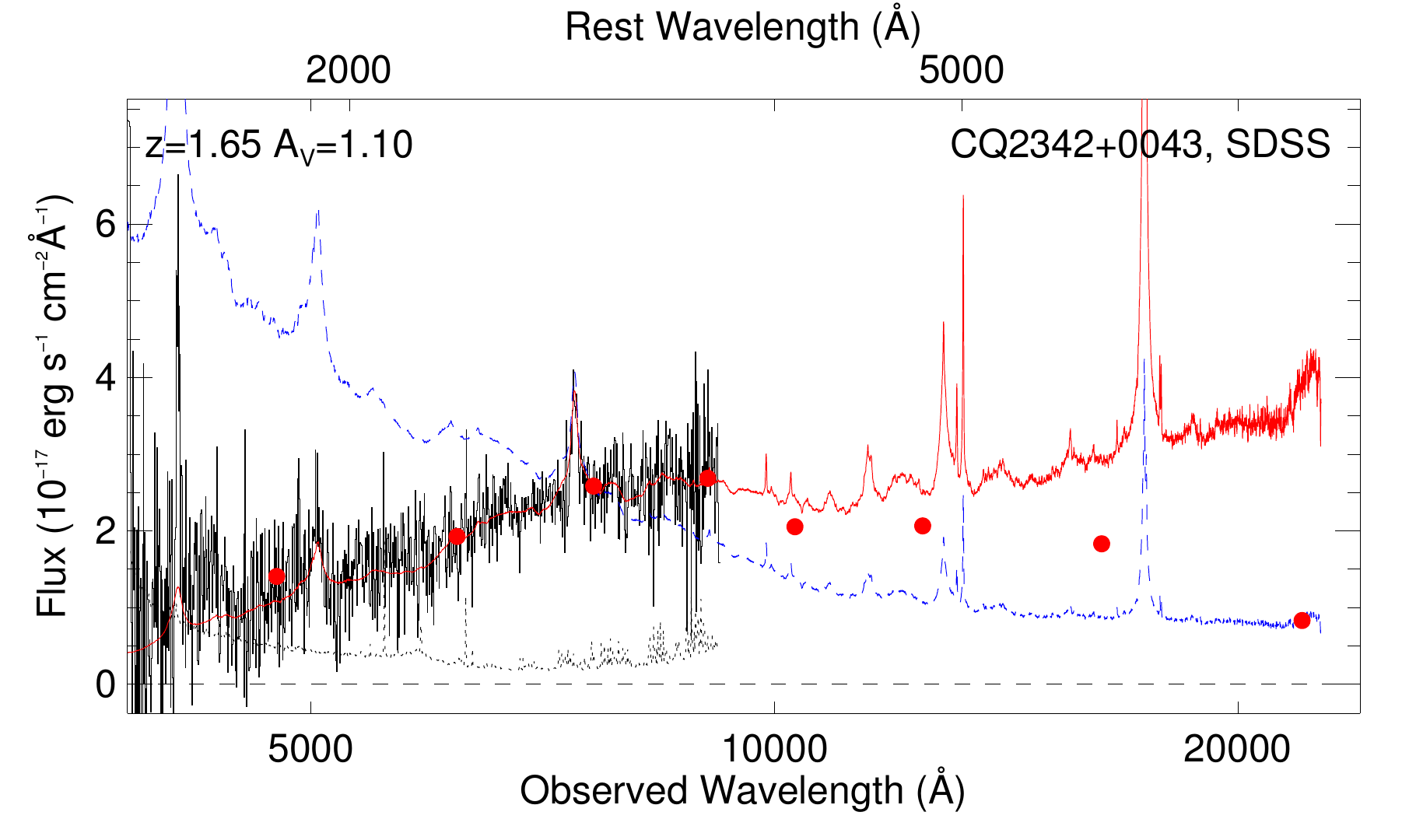}
\end{figure}
\begin{figure}
\plotone{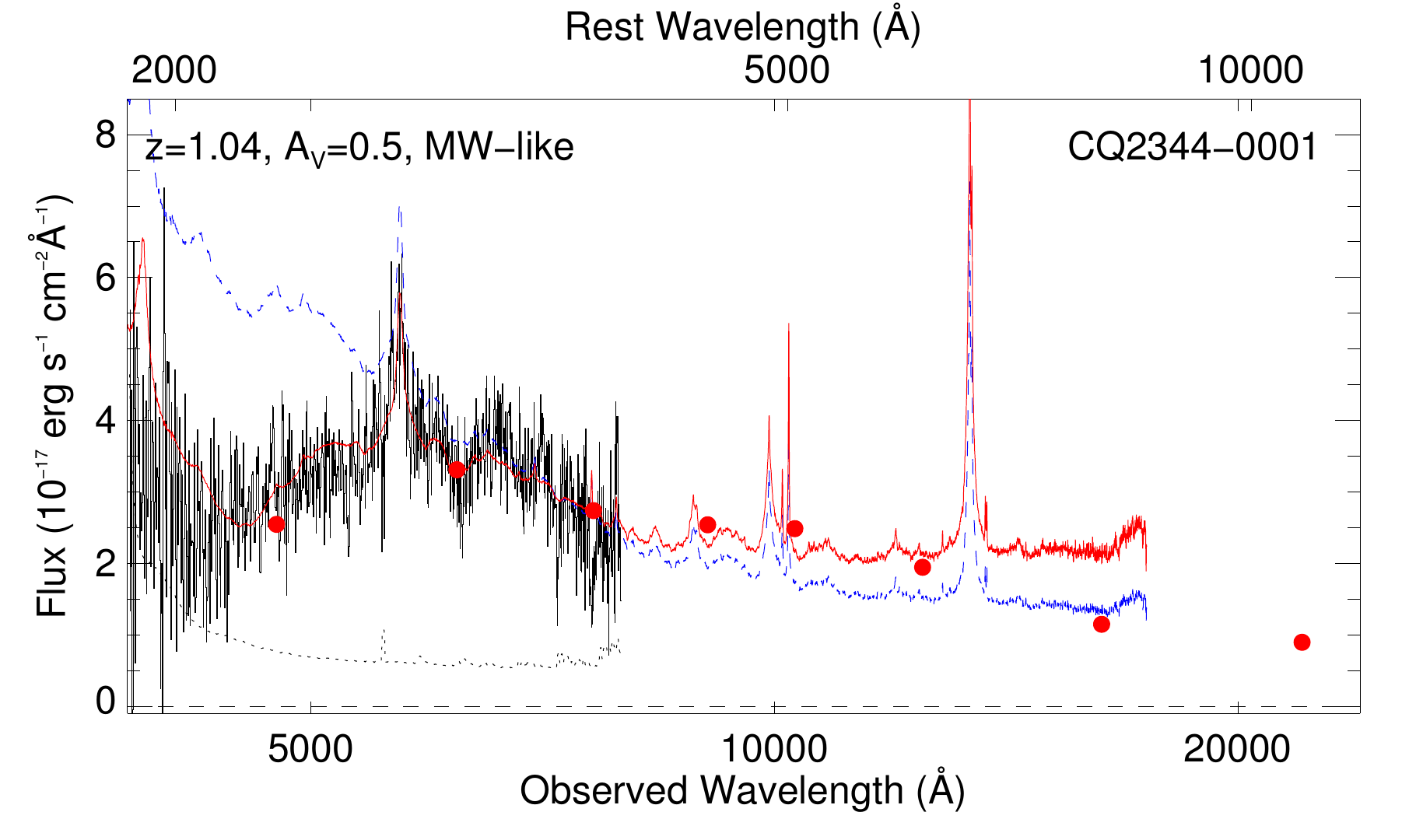}
\end{figure}
\begin{figure}
\plotone{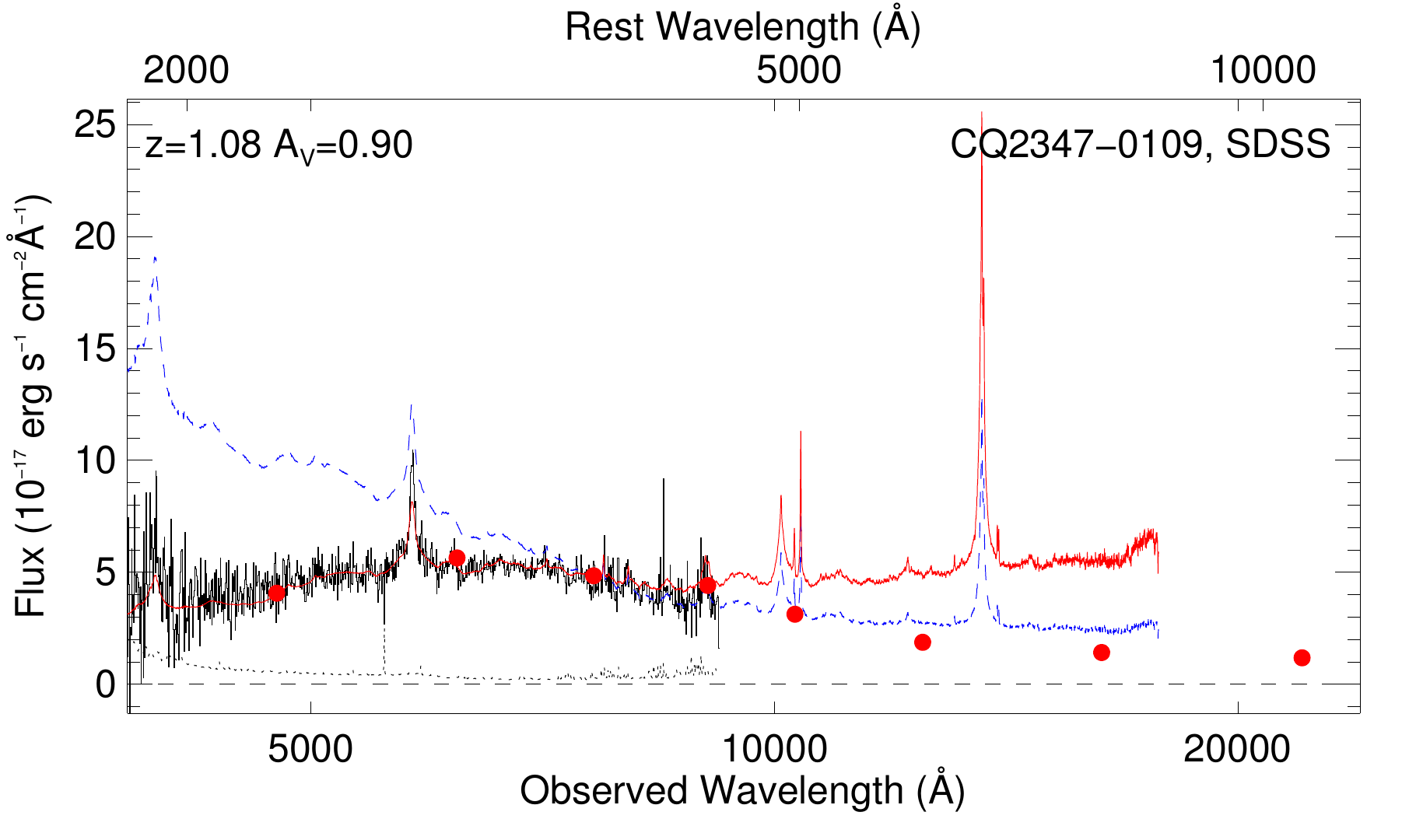}
\end{figure}
\begin{figure}
\plotone{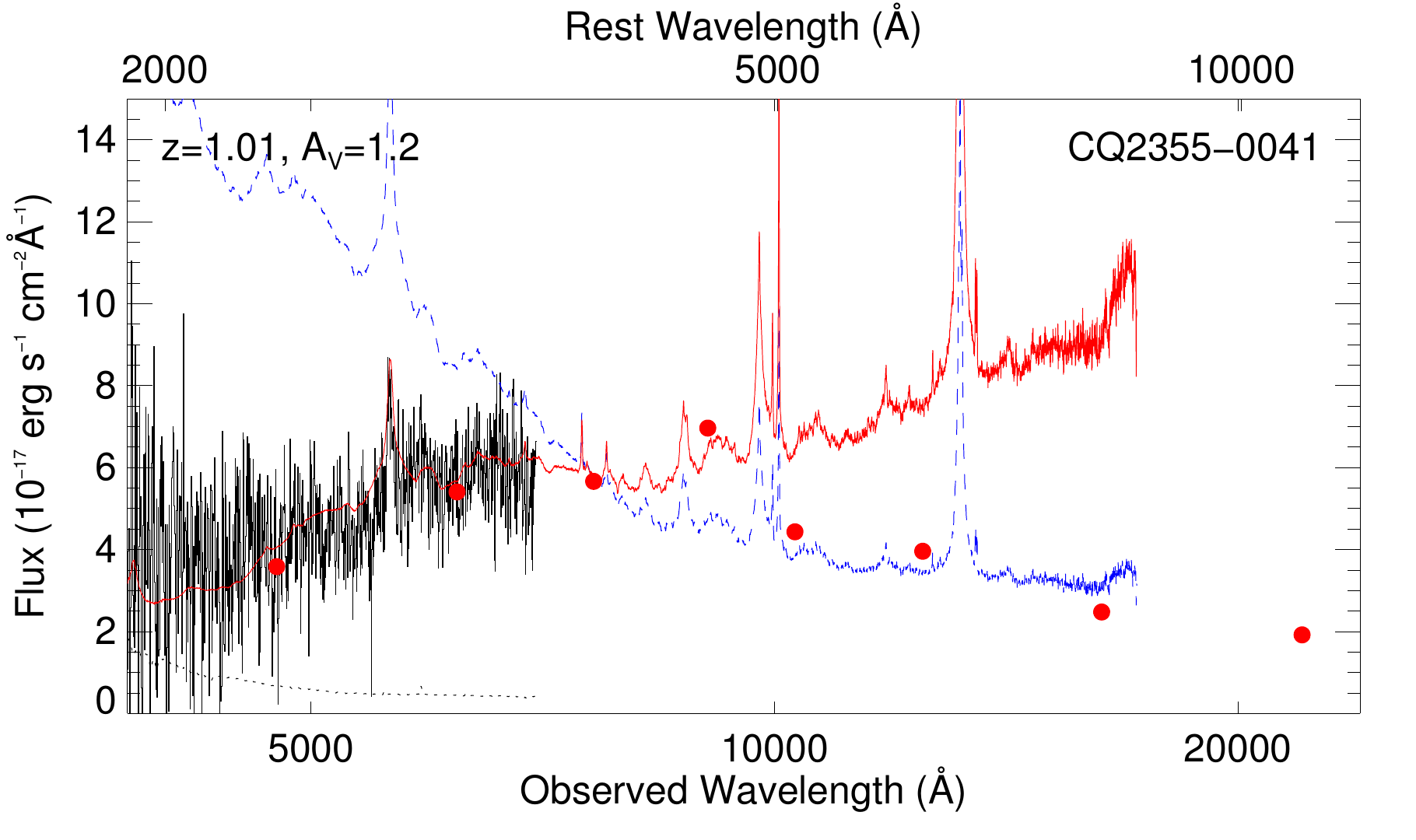}
\end{figure}
\clearpage
\begin{figure}
\plotone{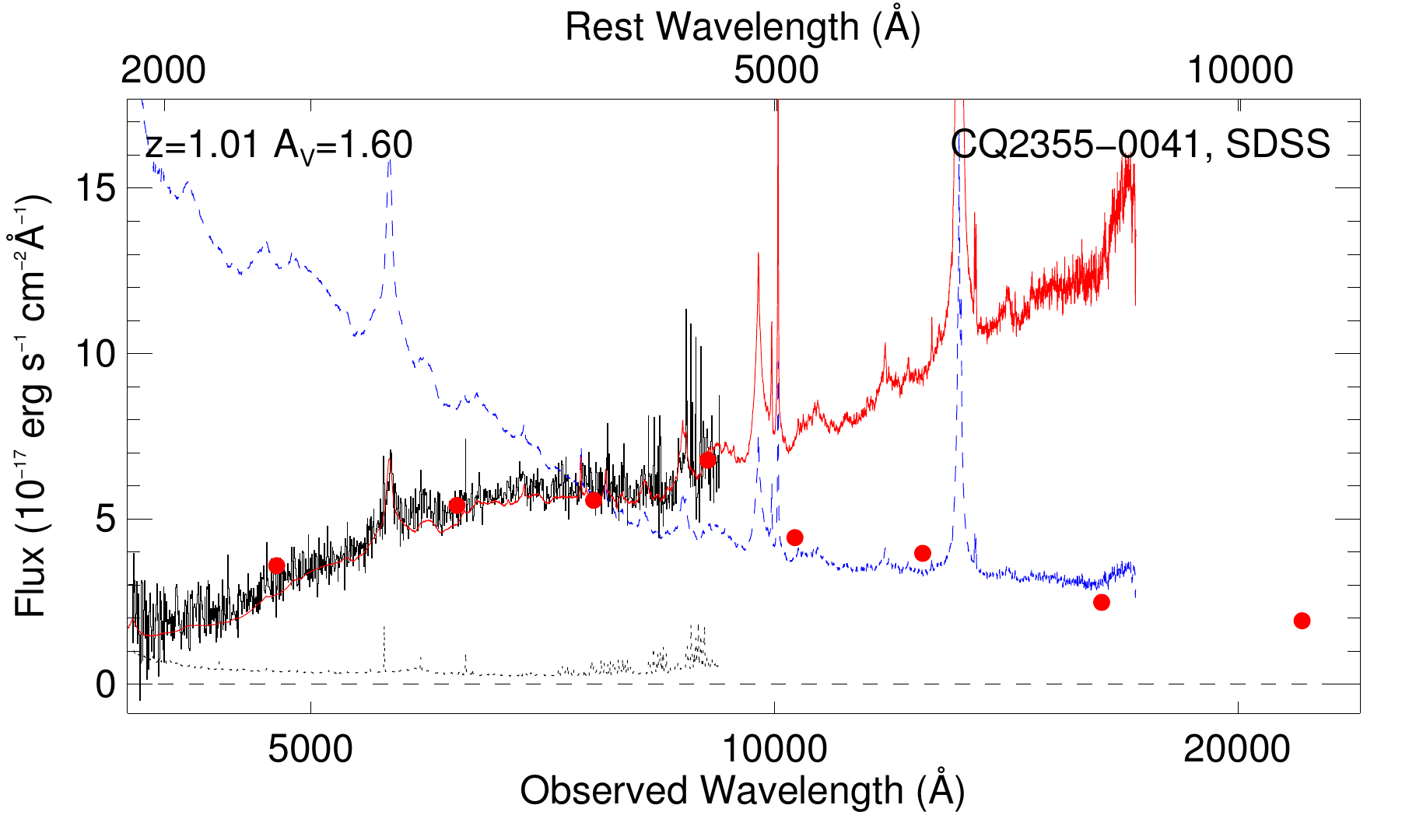}
\end{figure}
\begin{figure}
\plotone{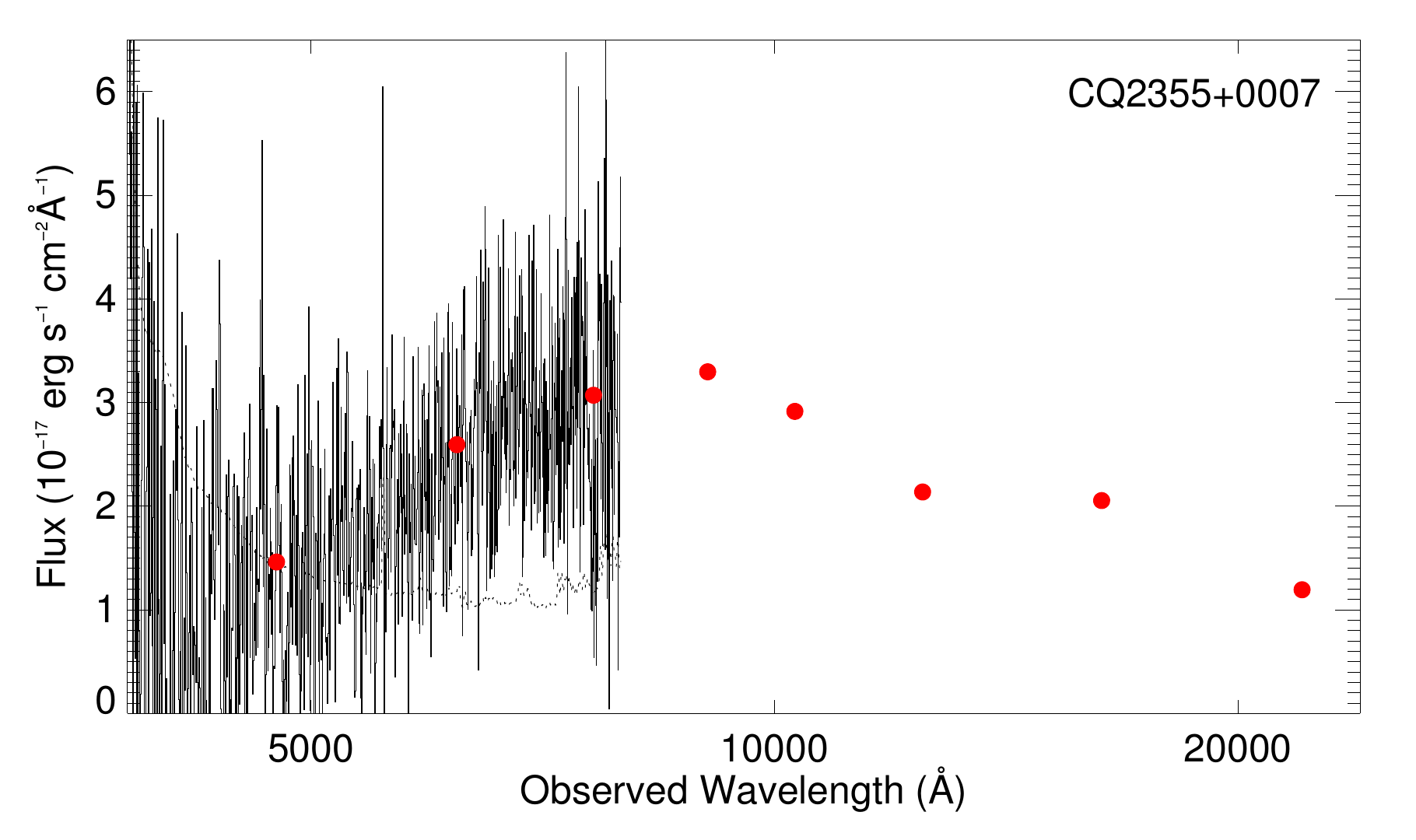}
\end{figure}
\begin{figure}
\plotone{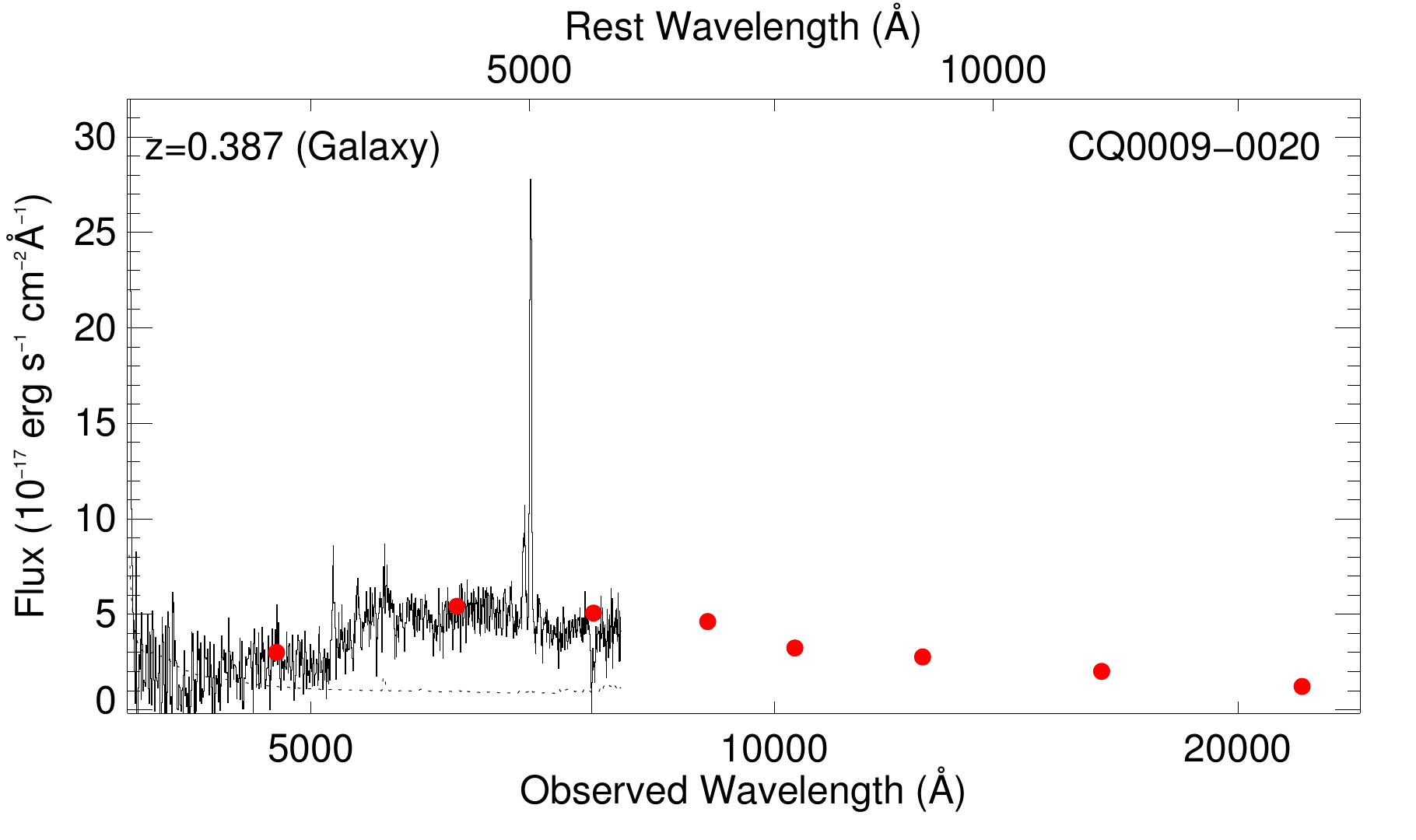}
\end{figure}
\begin{figure}
\plotone{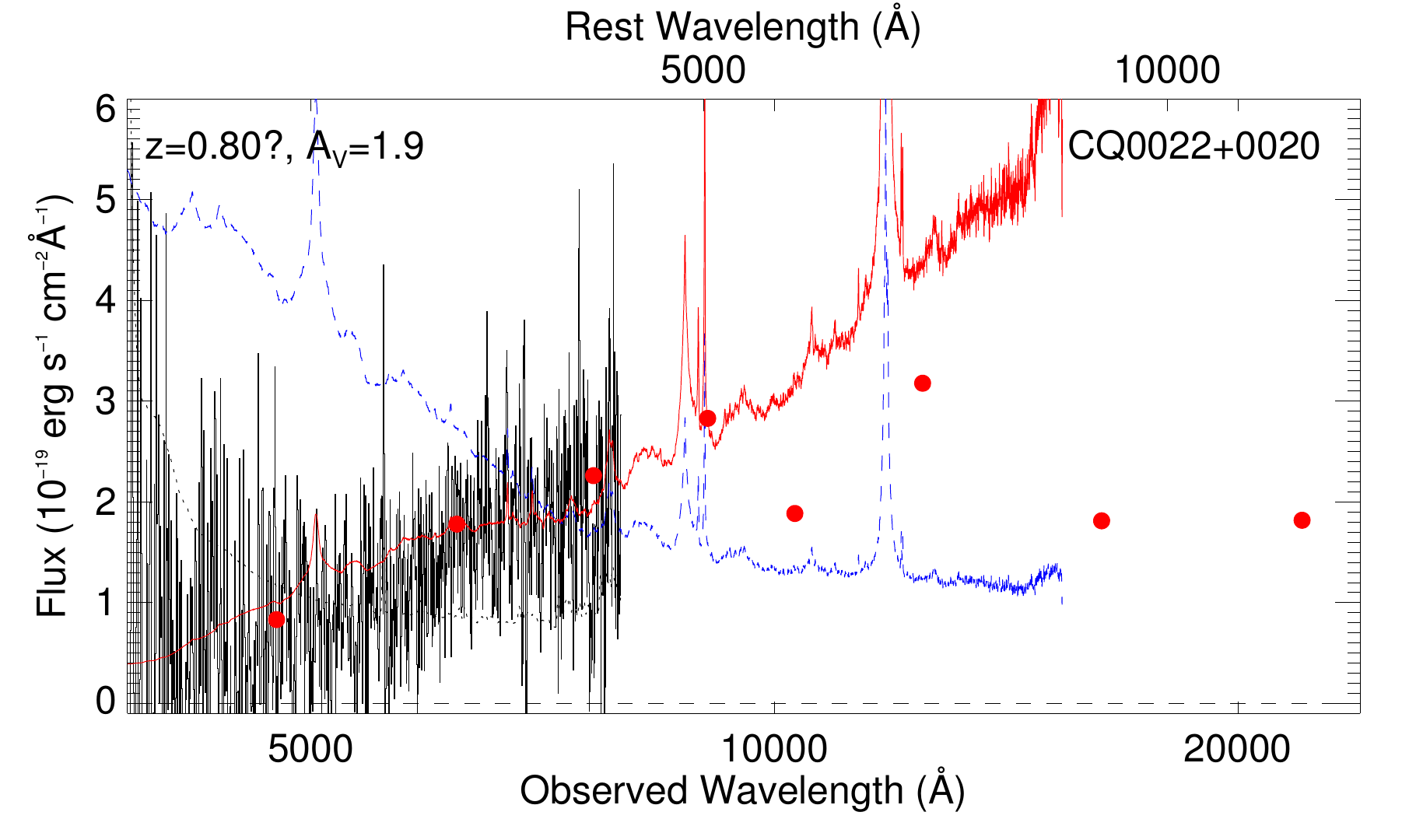}
\end{figure}
\begin{figure}
\plotone{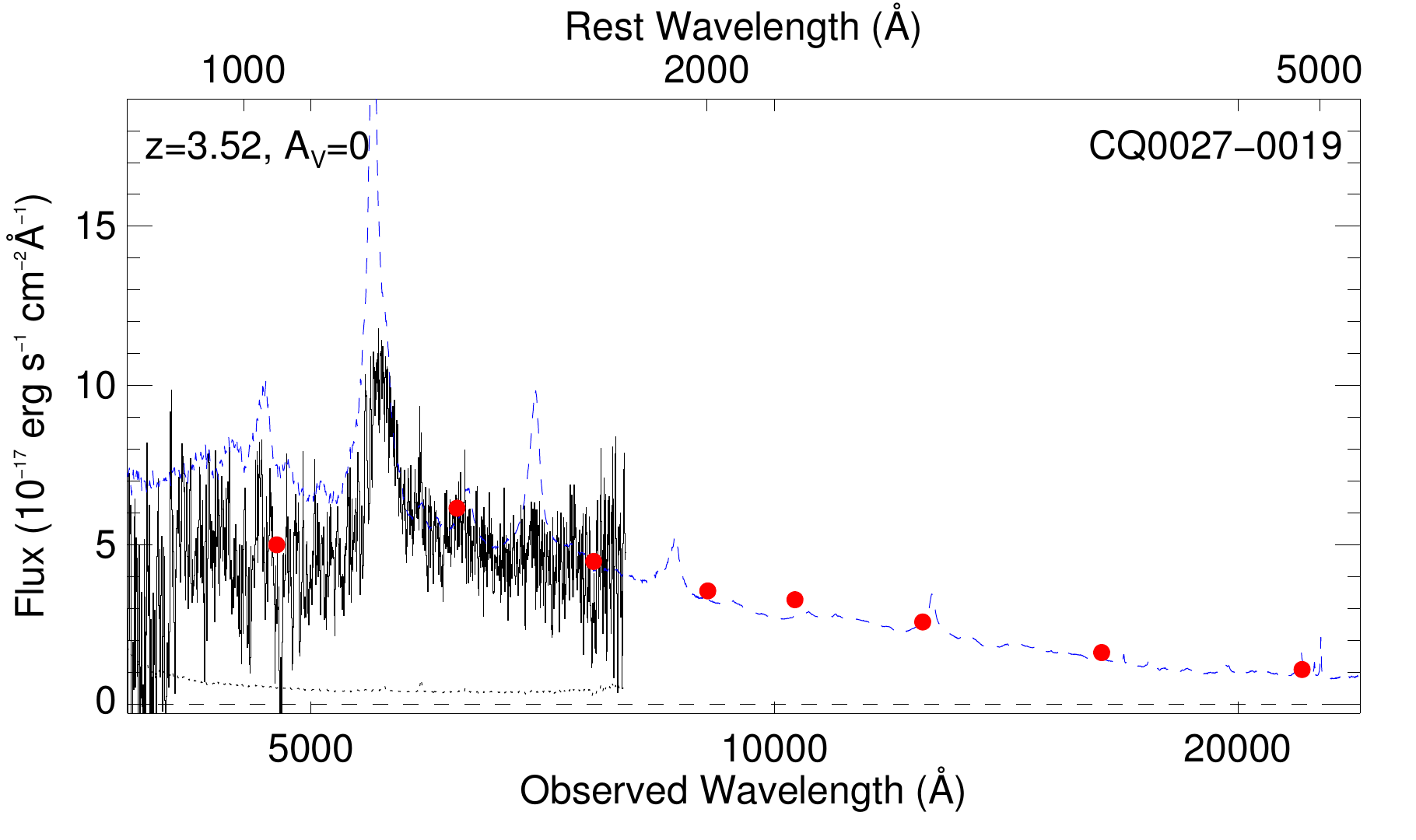}
\end{figure}
\begin{figure}
\plotone{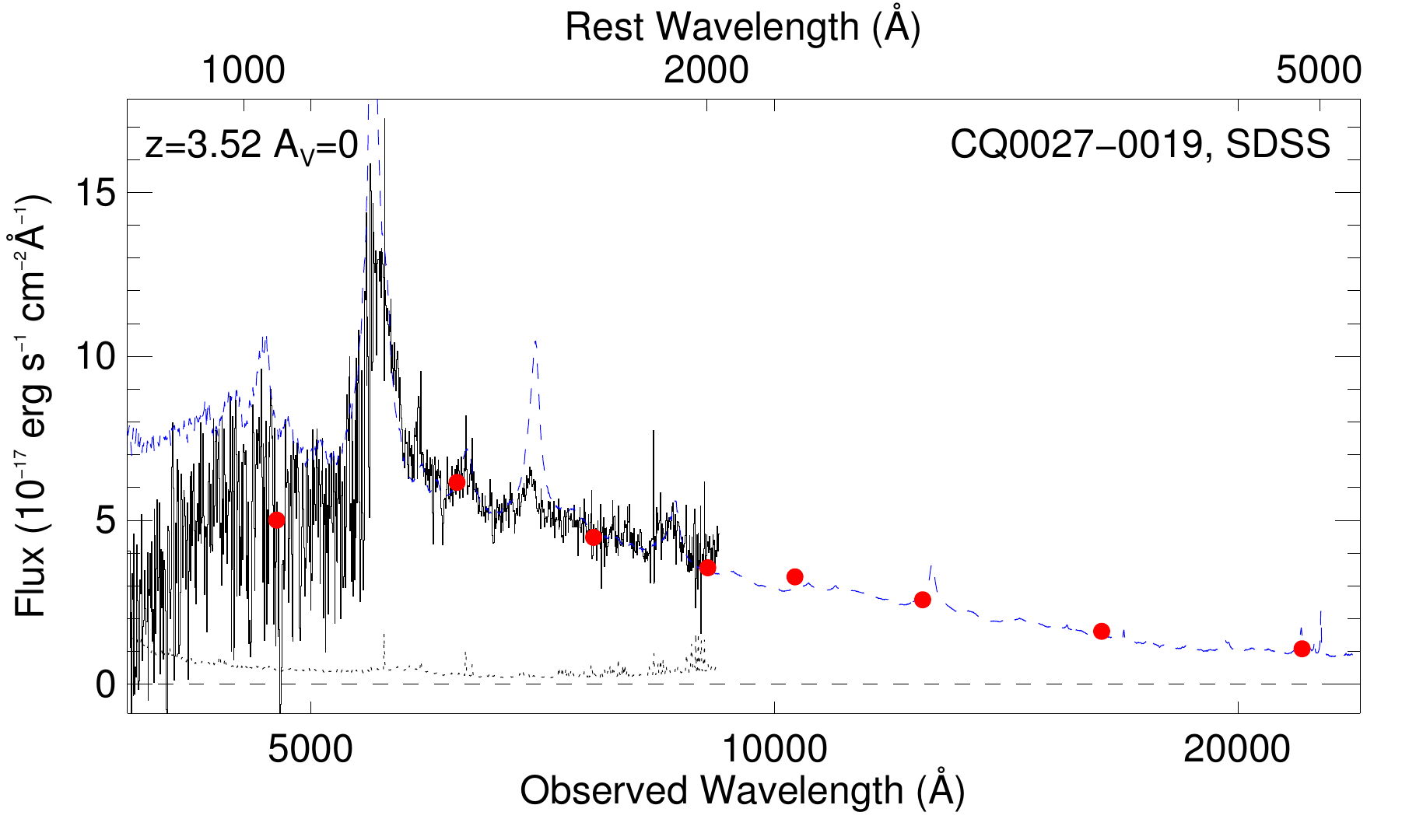}
\end{figure}
\begin{figure}
\plotone{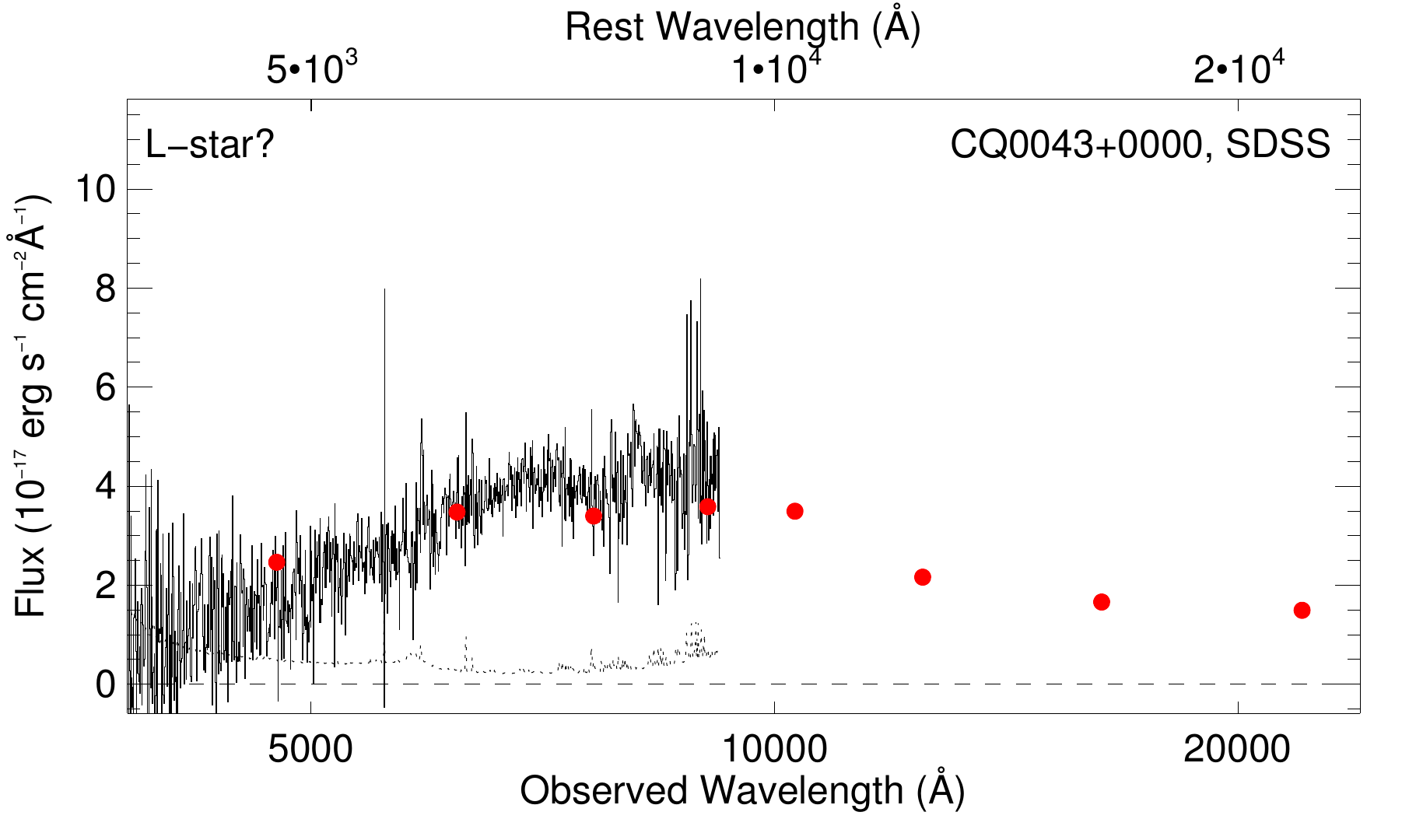}
\end{figure}
\begin{figure}
\plotone{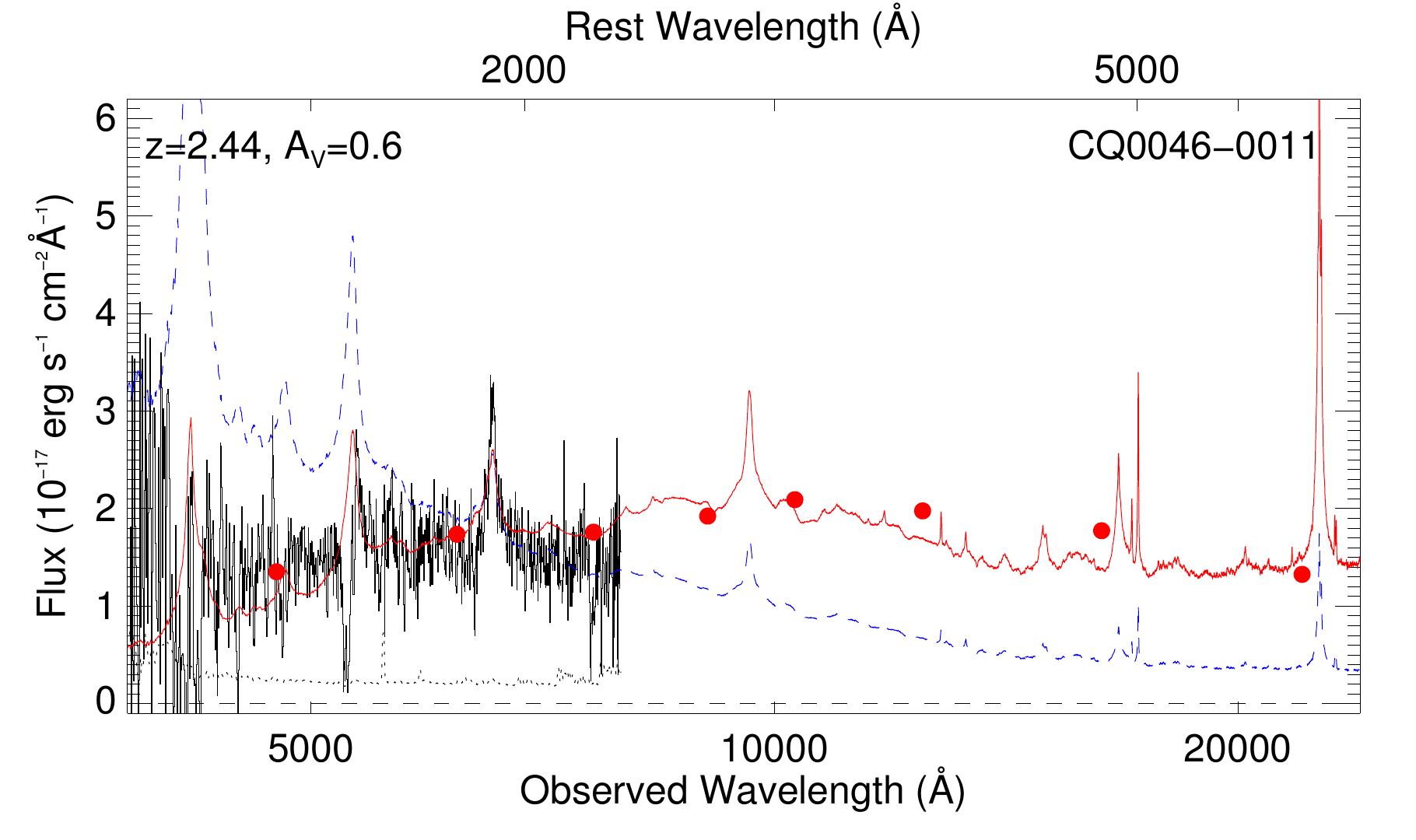}
\end{figure}
\begin{figure}
\plotone{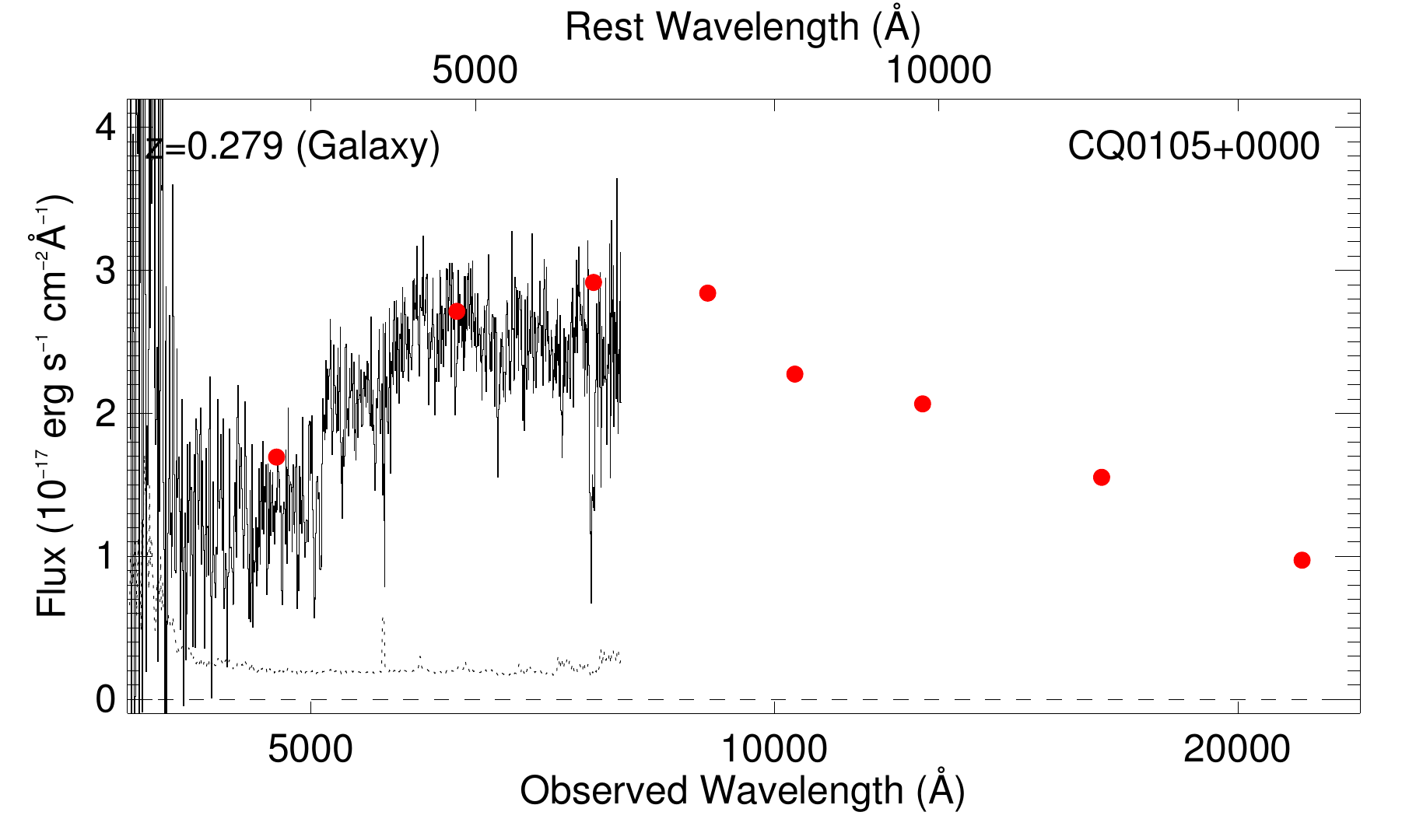}
\end{figure}
\begin{figure}
\plotone{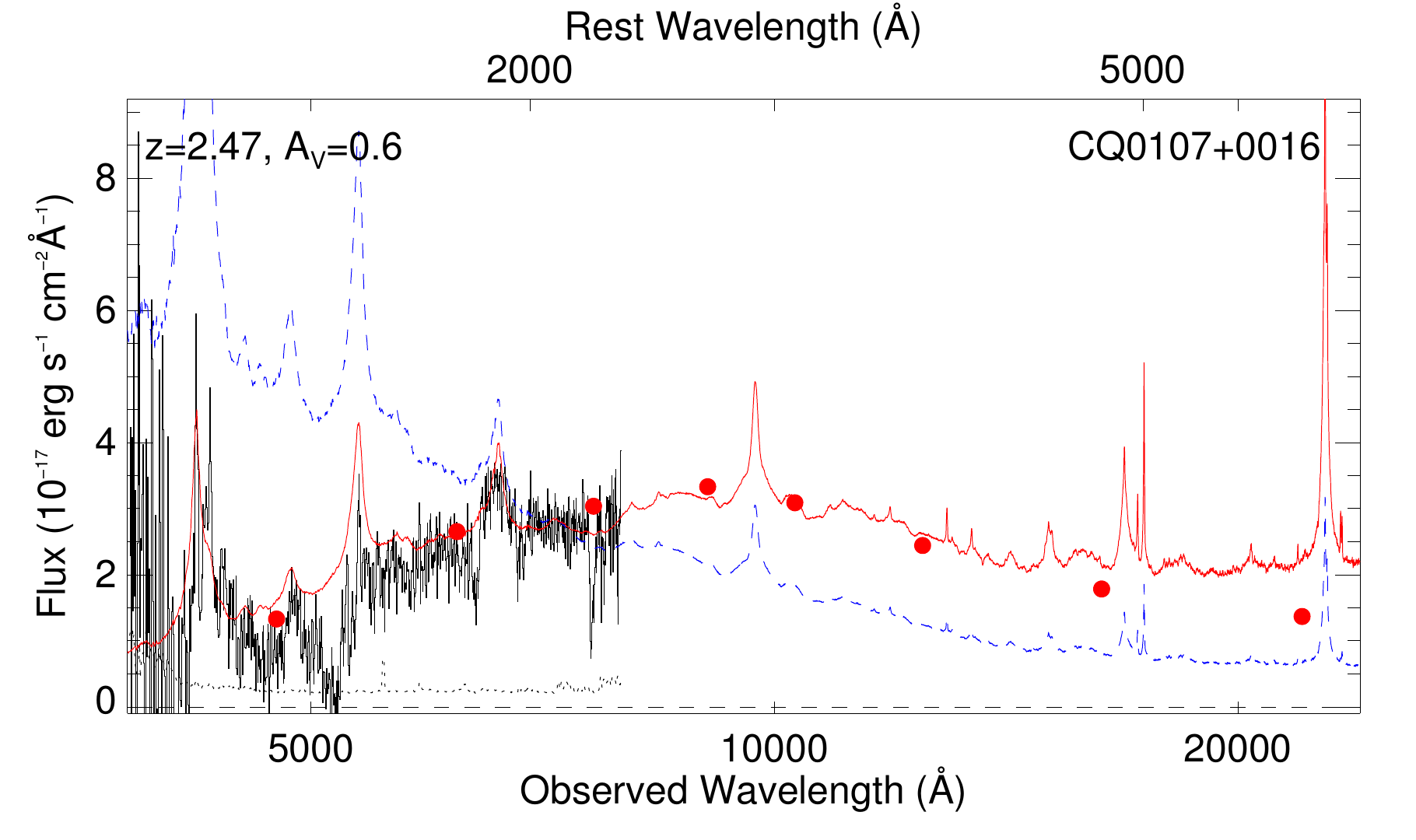}
\end{figure}
\begin{figure}
\plotone{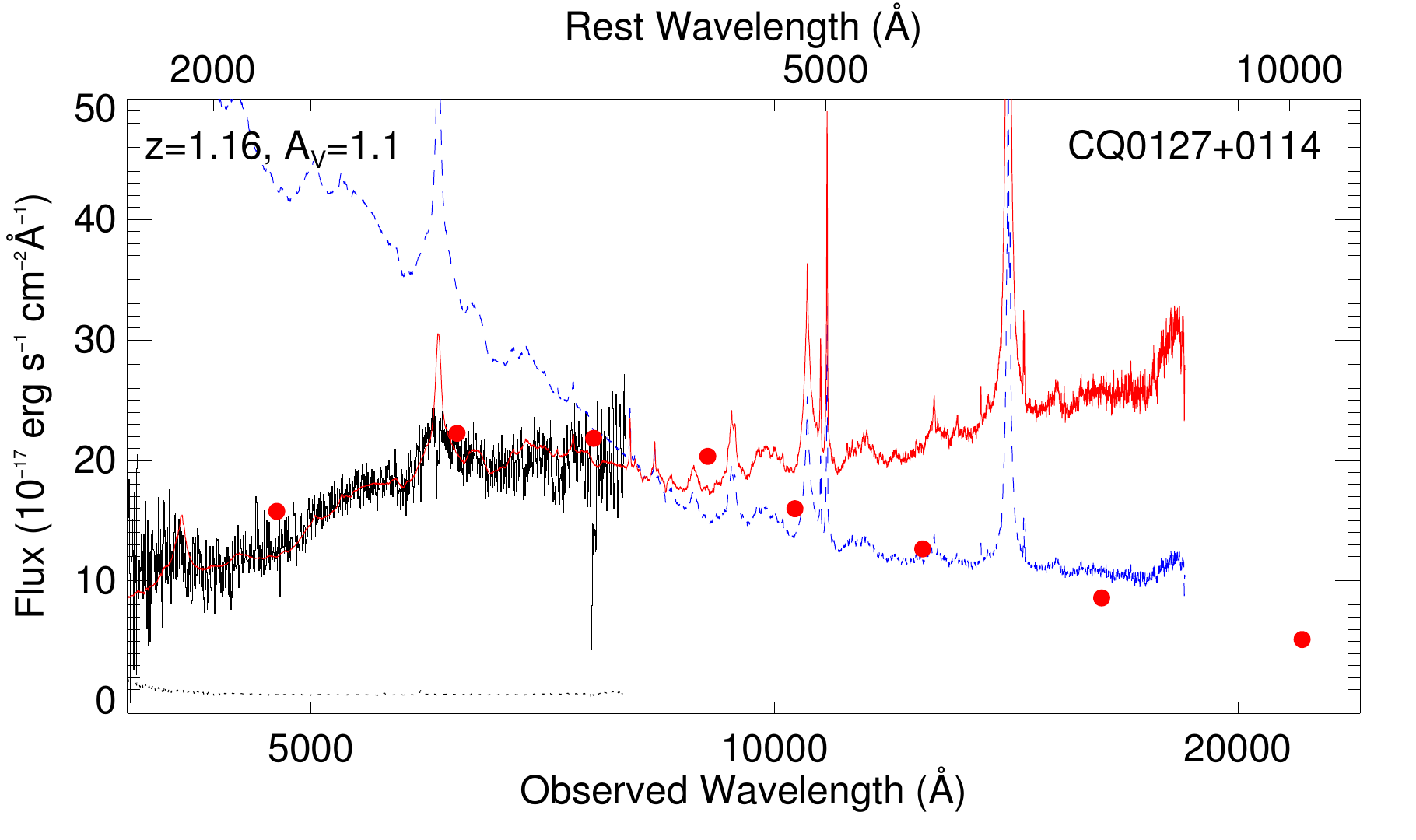}
\end{figure}
\begin{figure}
\plotone{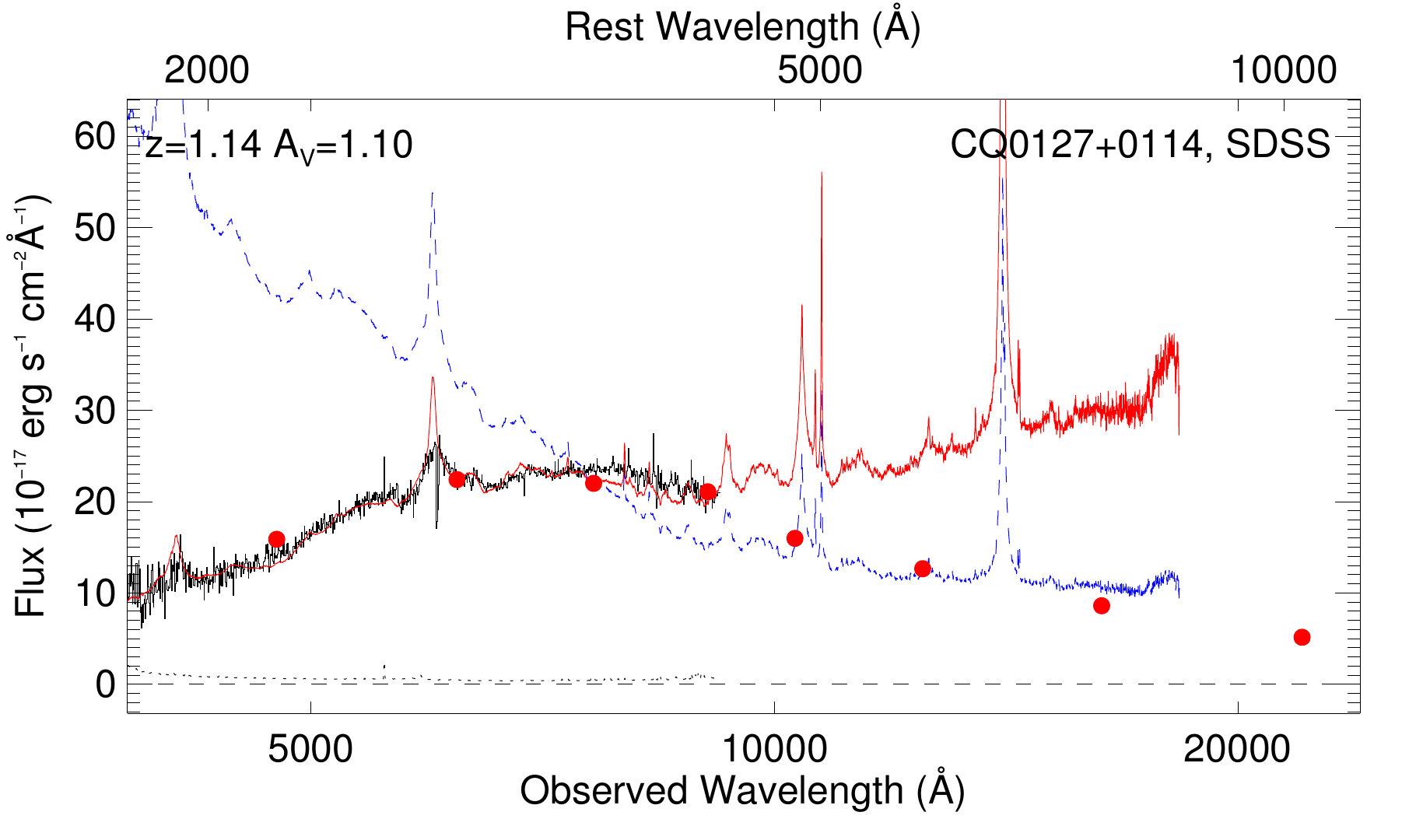}
\end{figure}
\begin{figure}
\plotone{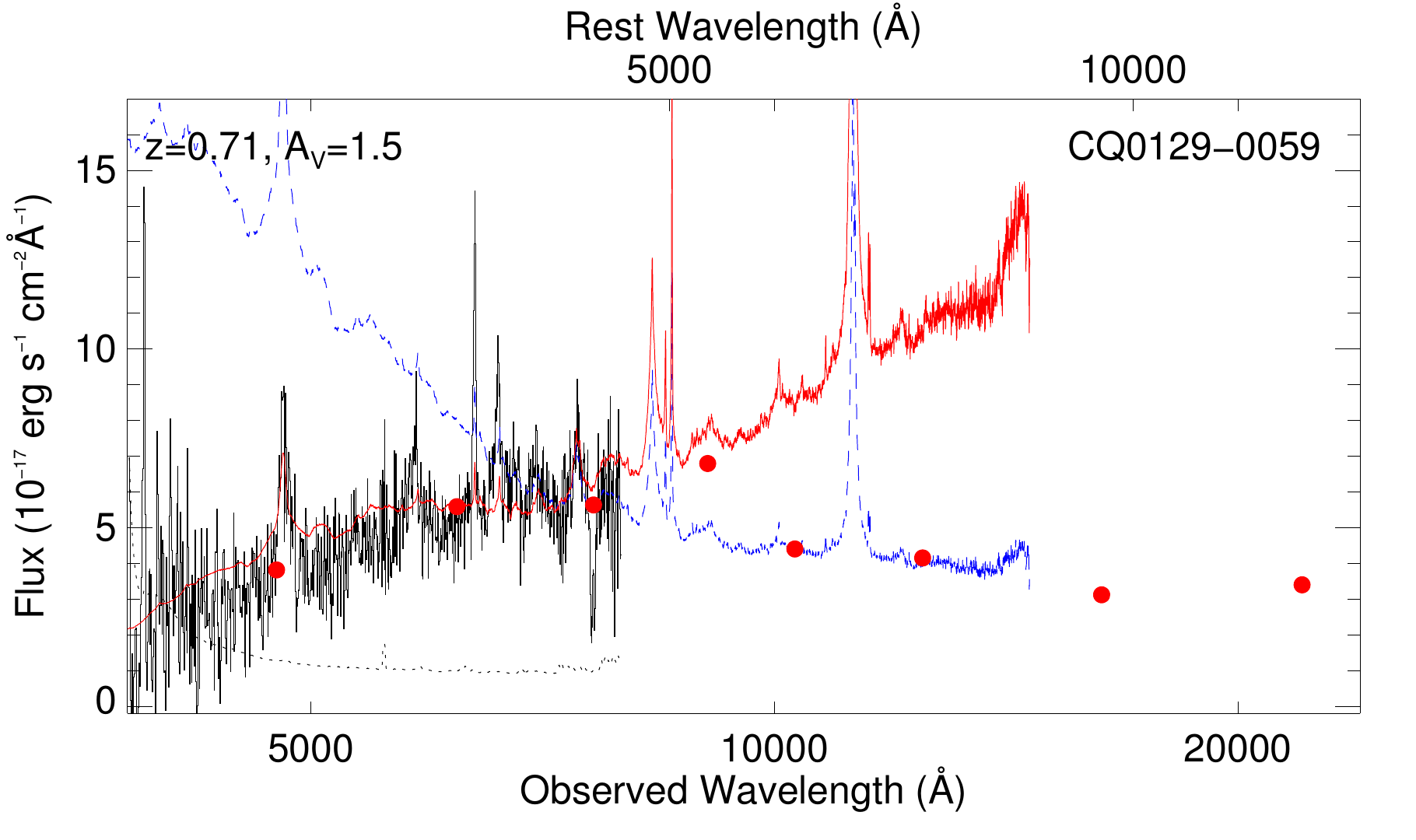}
\end{figure}
\begin{figure}
\plotone{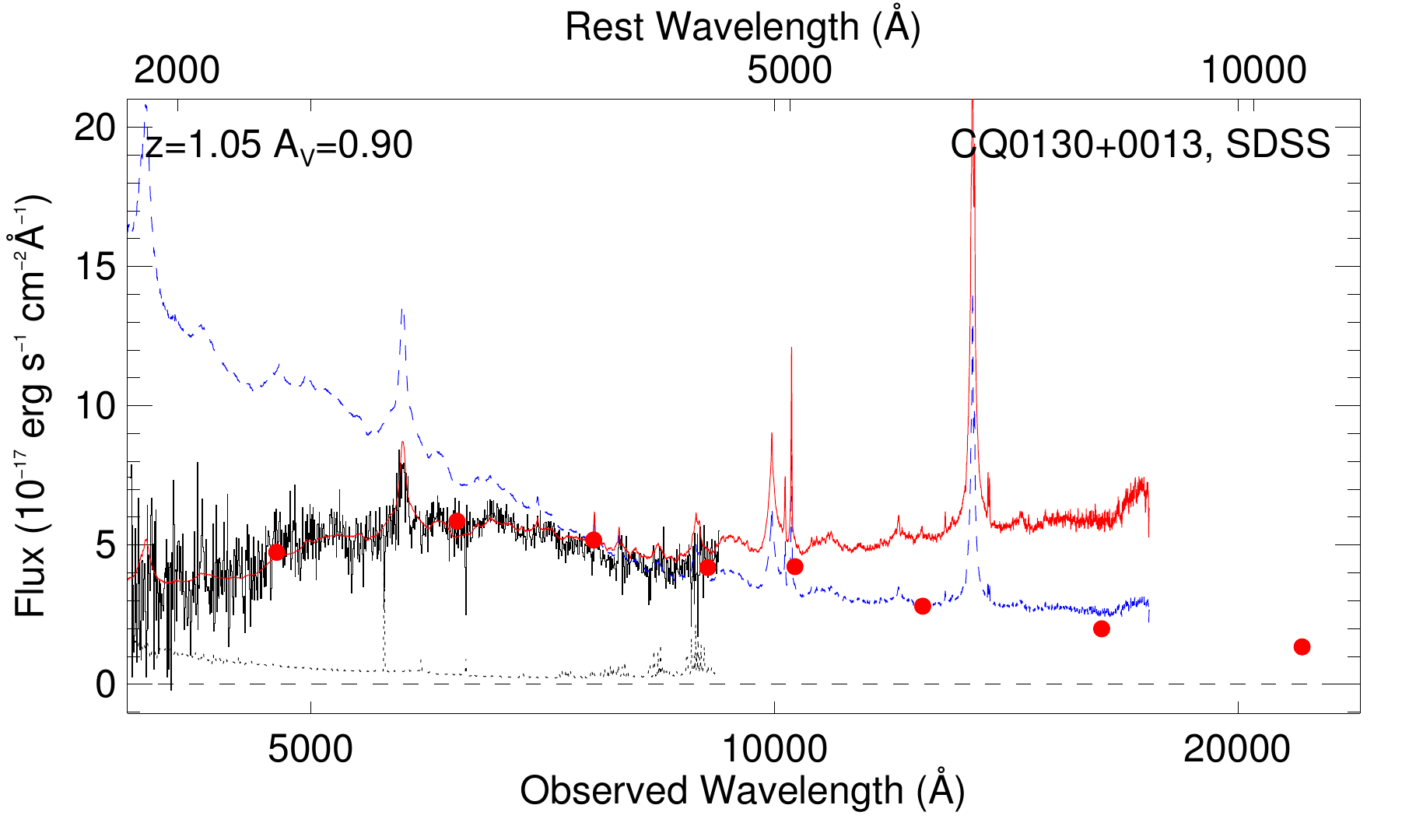}
\end{figure}
\begin{figure}
\plotone{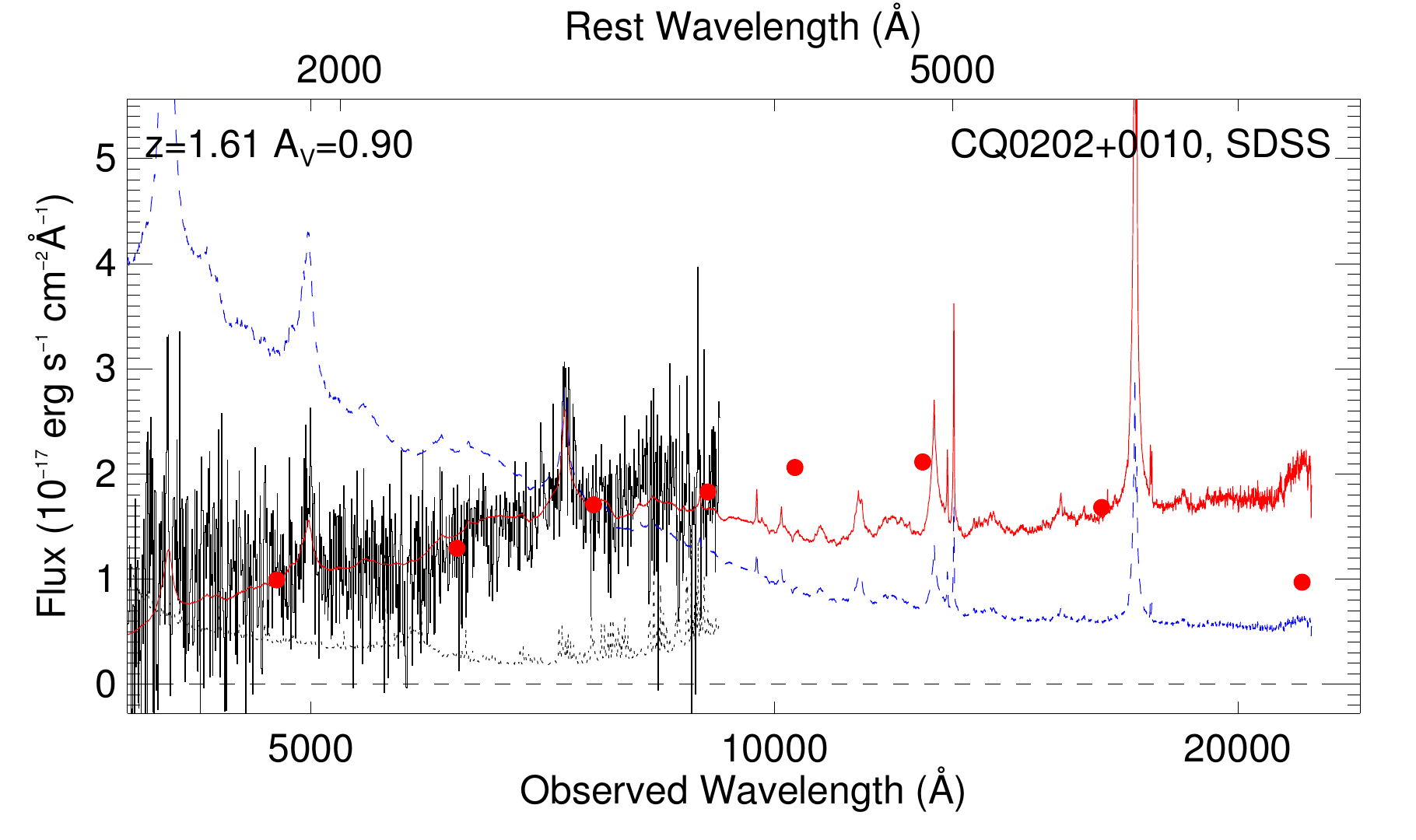}
\end{figure}
\begin{figure}
\plotone{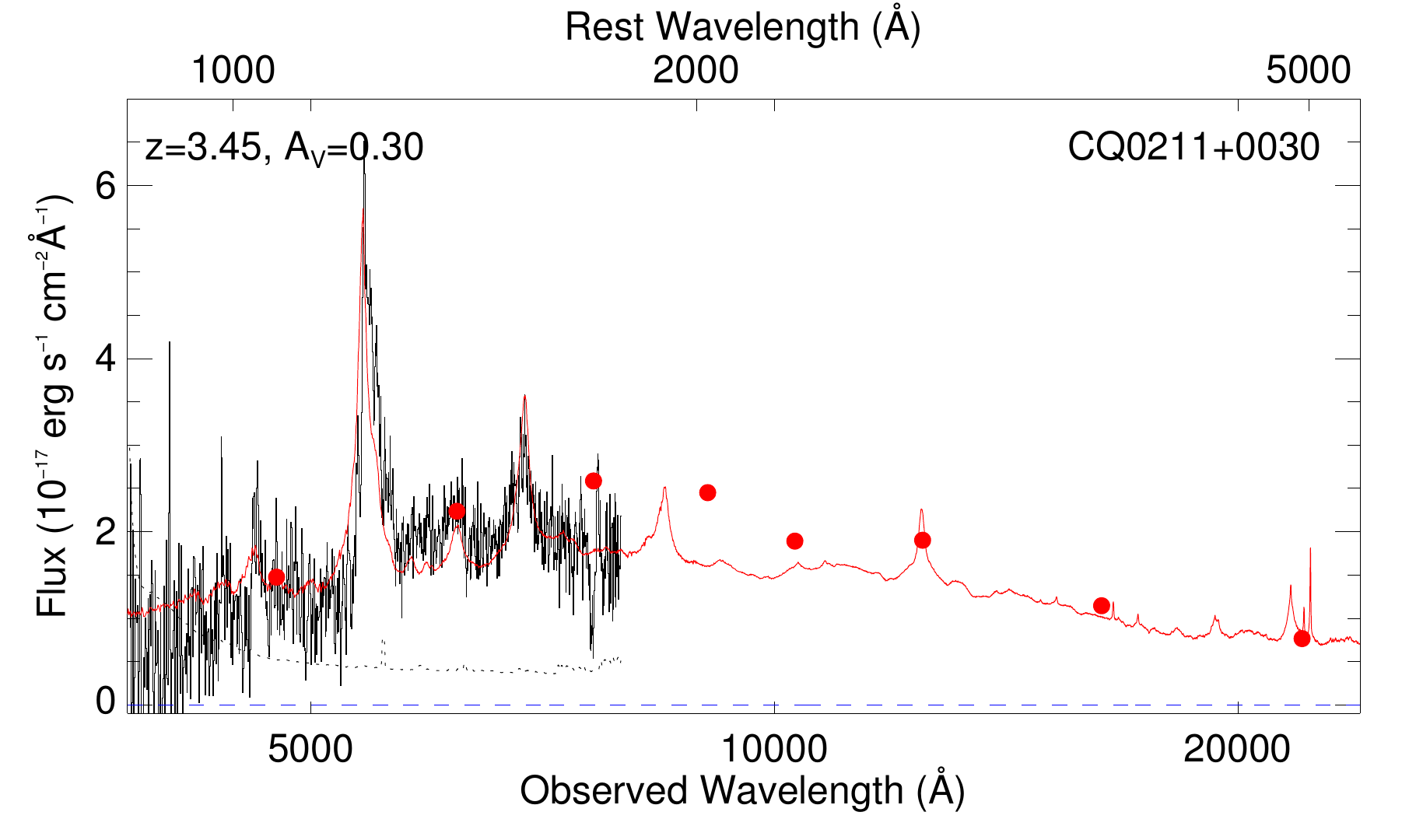}
\end{figure}
\begin{figure}
\plotone{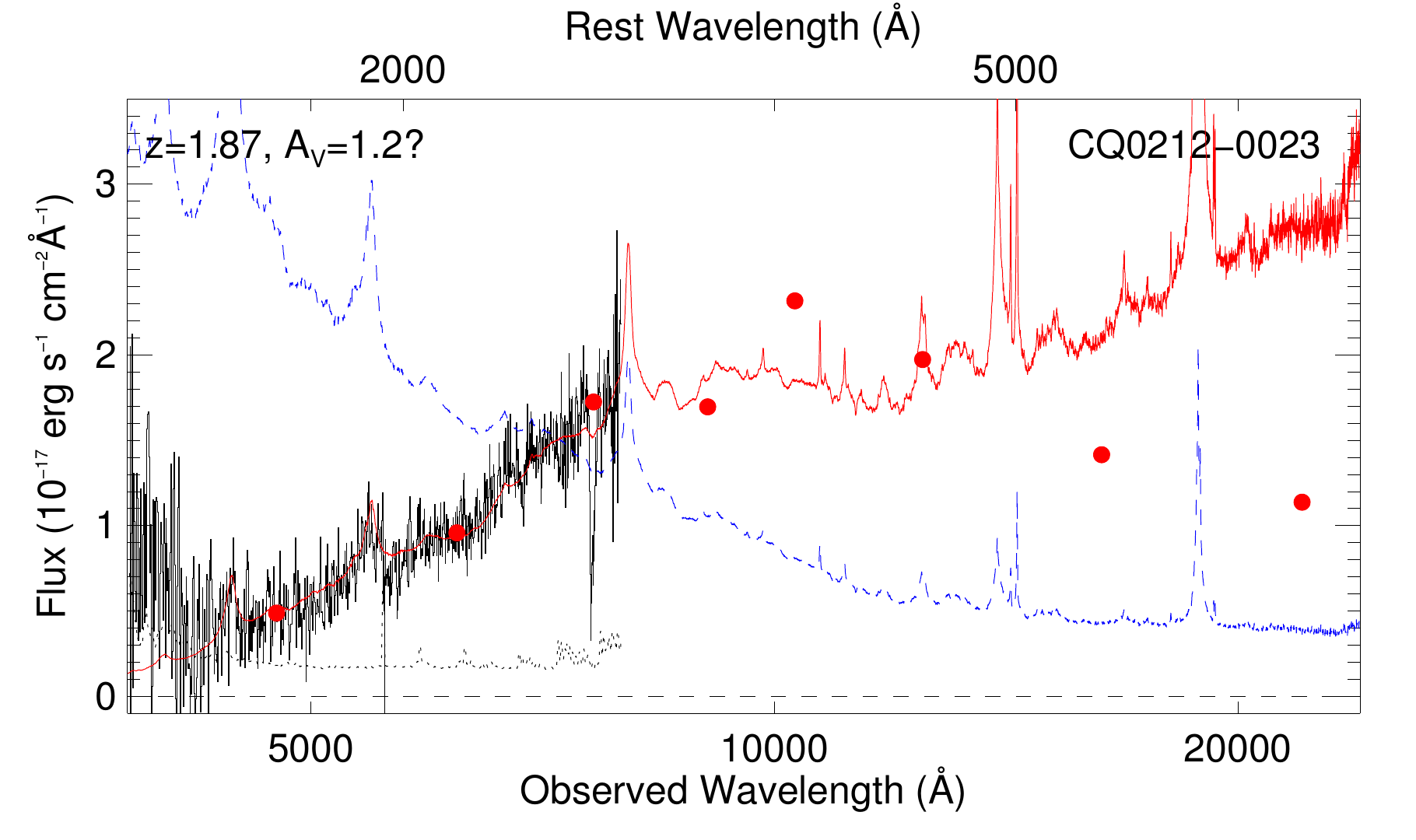}
\end{figure}
\begin{figure}
\plotone{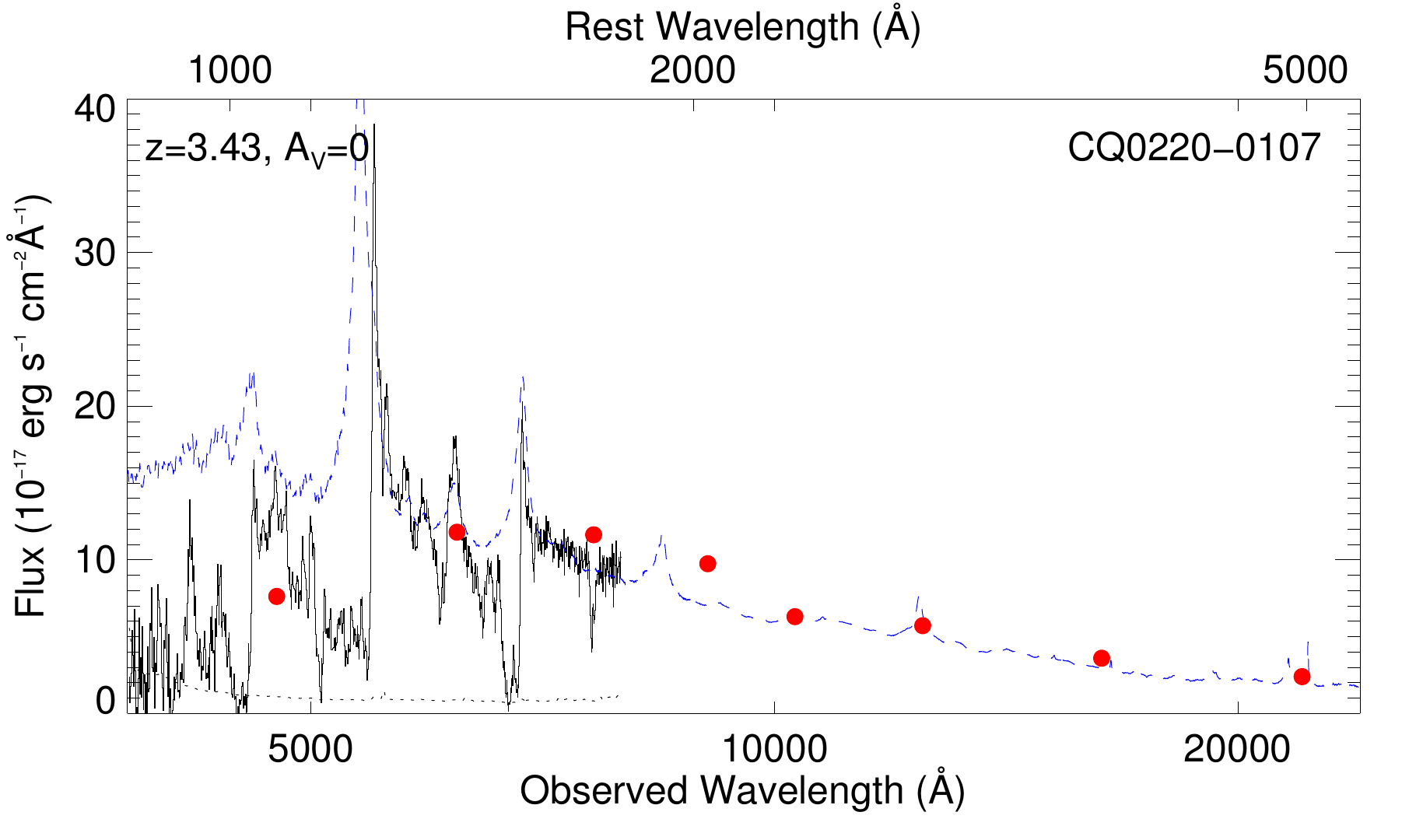}
\end{figure}
\clearpage
\begin{figure}
\plotone{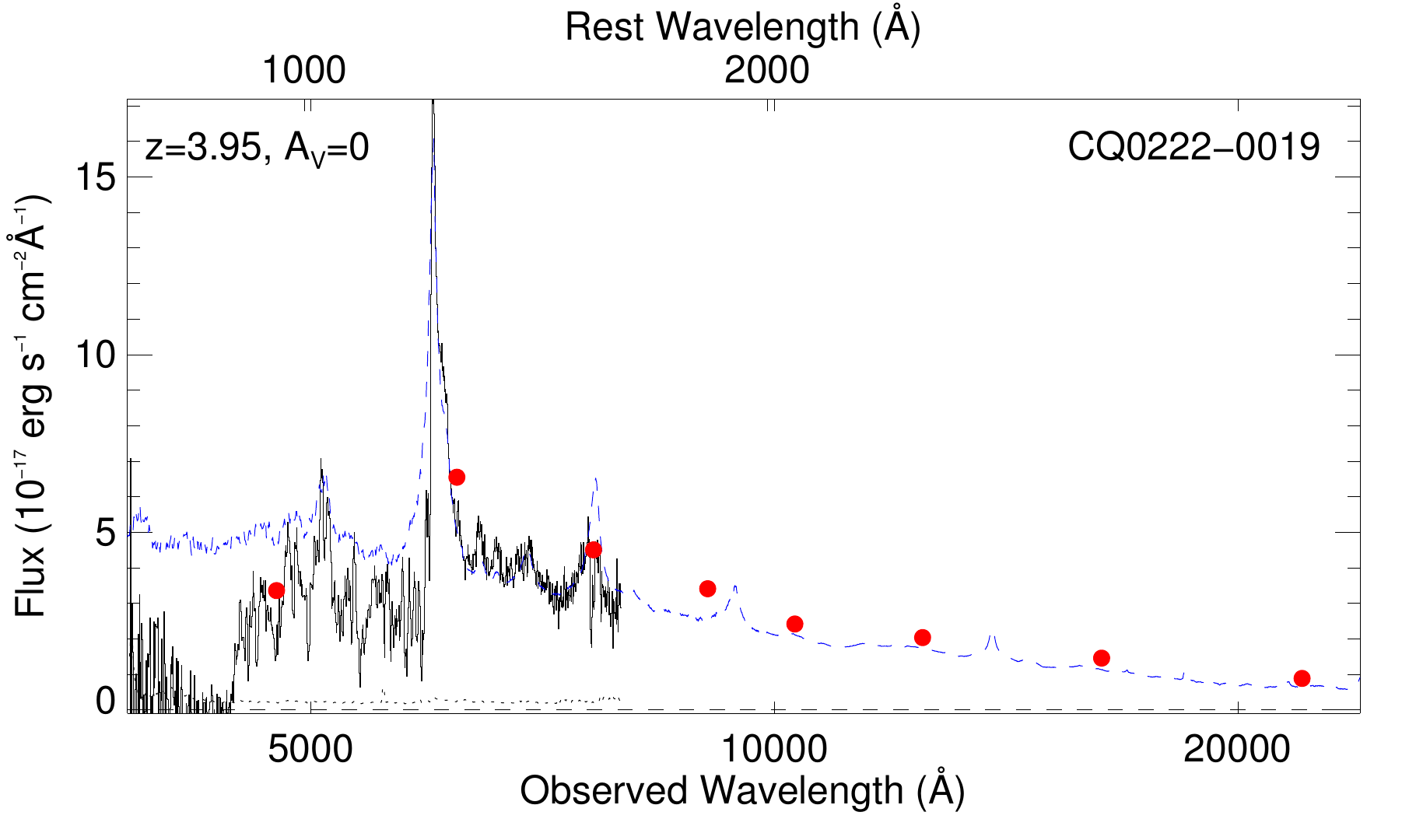}
\end{figure}
\begin{figure}
\plotone{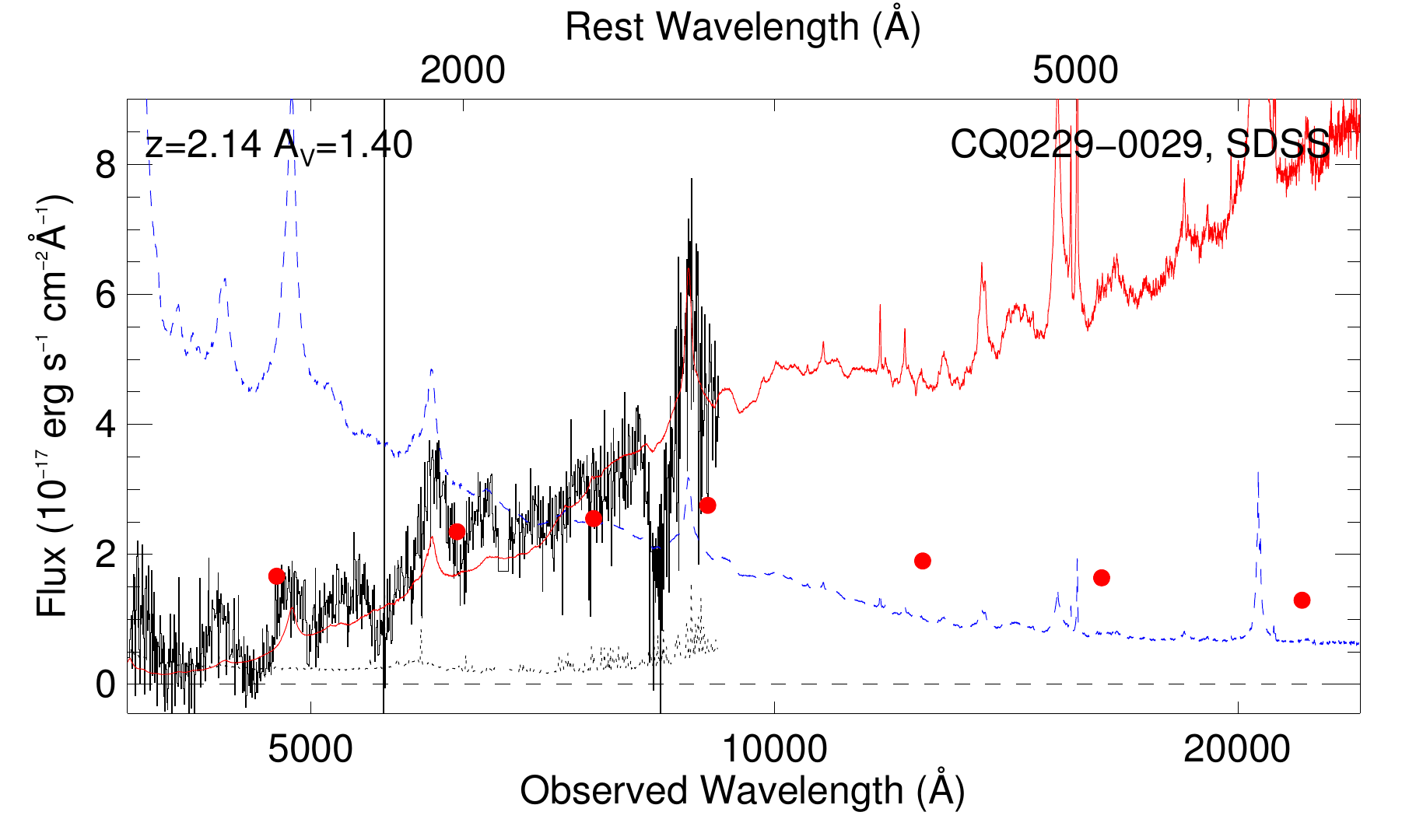}
\end{figure}
\begin{figure}
\plotone{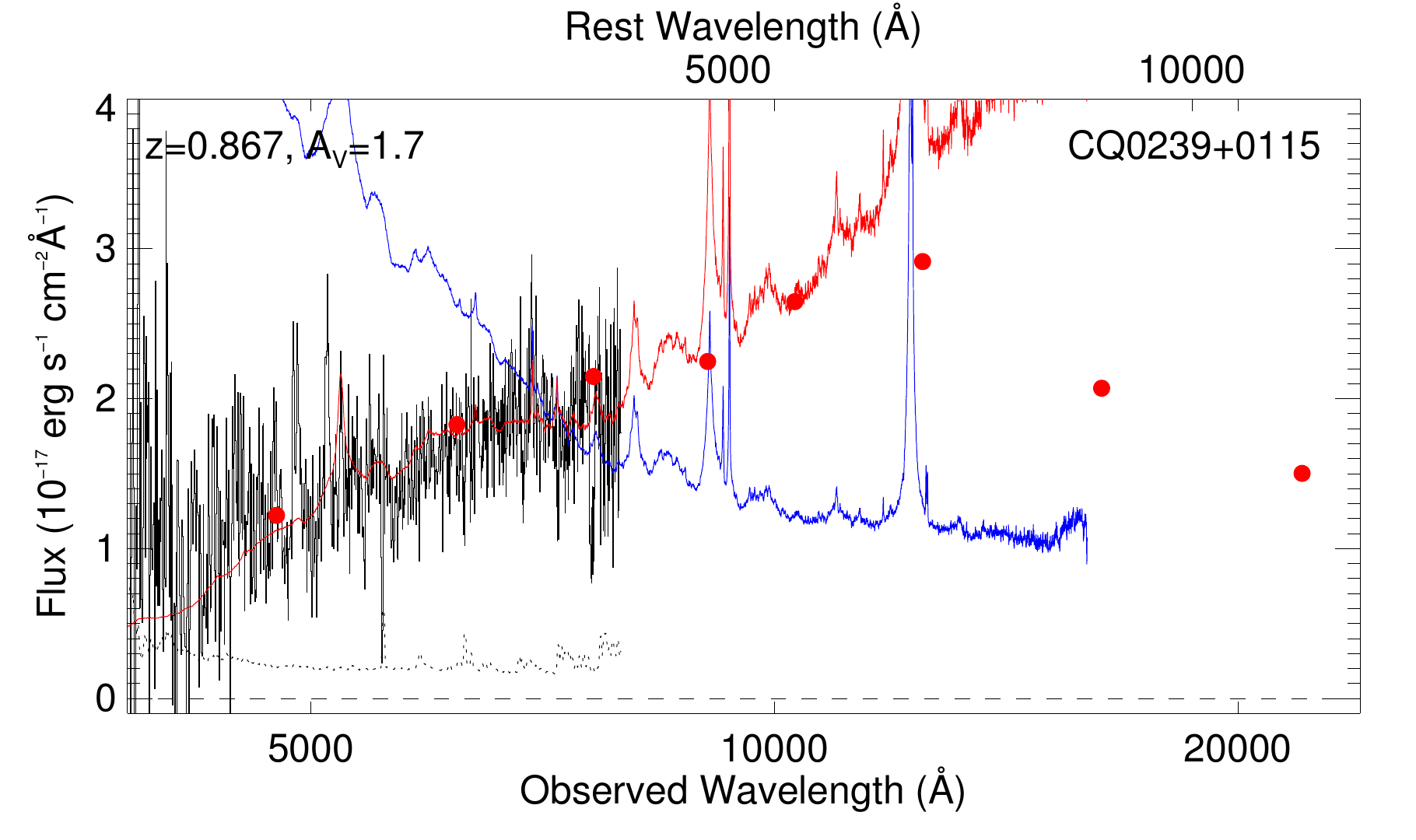}
\end{figure}
\begin{figure}
\plotone{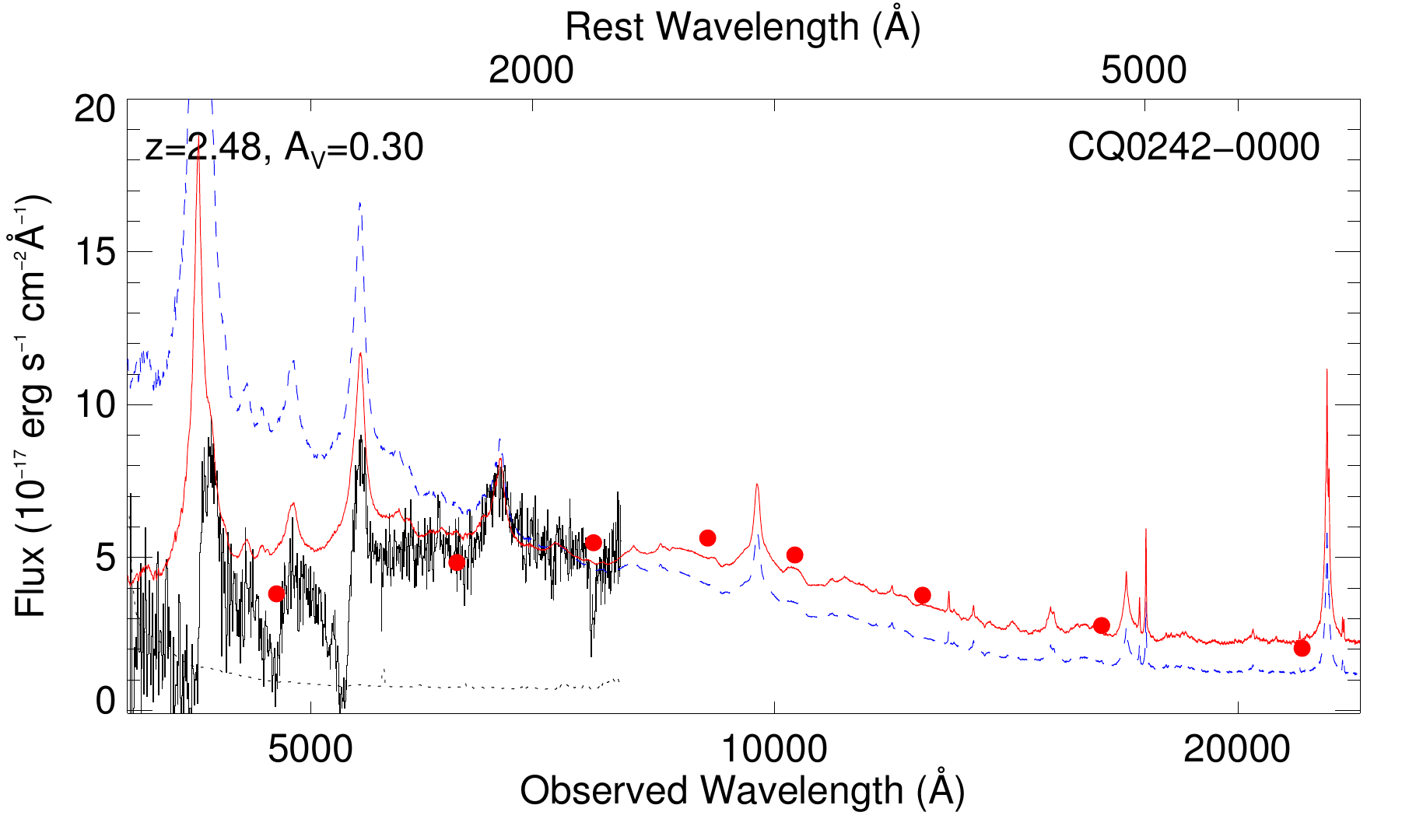}
\end{figure}
\begin{figure}
\plotone{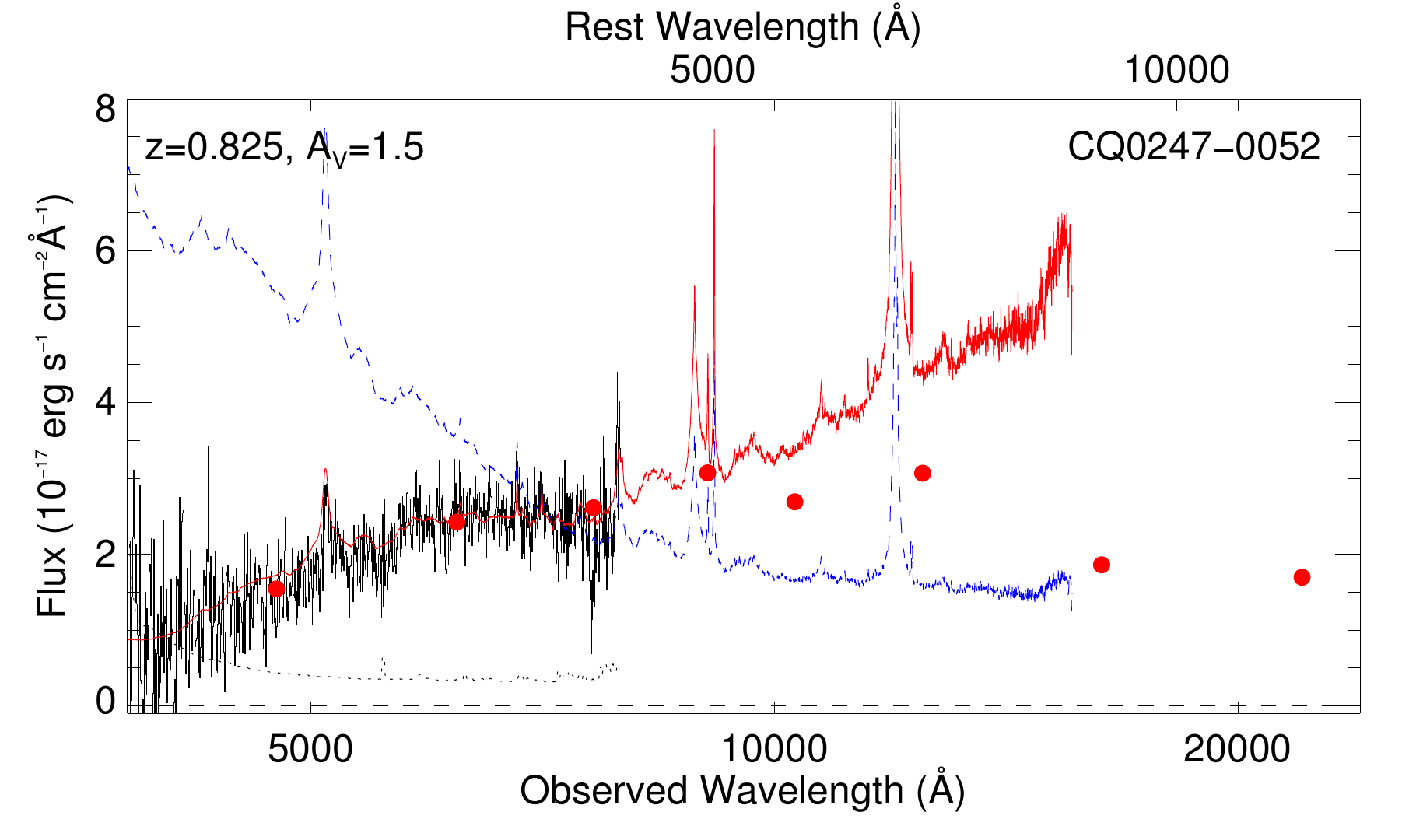}
\end{figure}
\begin{figure}
\plotone{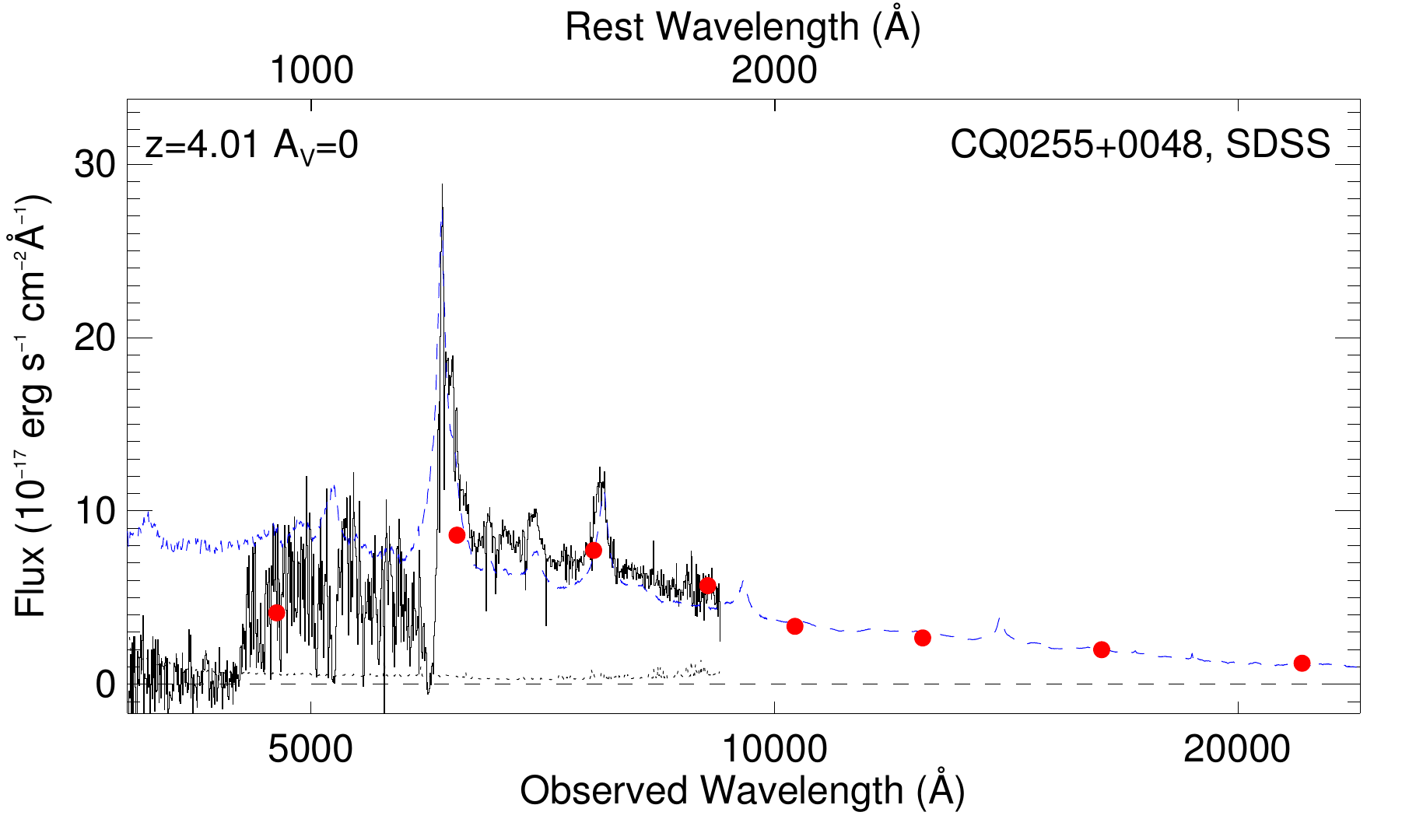}
\end{figure}
\begin{figure}
\plotone{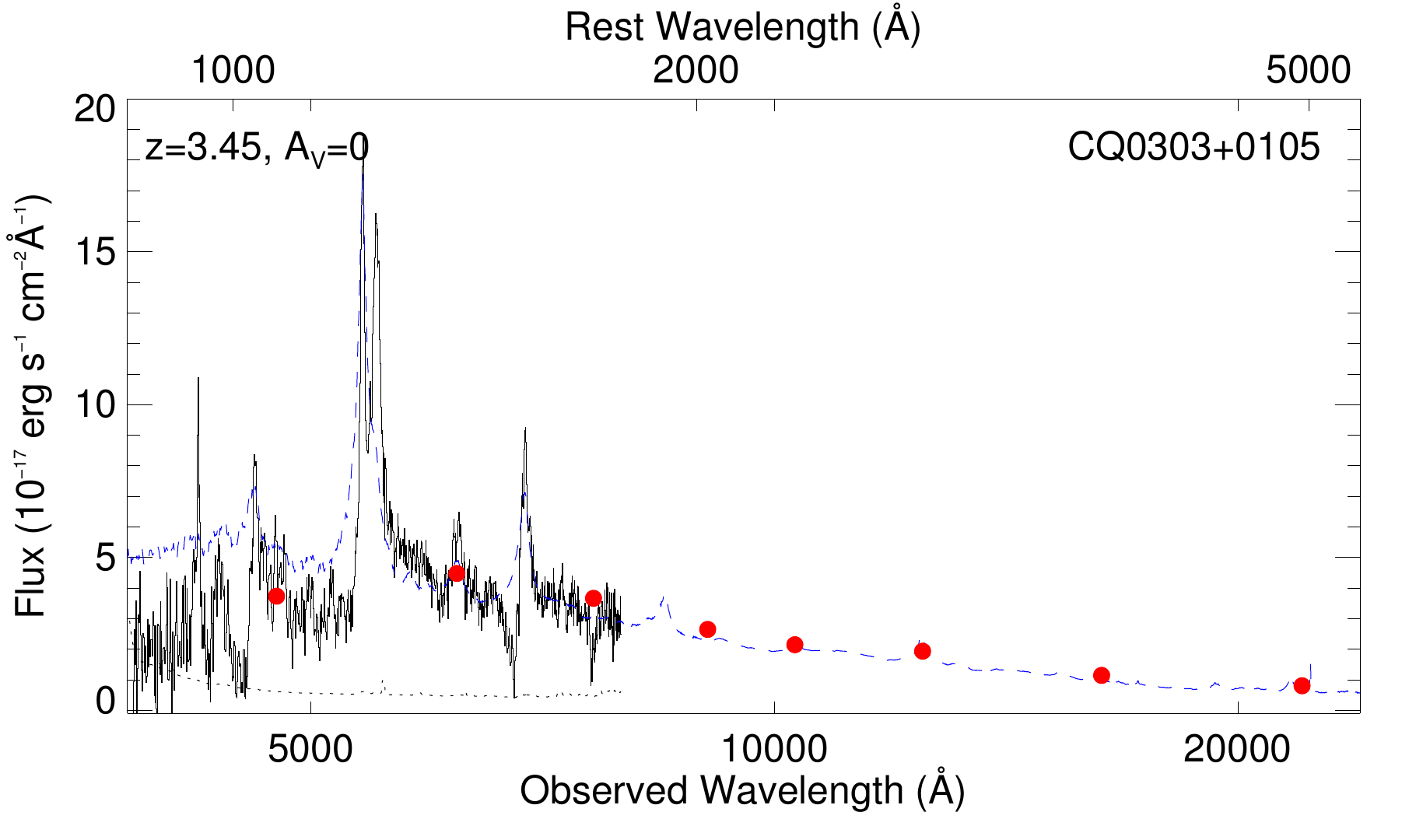}
\end{figure}
\begin{figure}
\plotone{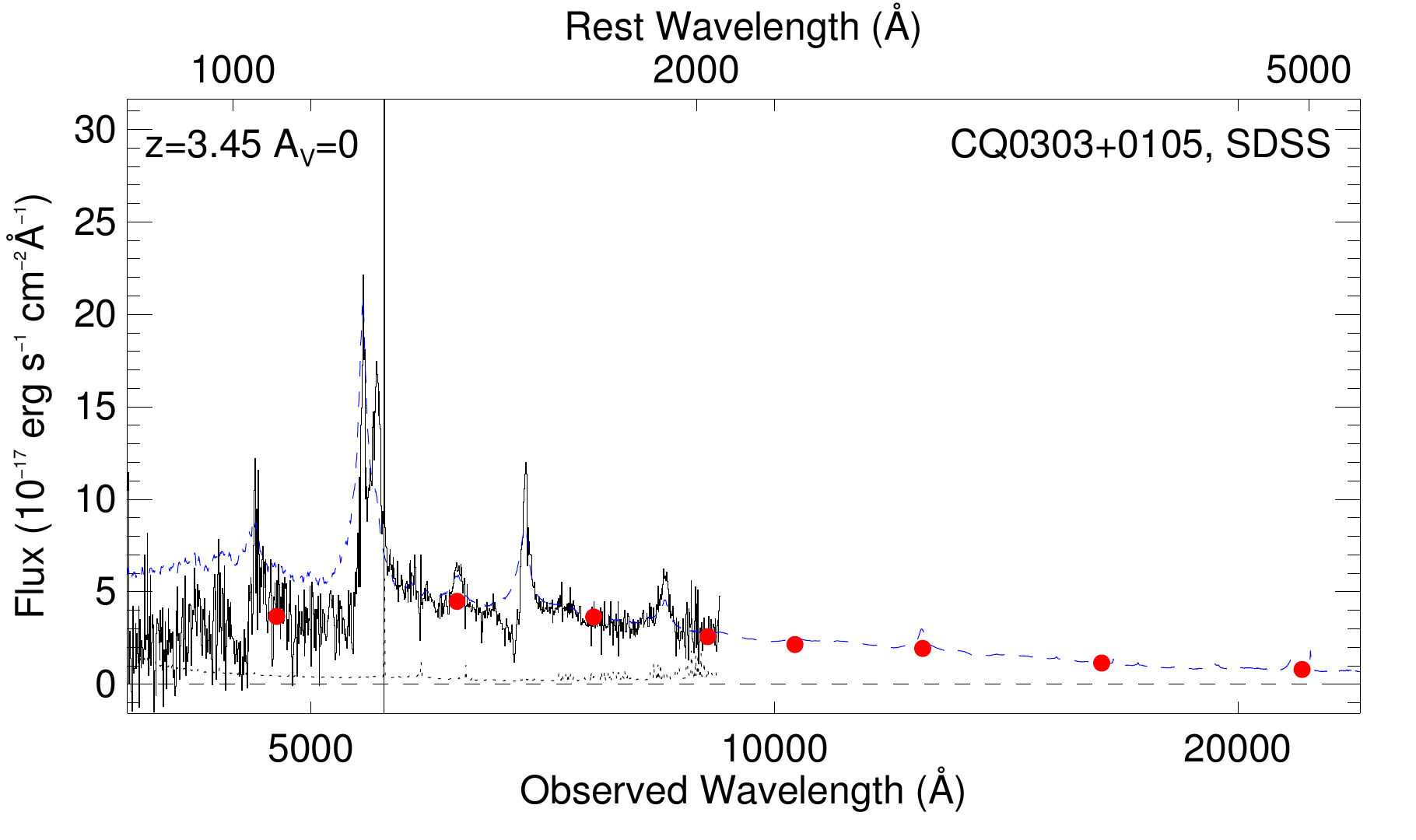}
\end{figure}
\begin{figure}
\plotone{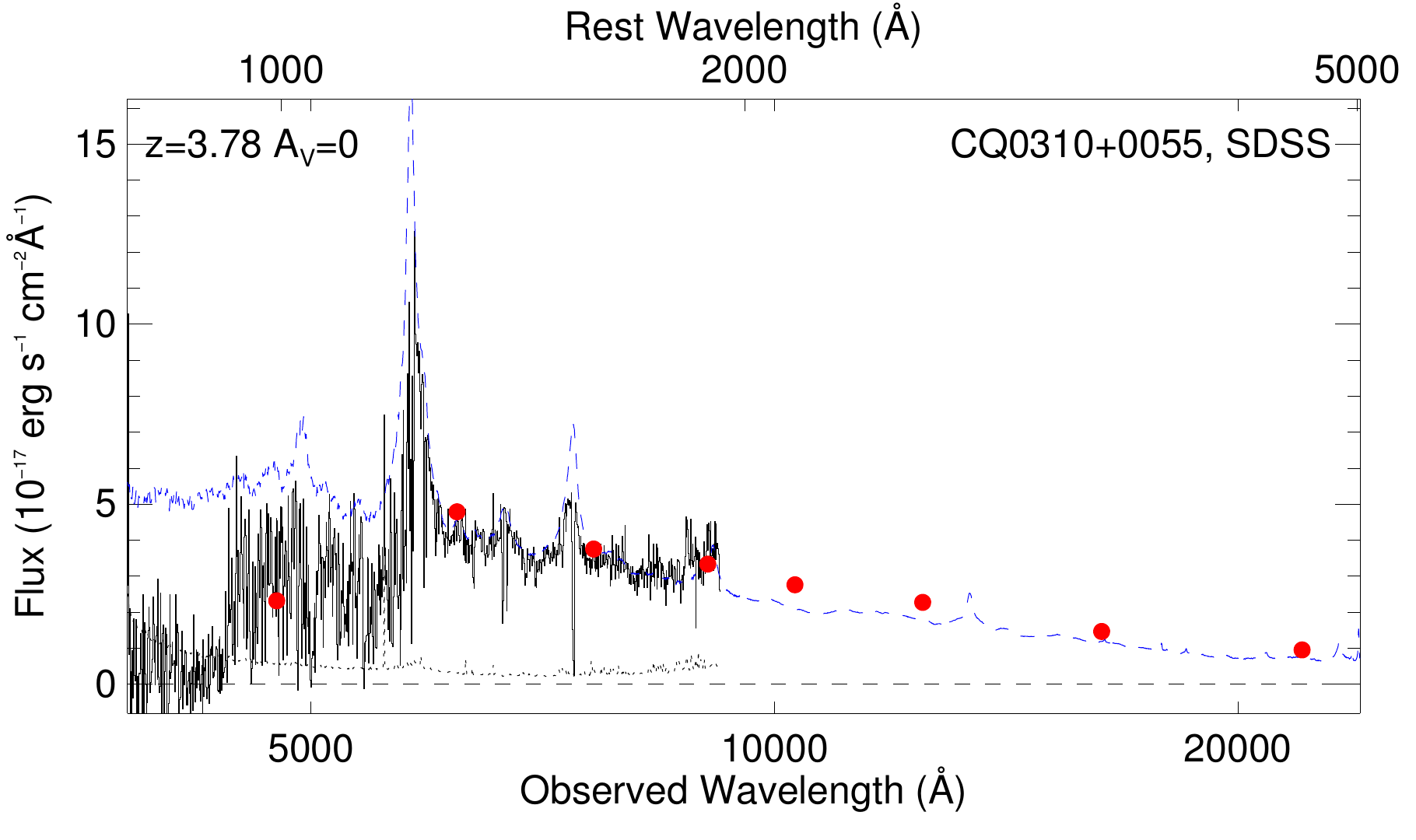}
\end{figure}
\begin{figure}
\plotone{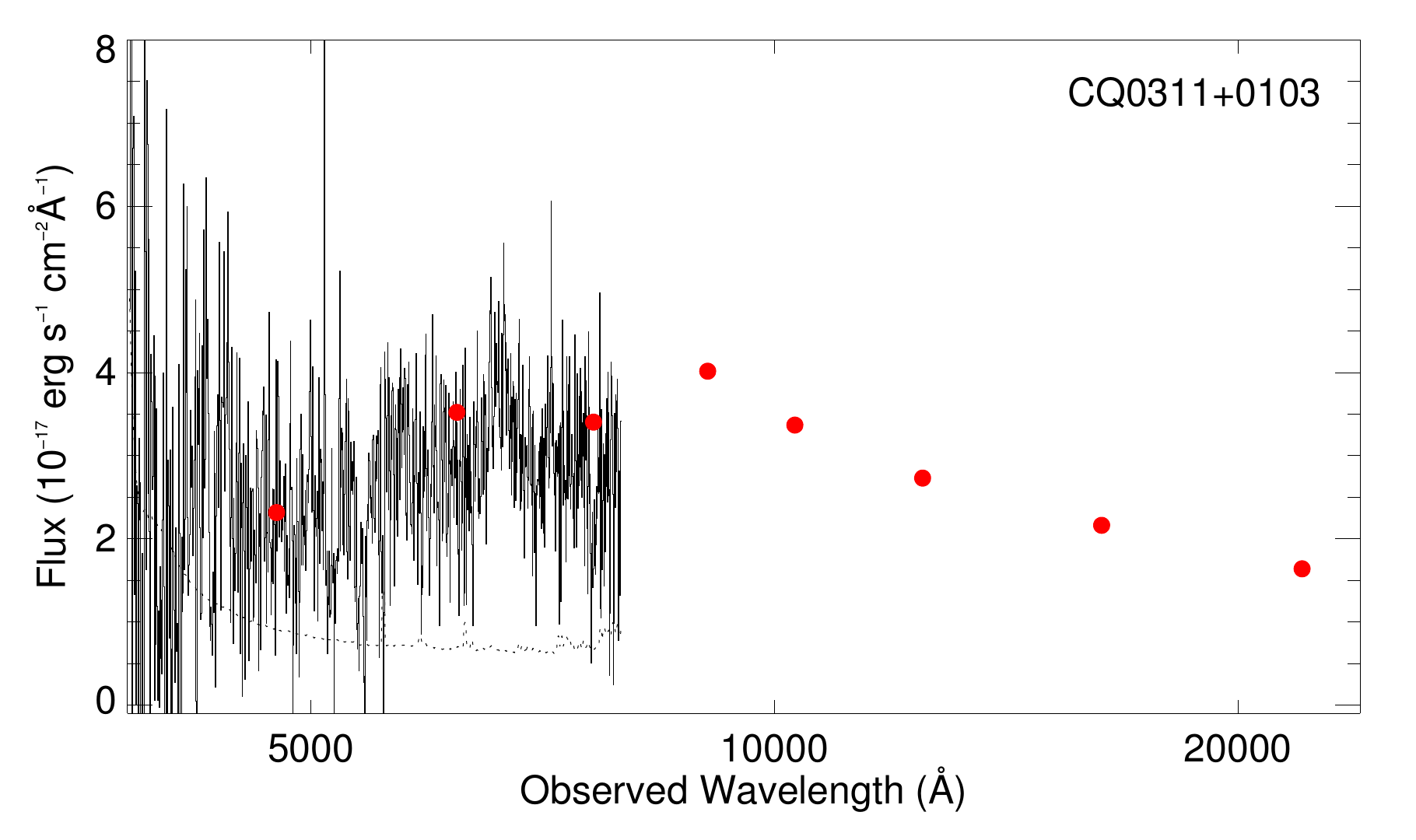}
\end{figure}
\begin{figure}
\plotone{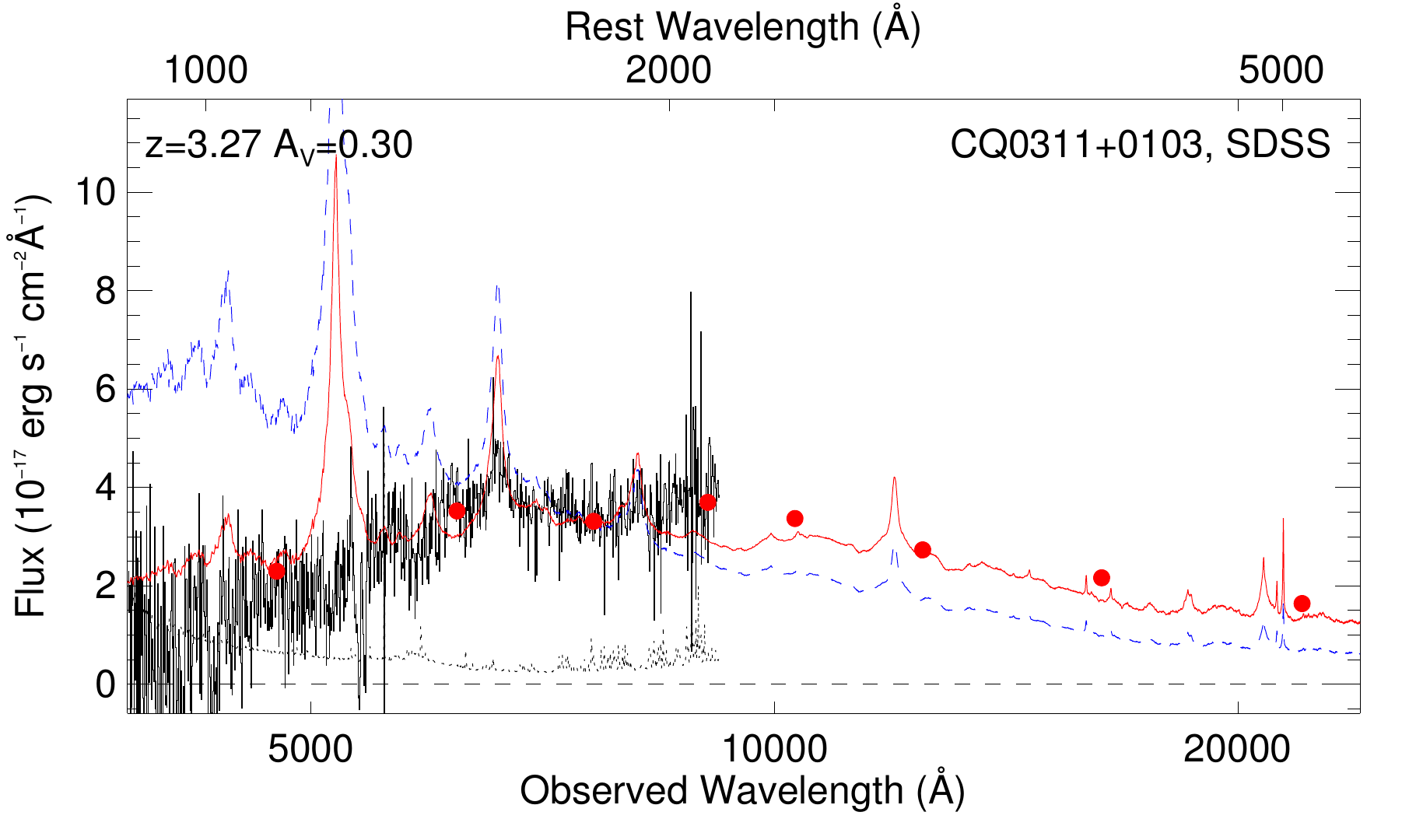}
\end{figure}
\begin{figure}
\plotone{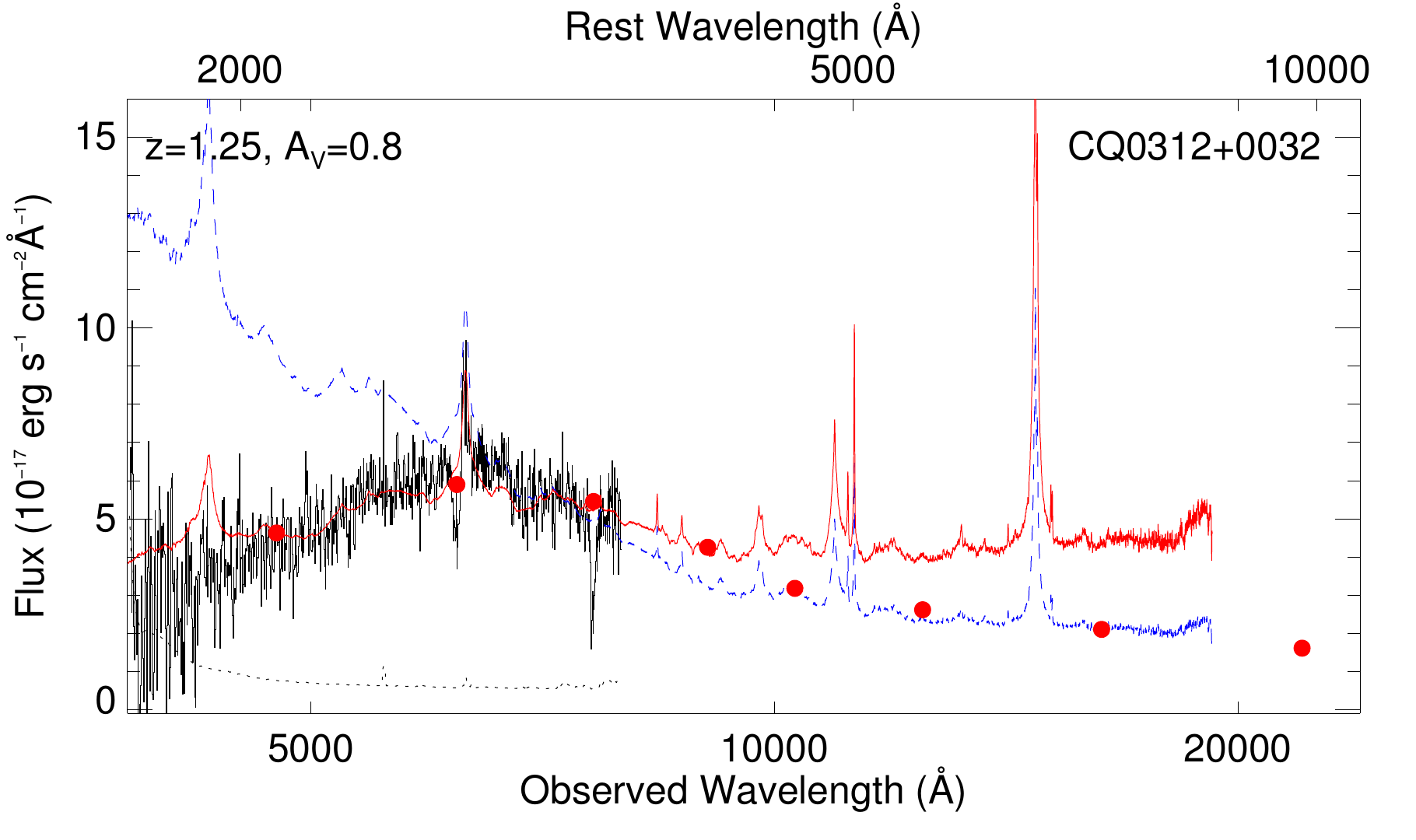}
\end{figure}
\begin{figure}
\plotone{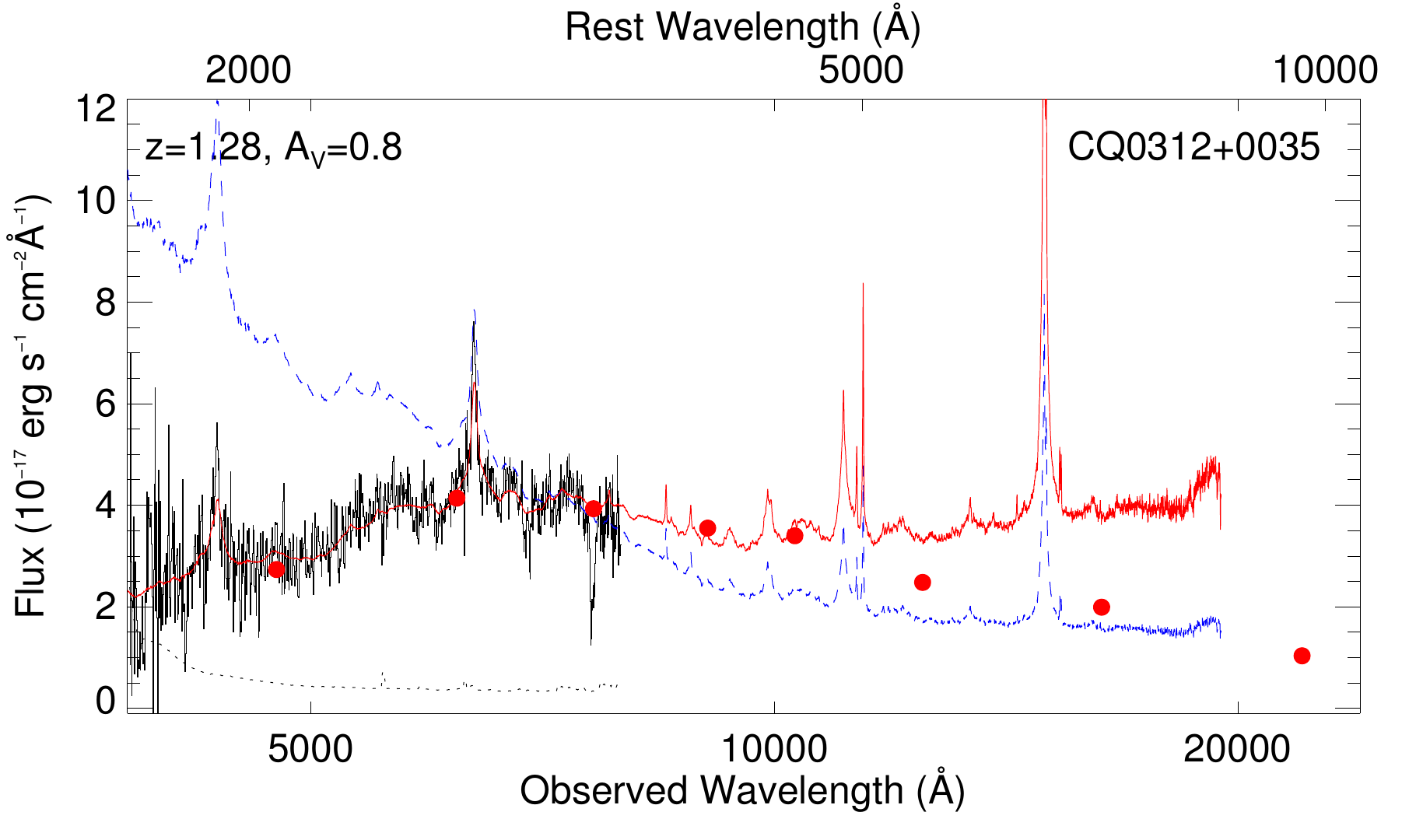}
\end{figure}
\begin{figure}
\plotone{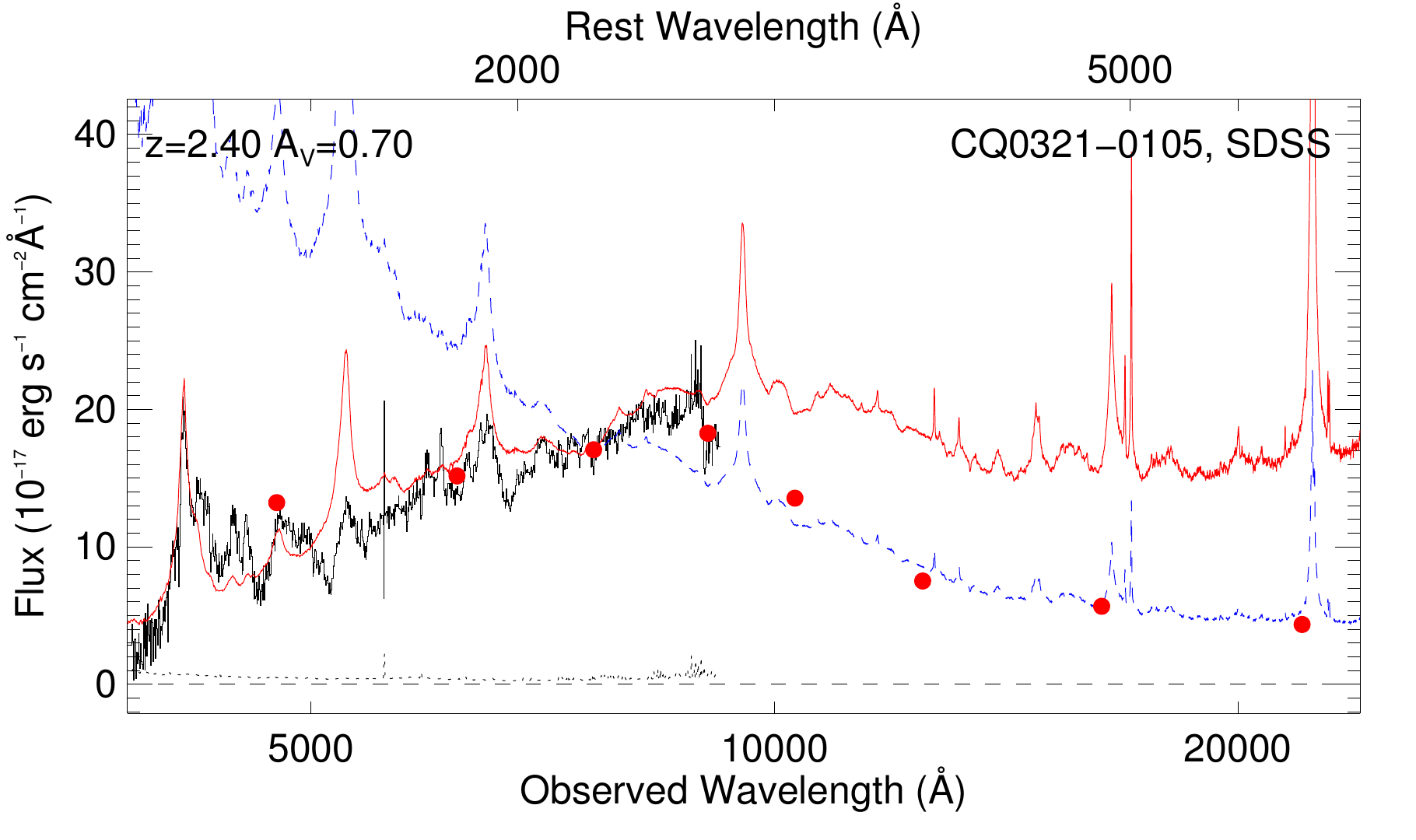}
\end{figure}
\begin{figure}
\plotone{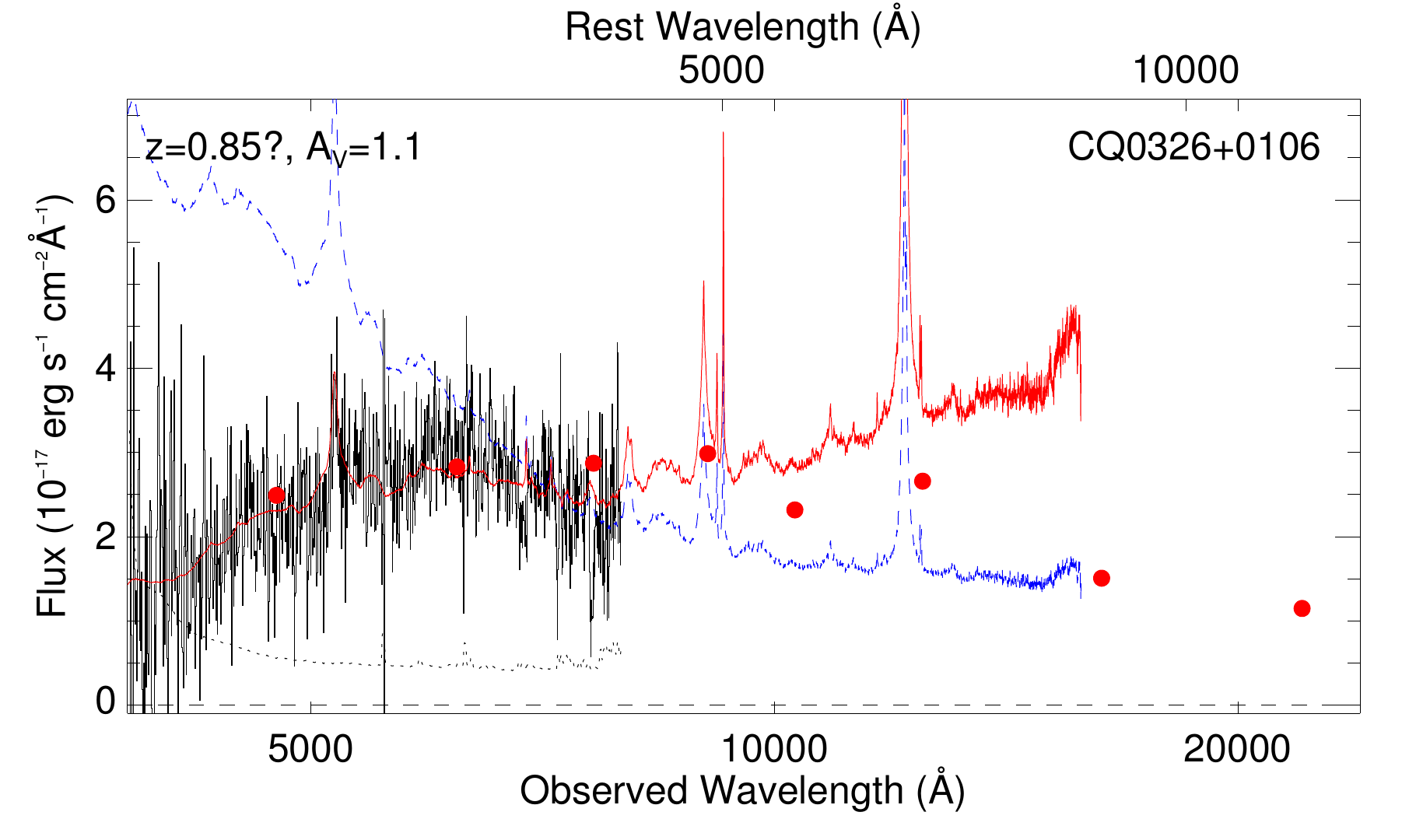}
\end{figure}
\begin{figure}
\plotone{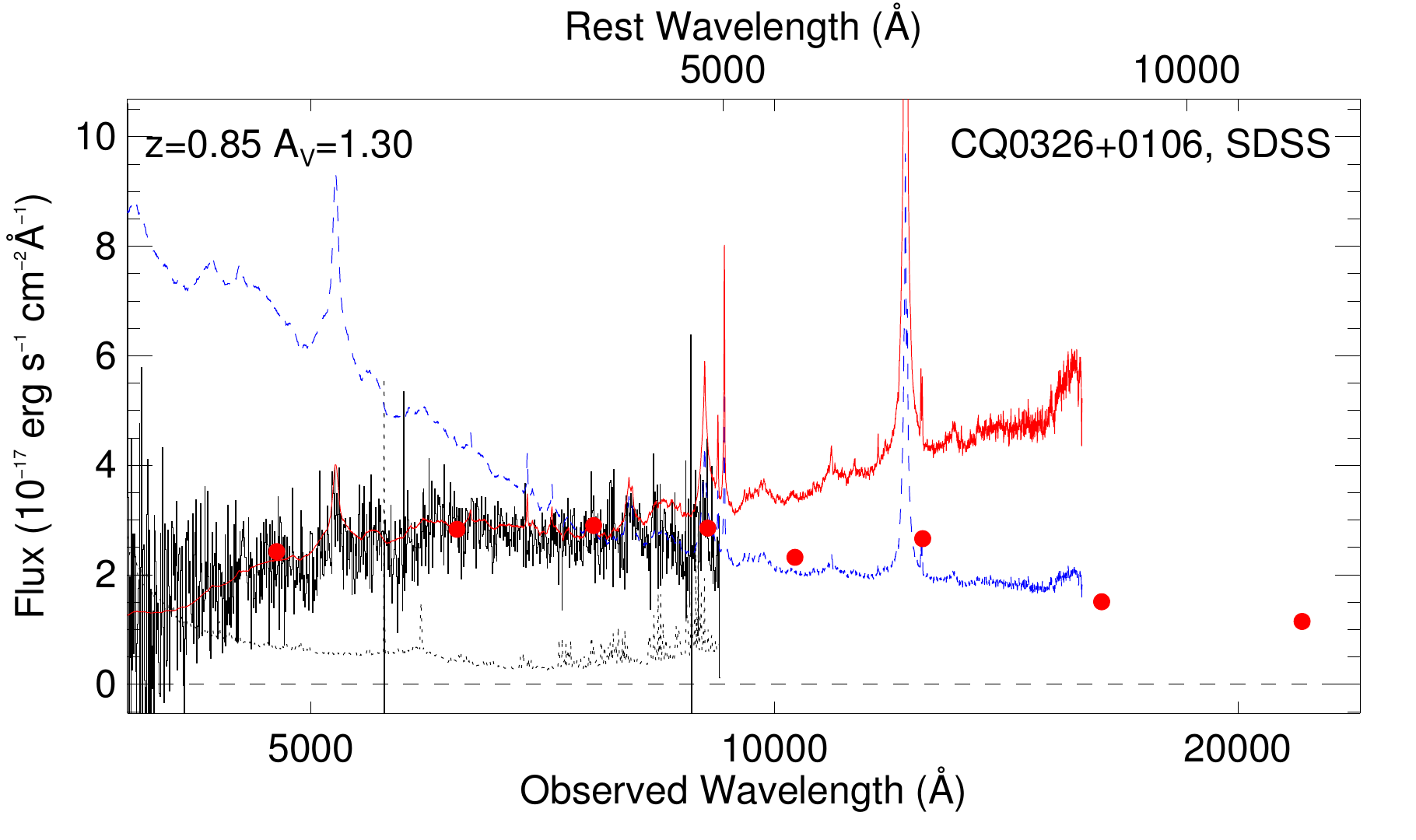}
\end{figure}
\begin{figure}
\plotone{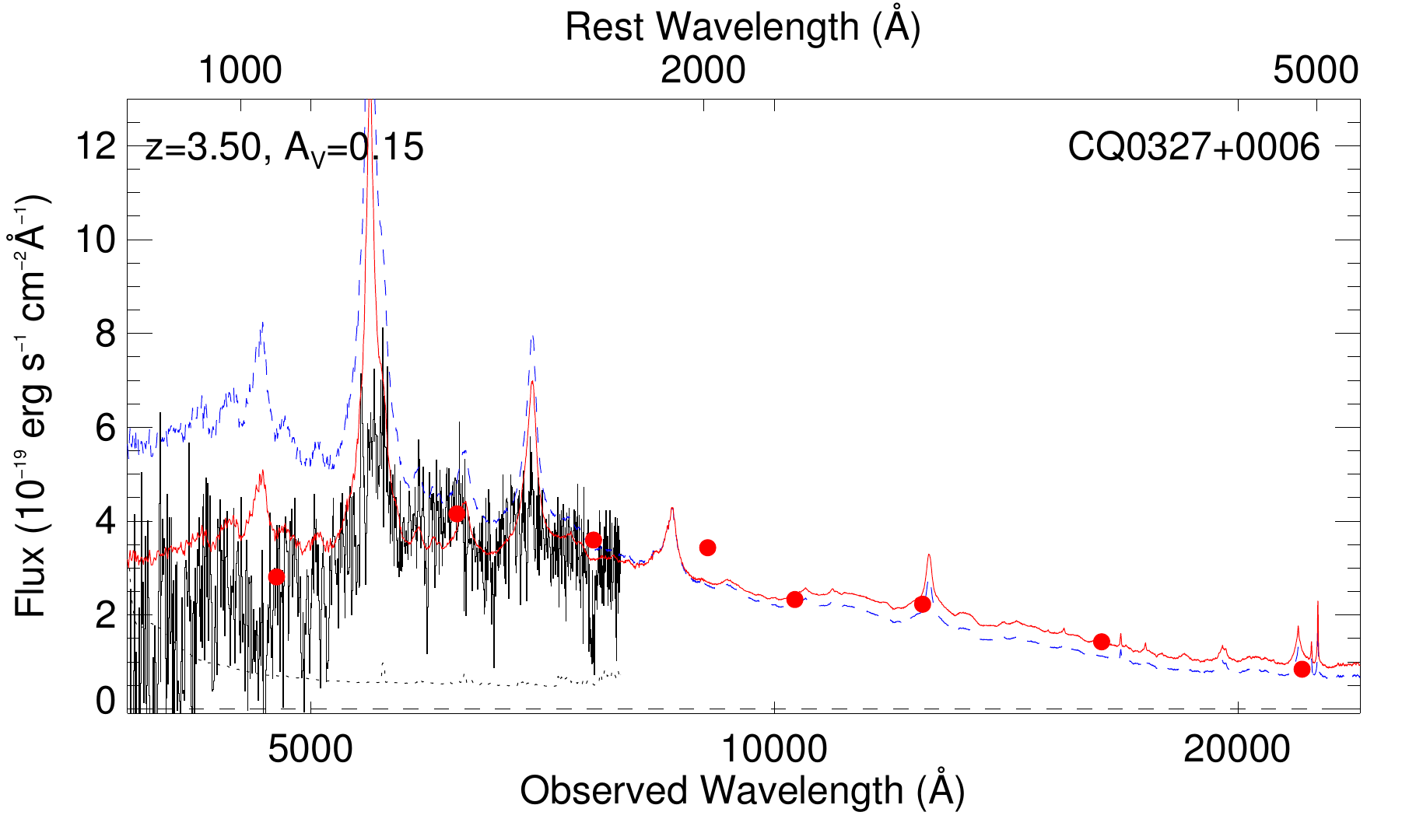}
\end{figure}
\begin{figure}
\plotone{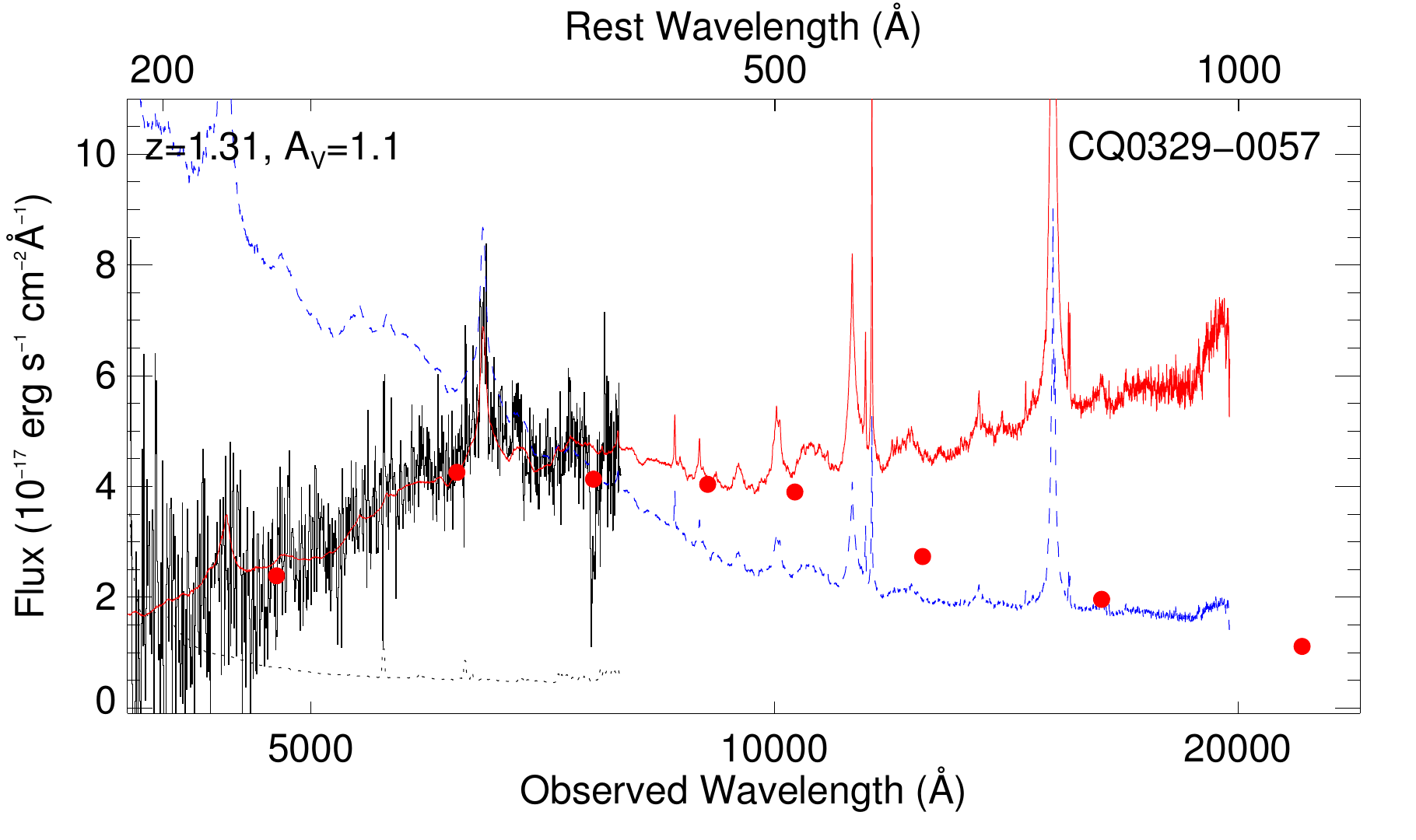}
\end{figure}
\clearpage
\begin{figure}
\plotone{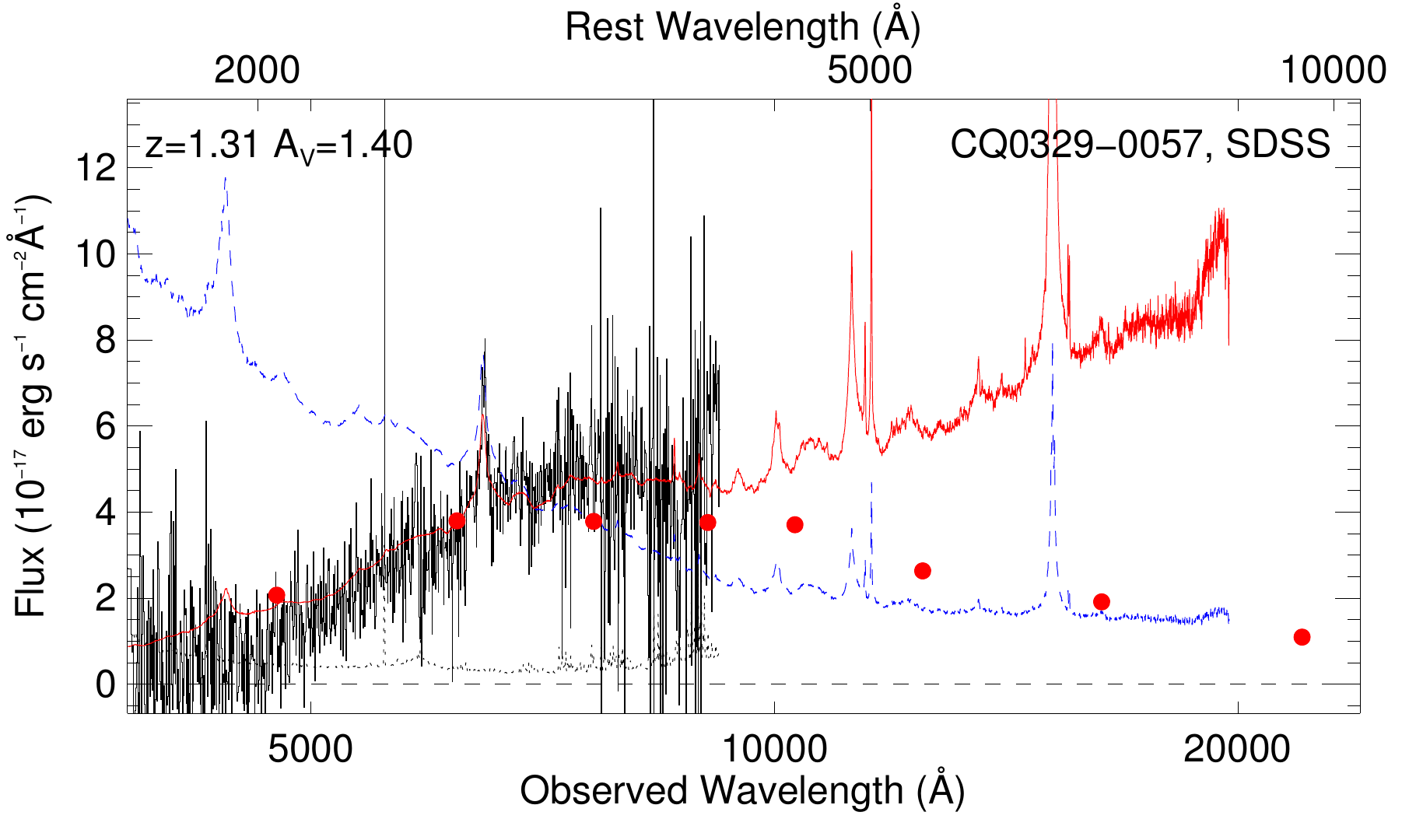}
\end{figure}
\begin{figure}
\plotone{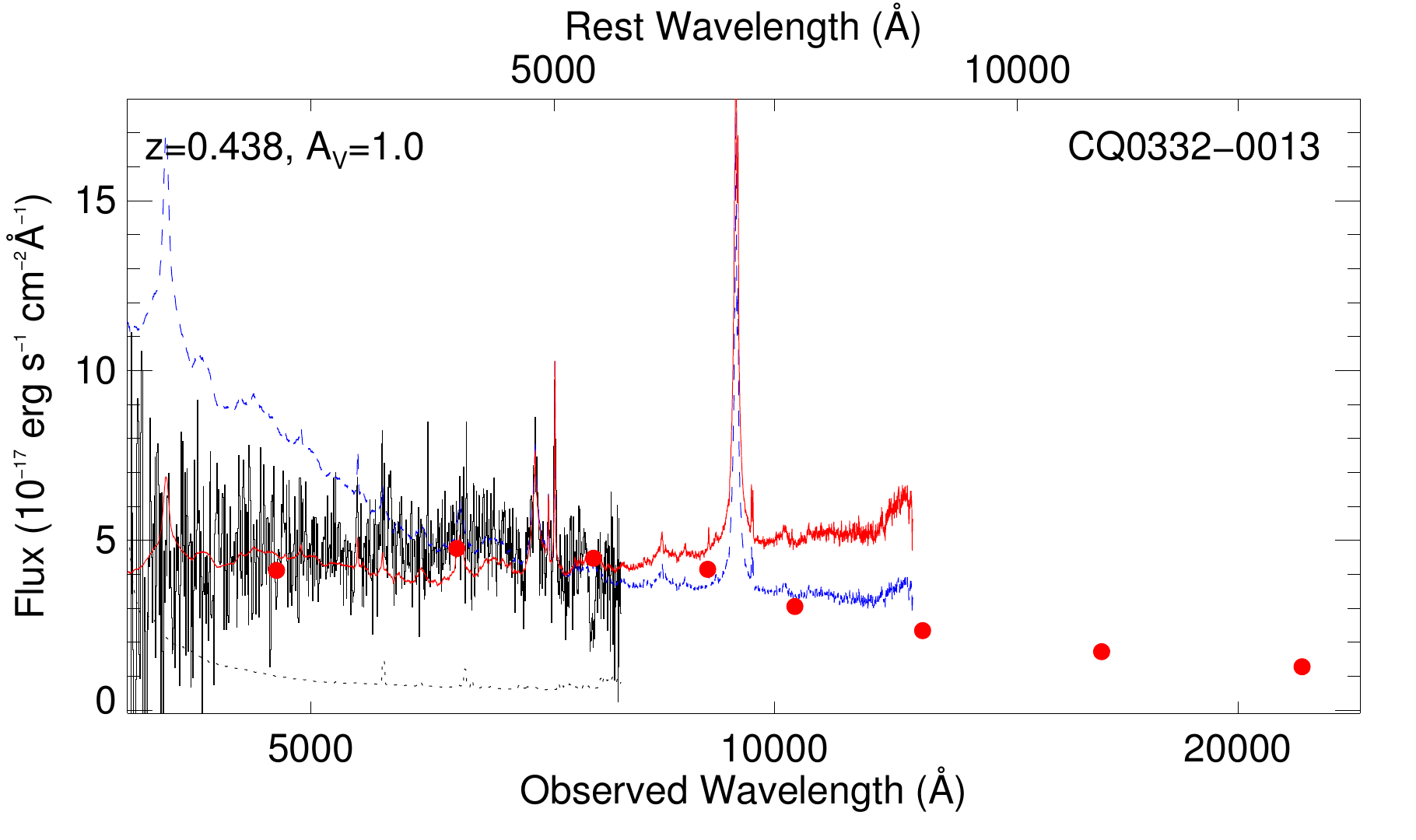}
\end{figure}
\begin{figure}
\plotone{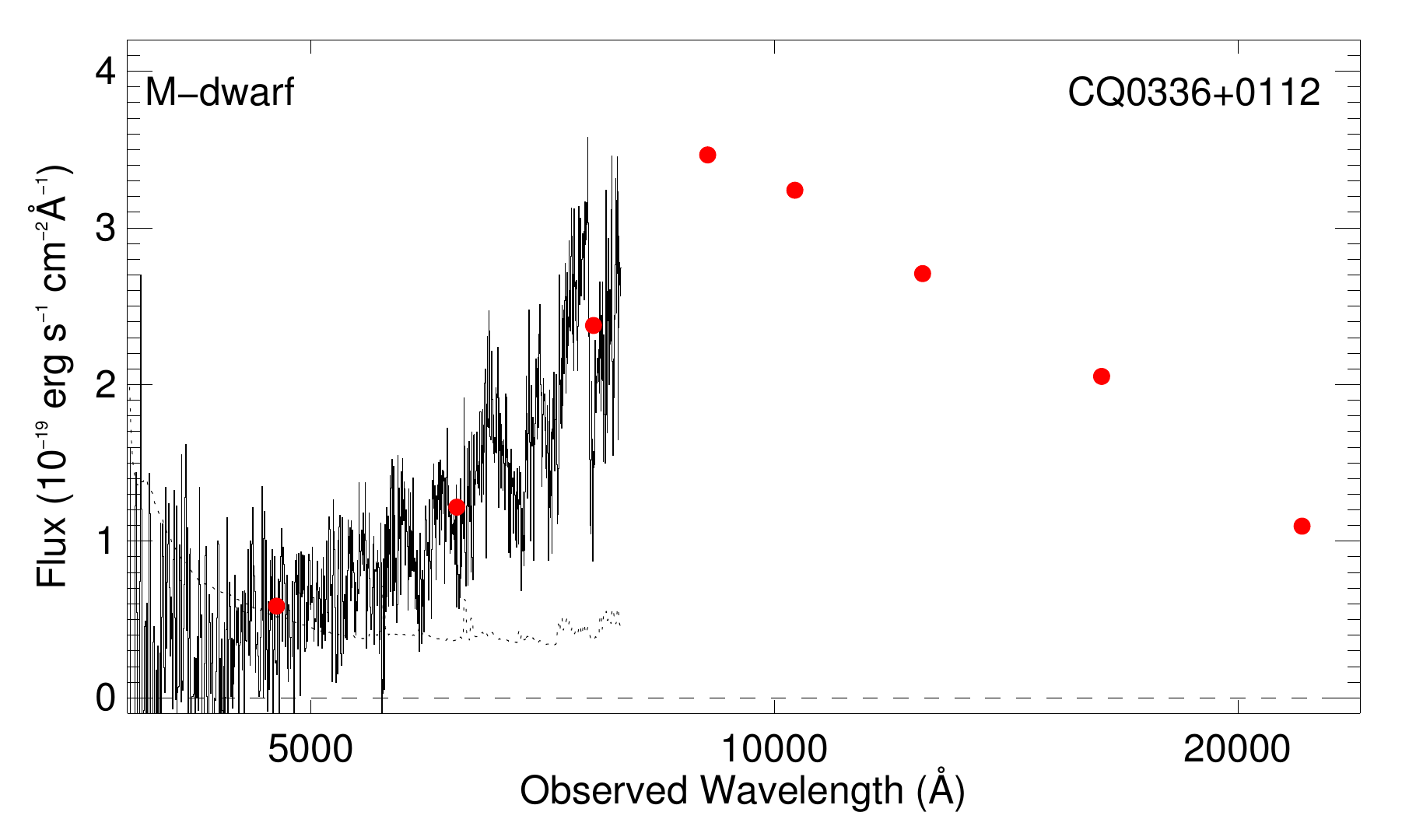}
\end{figure}
\begin{figure}
\plotone{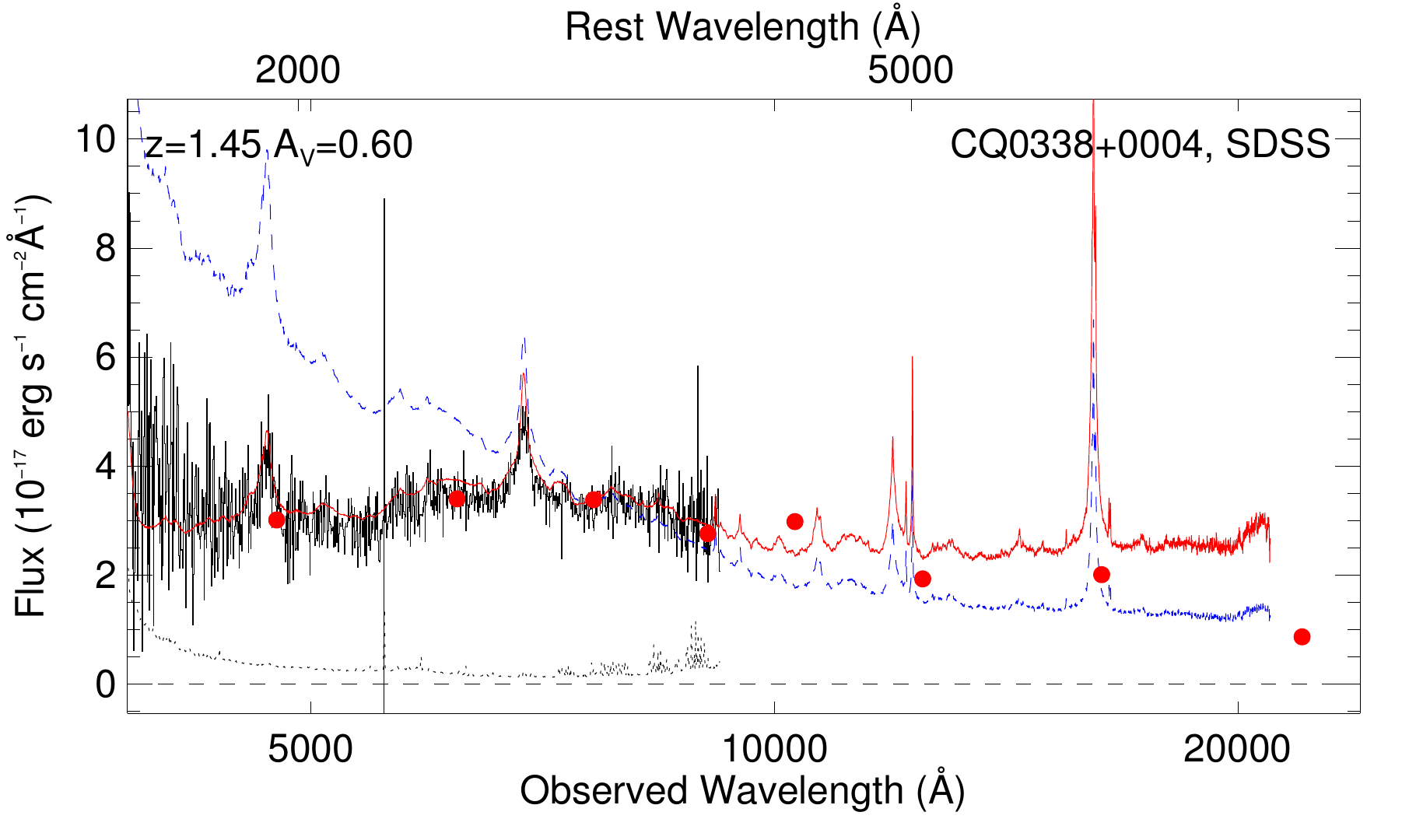}
\end{figure}
\begin{figure}
\plotone{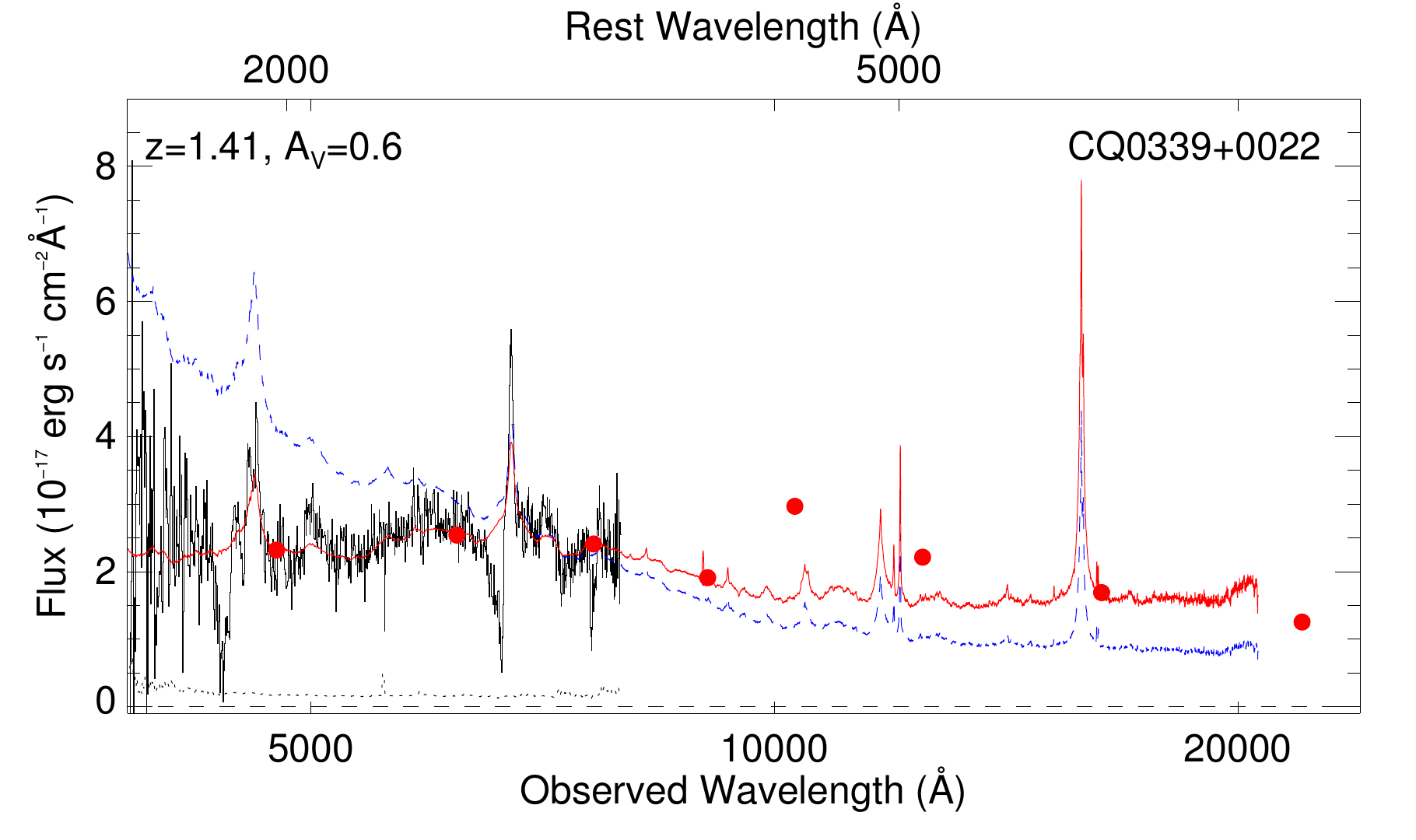}
\end{figure}
\begin{figure}
\plotone{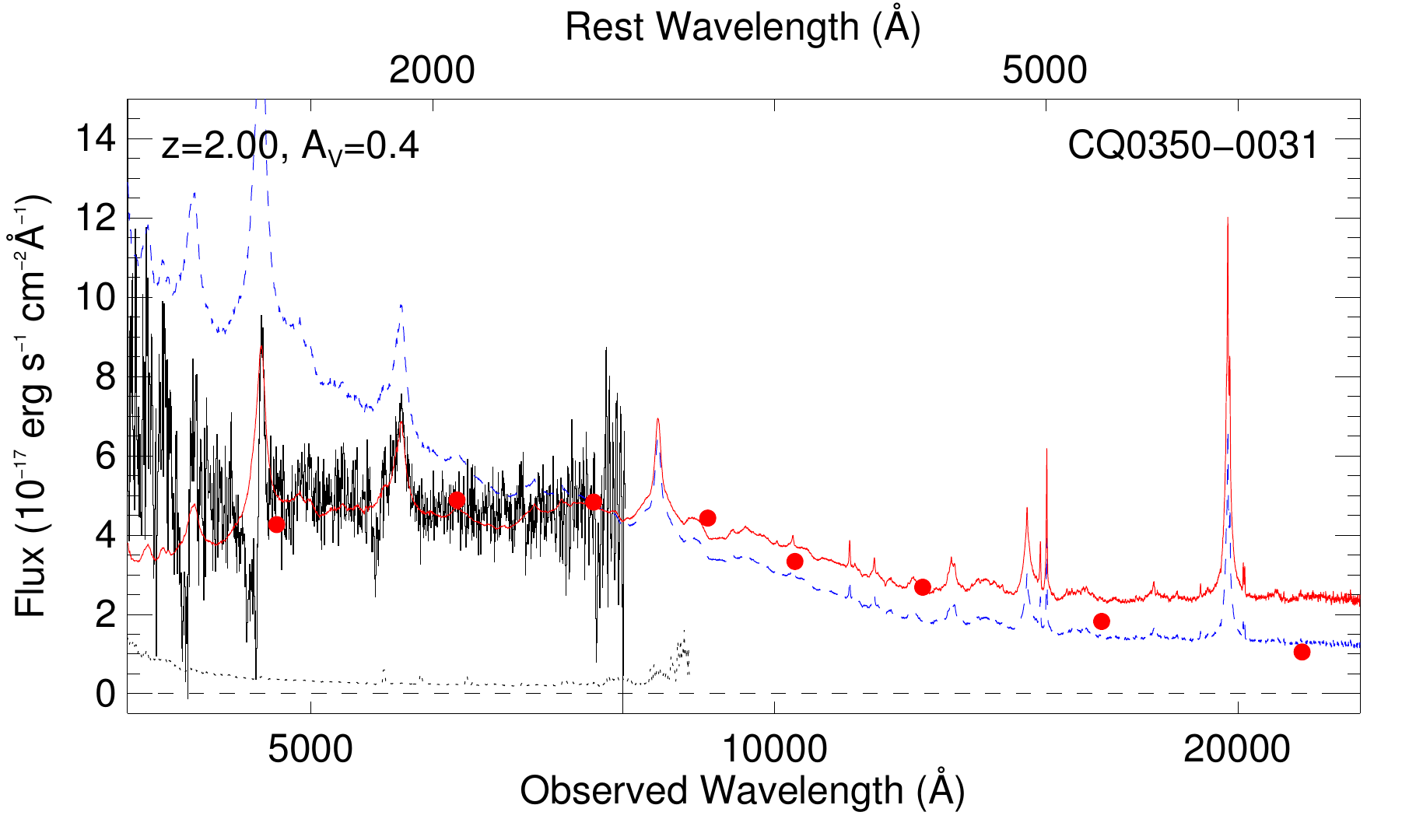}
\end{figure}
\begin{figure}
\plotone{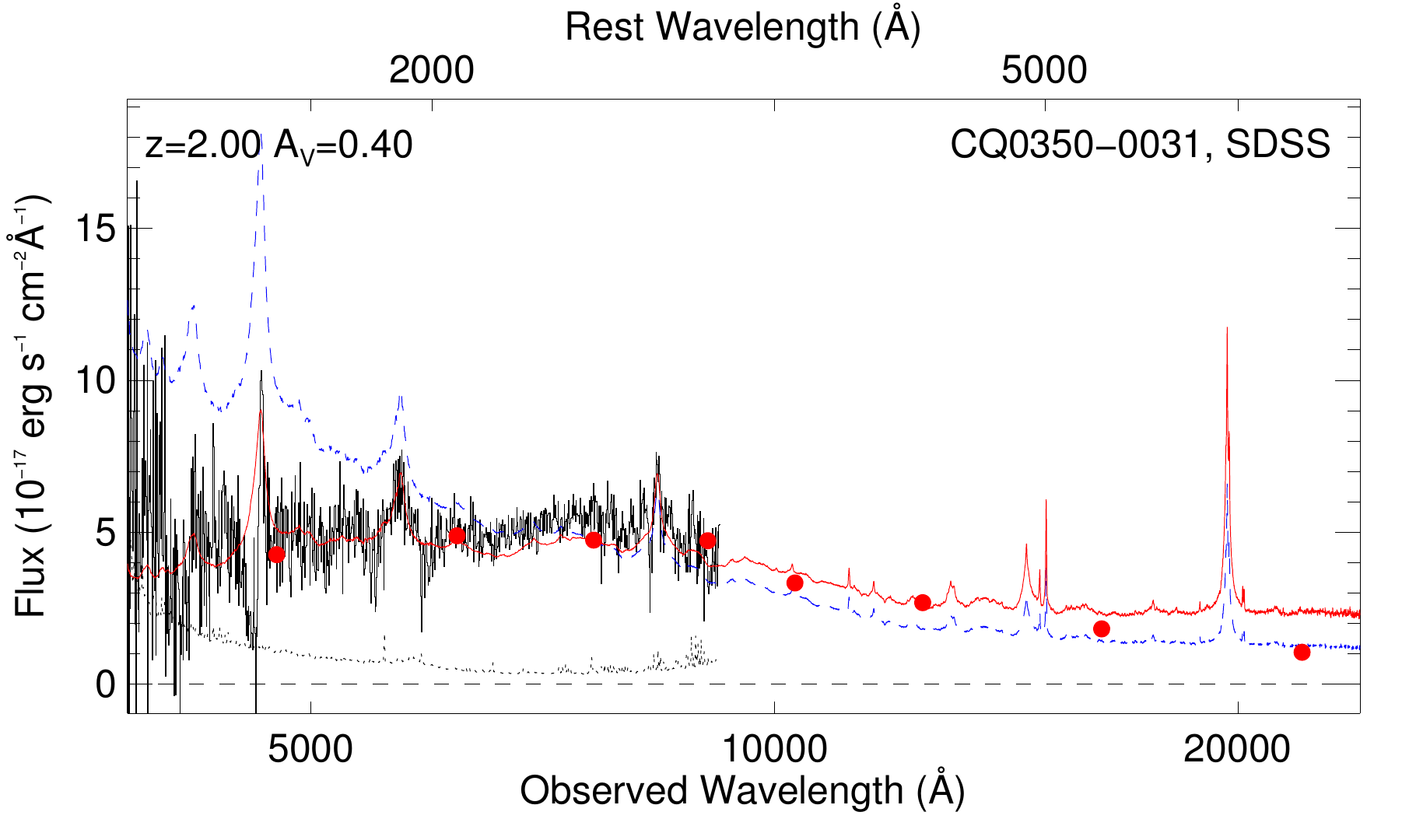}
\end{figure}
\begin{figure}
\plotone{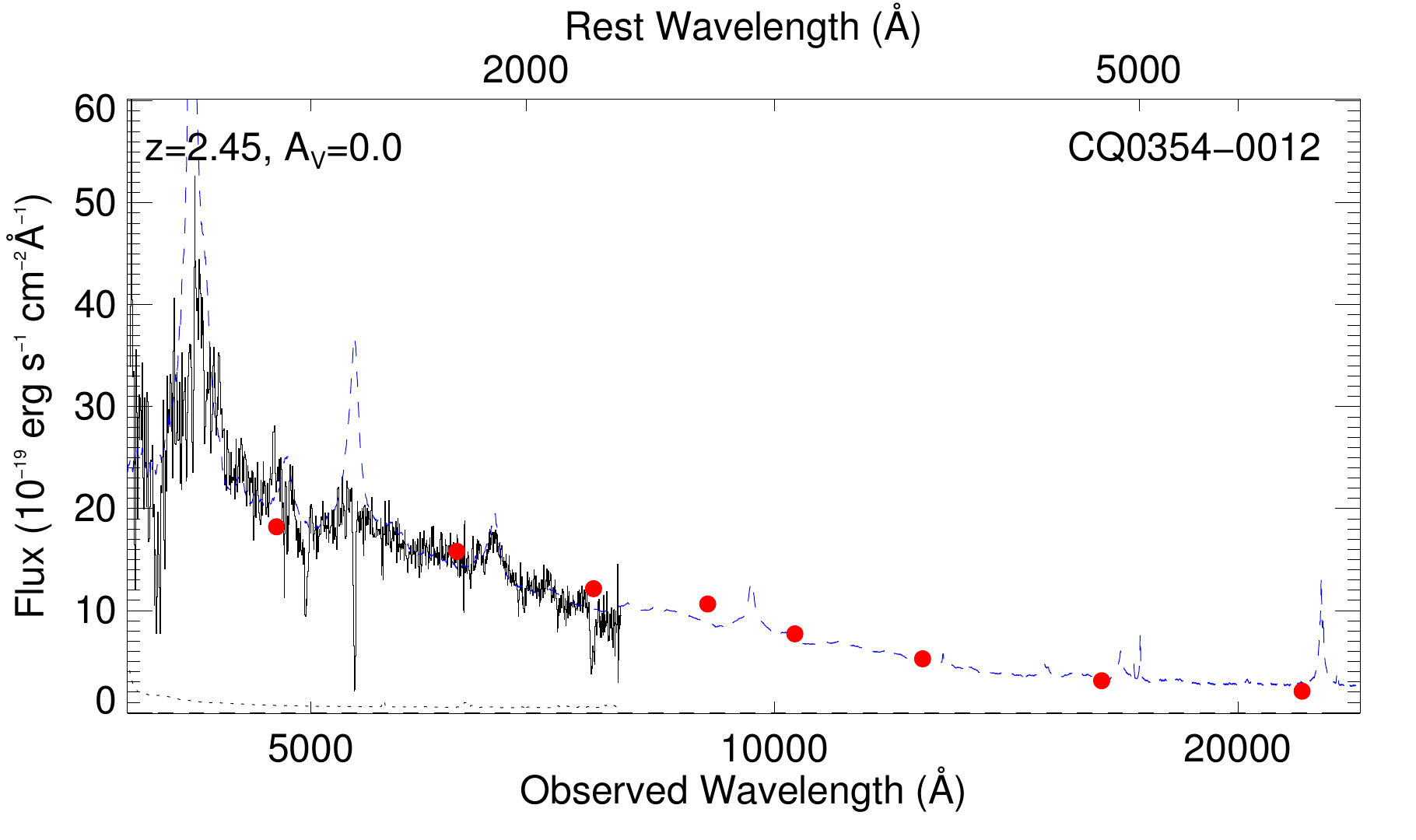}
\end{figure}
\begin{figure}
\plotone{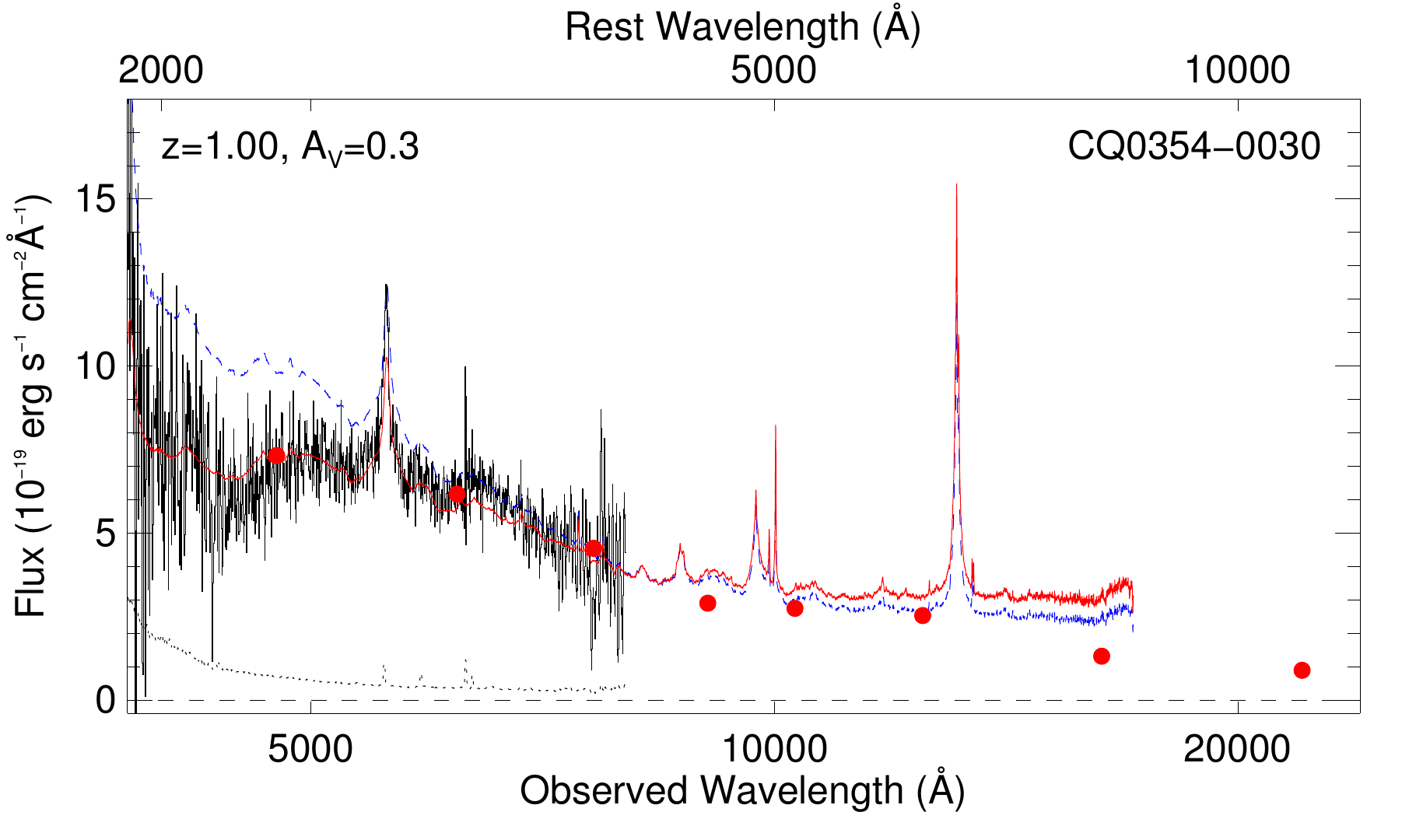}
\end{figure}
\begin{figure}
\plotone{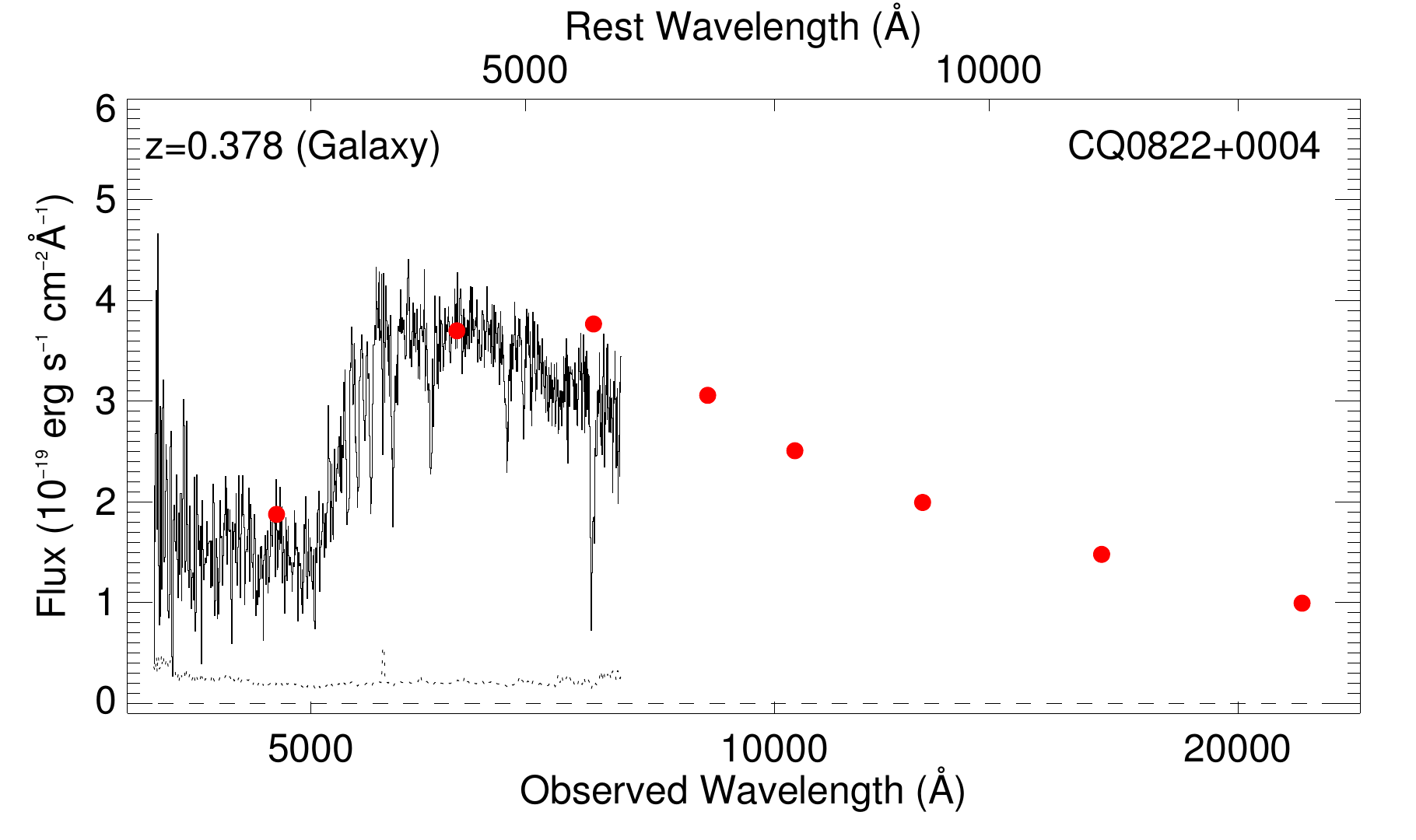}
\end{figure}
\begin{figure}
\plotone{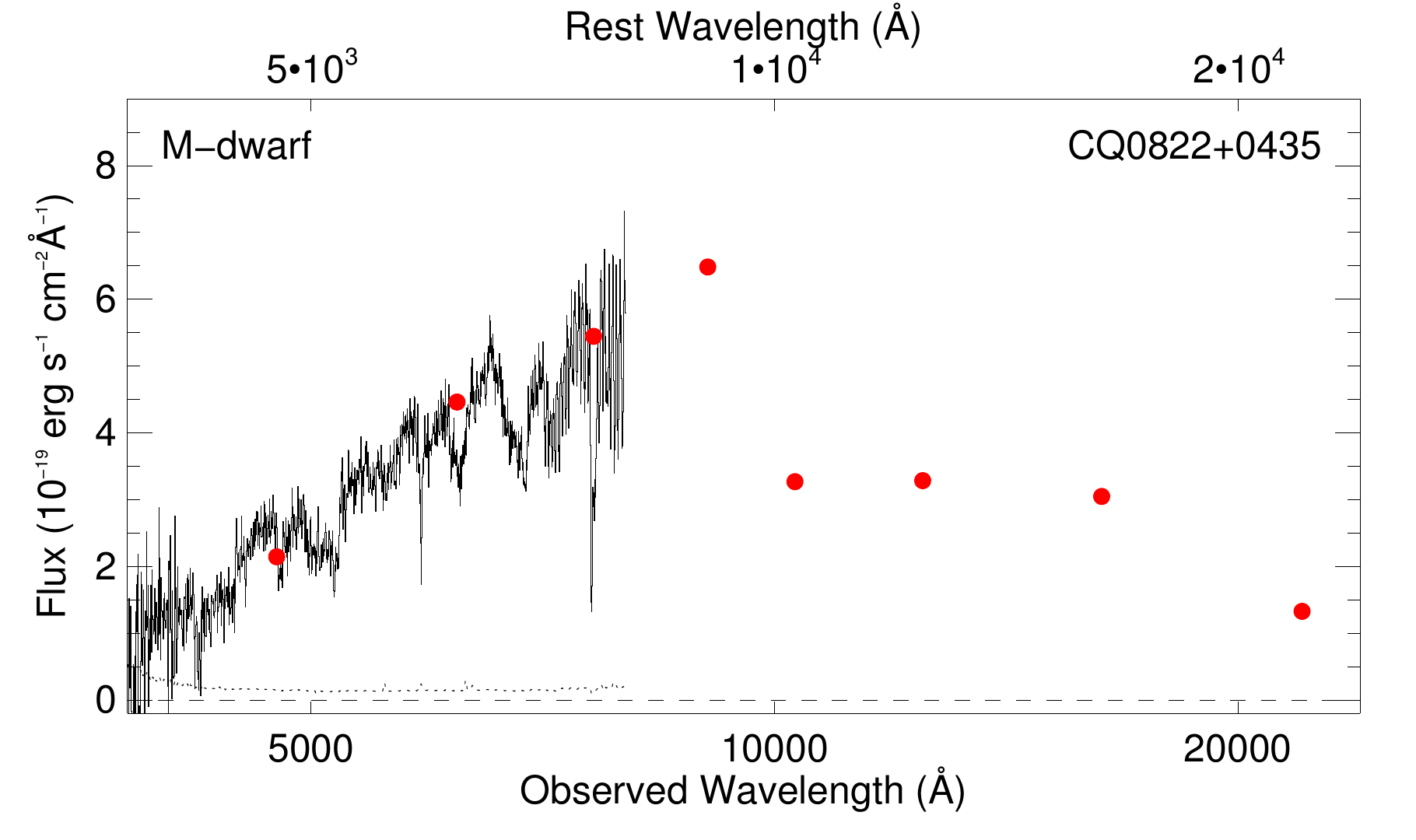}
\end{figure}
\begin{figure}
\plotone{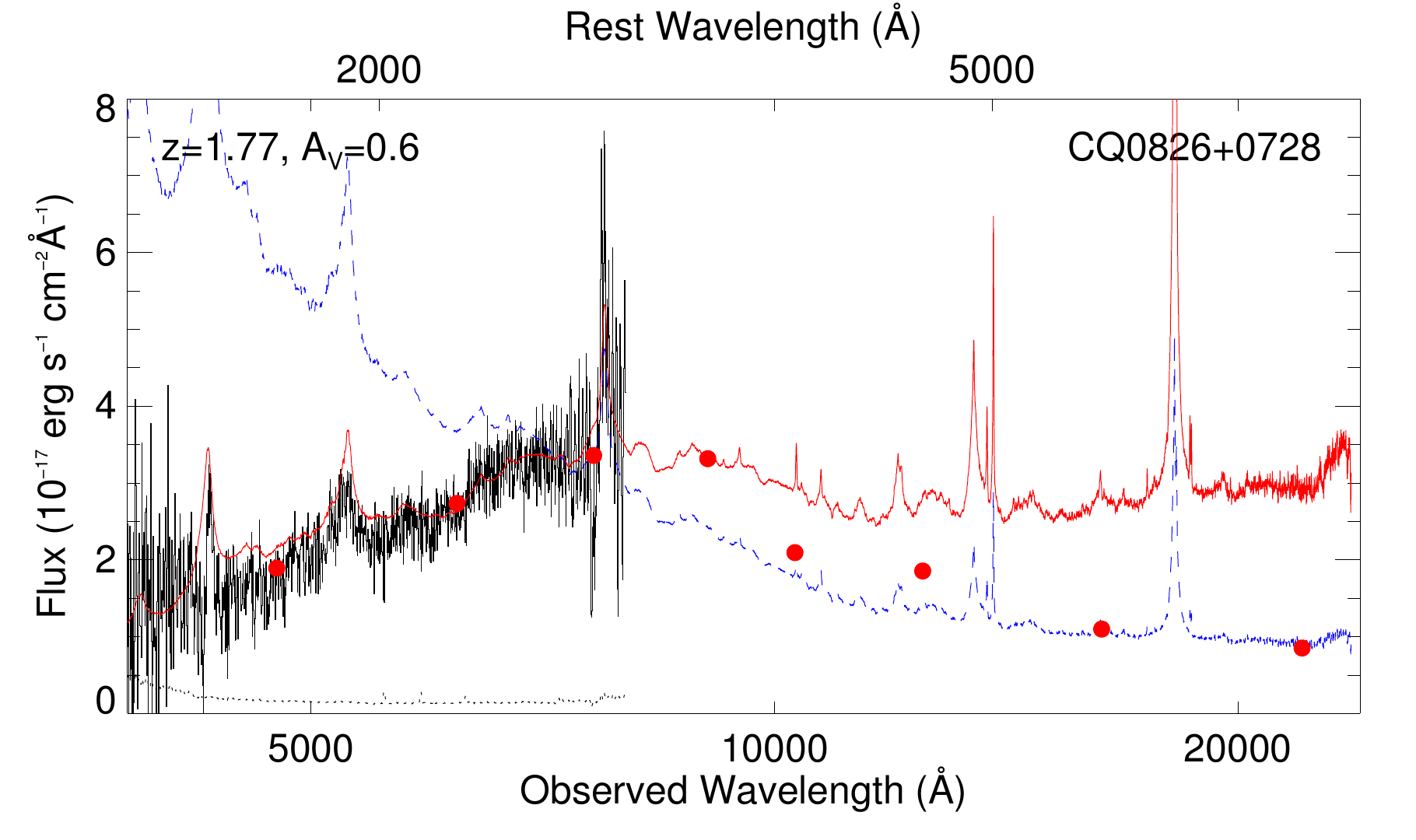}
\end{figure}
\begin{figure}
\plotone{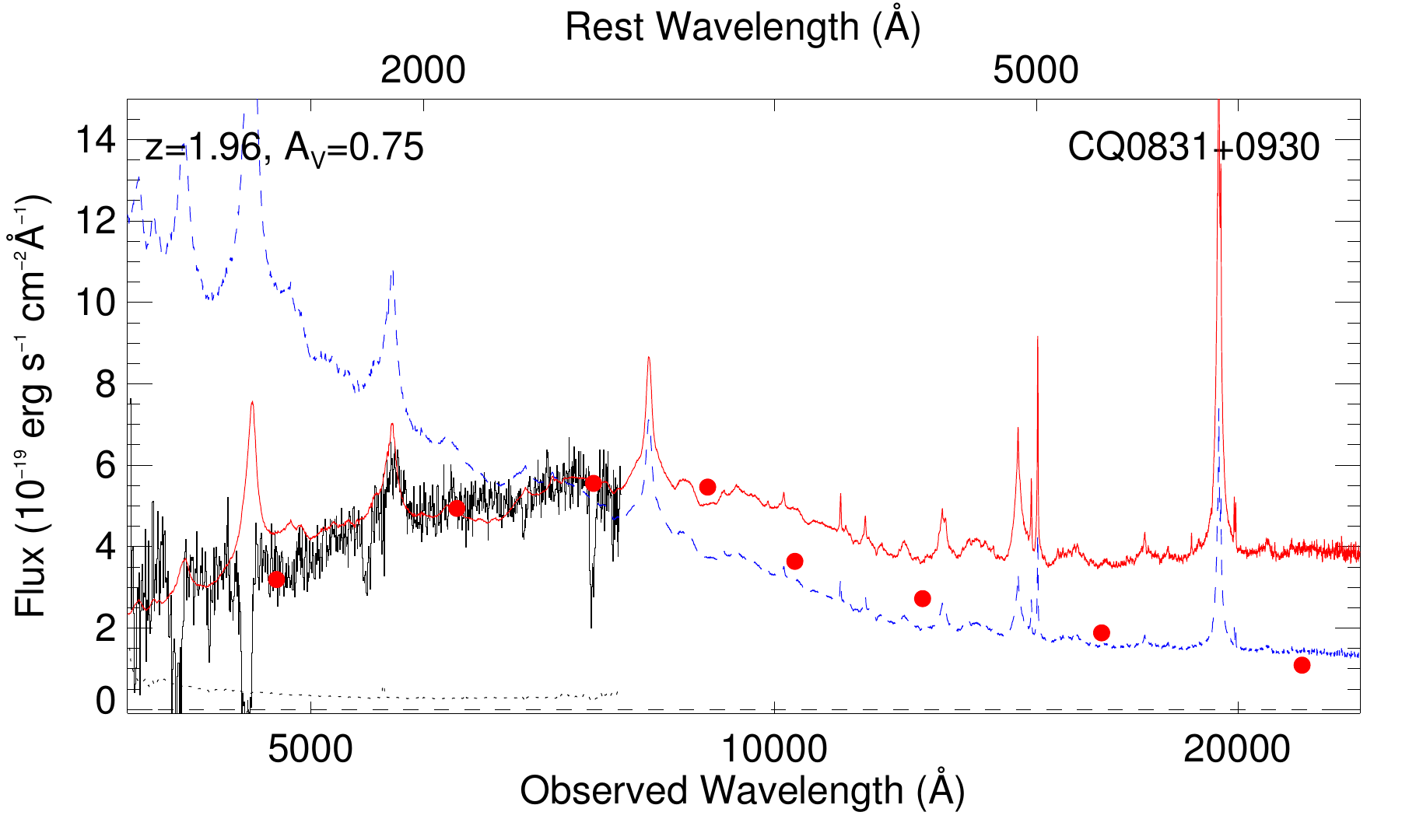}
\end{figure}
\begin{figure}
\plotone{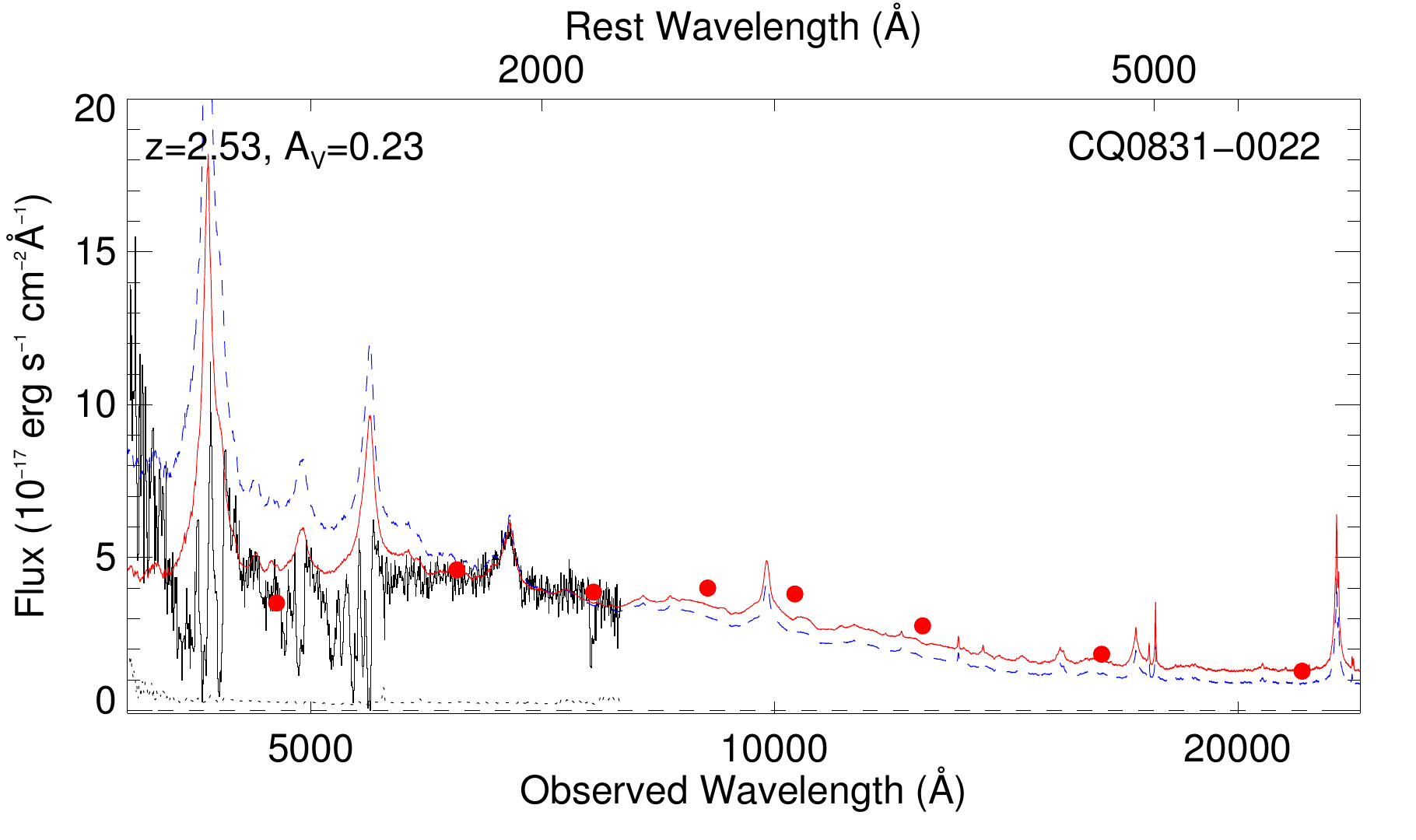}
\end{figure}
\begin{figure}
\plotone{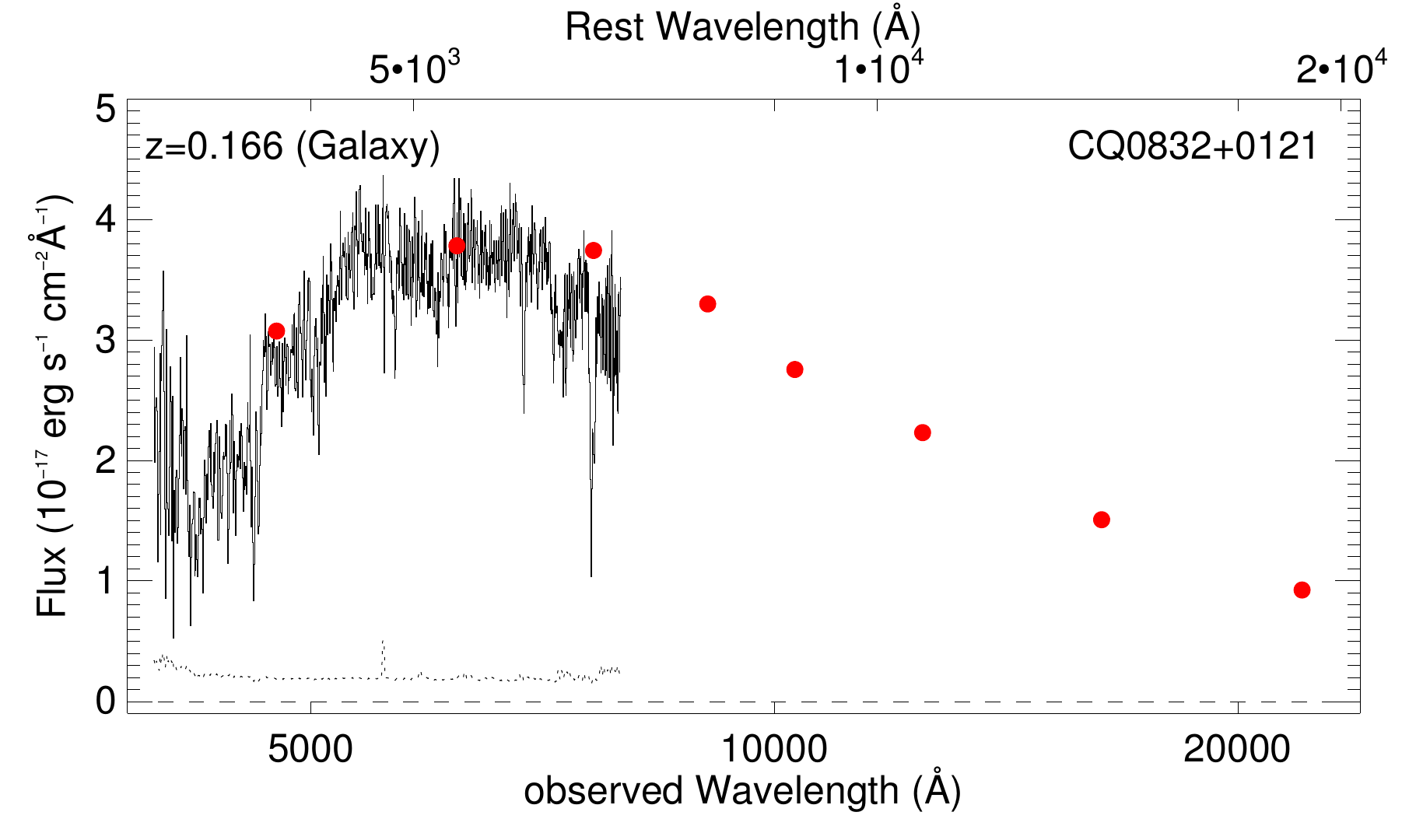}
\end{figure}
\begin{figure}
\plotone{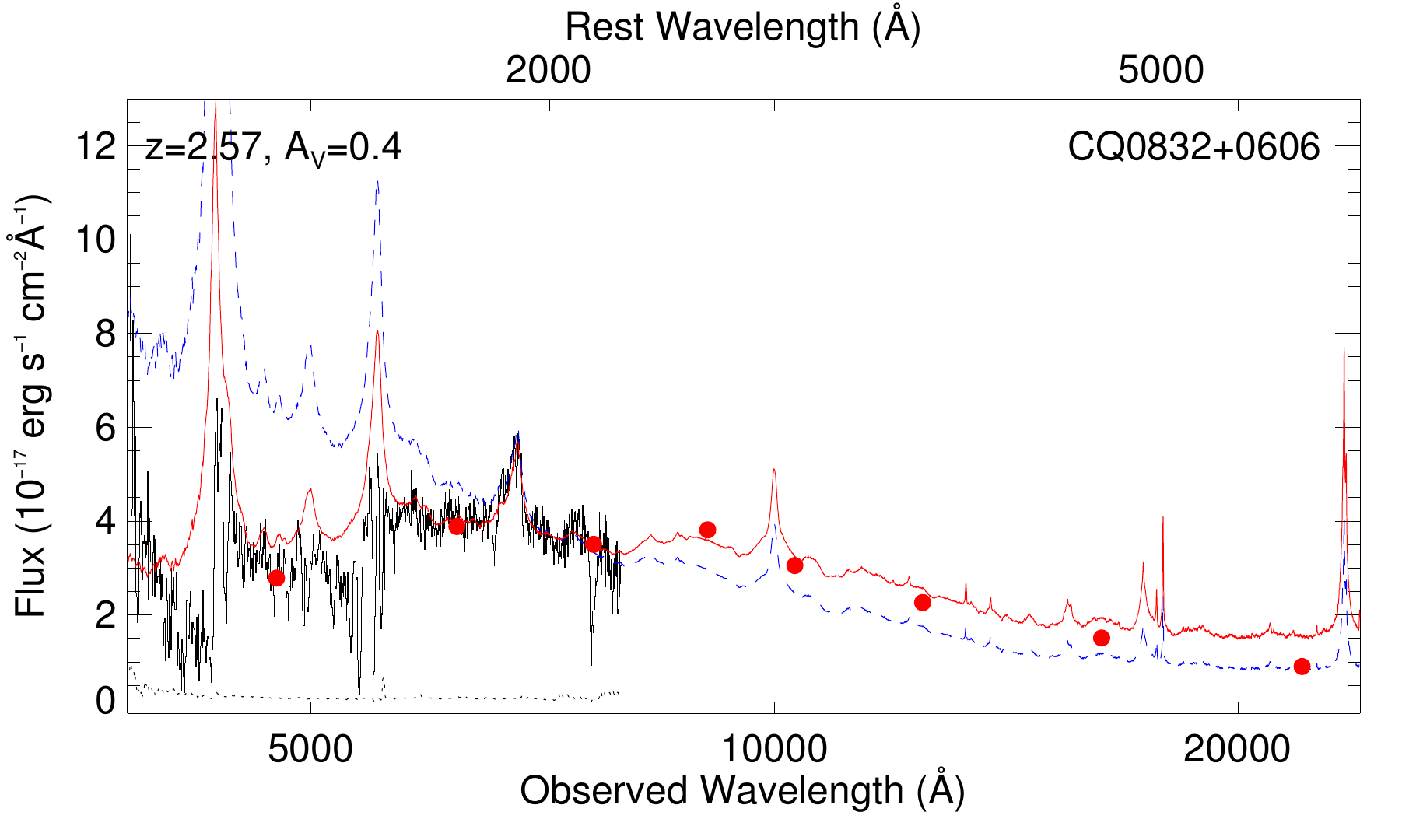}
\end{figure}

\section{Notes on individual objects}
\label{notes}

\subsection{CQ2143+0022 ($z=1.26$)}
This object is only observed by SDSS where a redshift
of $z=1.26$ is derived (presumably based on two emission lines interpreted as \ion{Mg}{2}
and \ion{C}{3}). The object is a dust-reddened QSO with an estimated 
amount of extinction corresponding to $A_\mathrm{V}=0.55$. The SMC-extinction 
reddened QSO template provides a good match to the SDSS spectrum and photometry,
but it does not match the UKIDSS photometry. This is a problem that is common
to most of our spectra of reddened QSOs. In SDSS the source is flagged QSO\_CAP.

\subsection{CQ2144+0045 ($z=0$)}
This is an M-dwarf. Based on the SDSS photometry the source is not flagged as a QSO.

\subsection{CQ2217+0033 ($z=0$)}
This is an M-dwarf. Based on the SDSS photometry the source is not flagged as a QSO.

\subsection{CQ2227+0022 ($z=2.23$)}
This is a reddened $z=2.23$ QSO observed both by us and by SDSS (SDSS infer a
redshift of $z=2.24$). The redshift is based on the detection of (narrow)
Ly$\alpha$ and (broad) \ion{C}{3} in emission and \ion{C}{4} in absorption. It
has an estimated extinction of $A_\mathrm{V} = 0.4$ (our spectrum) and 0.5
(SDSS spectrum).  In this case the extincted QSO template provides a reasonably
good match to the near-IR photometry out to around 5000 \AA \ in the rest
frame. Based on SDSS photometry this source is not flagged as a QSO.

\subsection{CQ2241+0115 (redshift unknown)}
The nature of this object could not be established. It has a narrow emission
line at 5258 \AA, a tentative narrow line at 6970 \AA \ and a broad emission
feature around 6400 \AA, but we have not been able to establish the nature 
of these features. The near-IR photometry shows evidence for an emission
line in the $H$ band and possibly also in the $Y$ band. Based on the 
SDSS photometry this source is flagged as QSO\_HIZ.

\subsection{CQ2241$-$0012 (redshift unknown)}
This spectrum is very complex with a mix of broad and narrow absorption lines
over the full spectral range covered by the instrument. The colors of the
object are peculiar, in particular $r-i=1.17$ and $Y-J=1.26$ are extreme values
(see Fig.~\ref{colplot}.  It is most likely an extreme BAL QSO at $z>2.5$
\citep[e.g.,][]{Hall02}. Based on the 
SDSS photometry this source is flagged as QSO\_HIZ.
The object is also included
in SDSS DR9 (released after our paper was submitted) where a redshift of
0.699 is derived. This redshift appears to be wrong.

\subsection{CQ2254$-$0001 ($z=3.69$)}
This is a relatively normal QSO at redshift of $z=3.69$ with no evidence for
excess reddening. Based on SDSS photometry it is flagged as QSO\_HIZ. 
It entered our selection due to its relatively high redshift.
The spectrum shows relatively weak Ly$\alpha$, probably due to associated
absorption also seen in Ly$\beta$, Lyman-limit absorption and in the \ion{C}{4}
line. The \ion{N}{5} line is very strong. The object is also included
in SDSS DR9 (released after our paper was submitted) where a redshift of
3.707 is derived.

\subsection{CQ2306+0108 ($z=3.65$)}
This is a QSO at a relatively high redshift of $z=3.65$ with no evidence for
extinction.  The spectrum shows somewhat narrower emission lines compared to
the template spectrum. In particular, Ly$\alpha$ and \ion{N}{5} are clearly
separated.  The source is also observed spectroscopically by SDSS, where a
redshift of $z=3.64$ is derived.  The spectrum displays a strong Ly$\alpha$
absorber at around 5300 \AA \ that could be damped.  There is also Lyman-limit
absorption at $z_{abs} = z_{em}$. Based on the SDSS photometry this object
is flagged as STAR\_CARBON and QSO\_HIZ.

\subsection{CQ2316+0023 ($z \approx 2.1$?)}
This is a BAL QSO, but we have not been able to establish the precise redshift.
Given the positions of the main absorption troughs around 5600 \AA \ and 4600
\AA \ we estimate that it is likely close to $z=2.1$ if these are from
\ion{Si}{4}, \ion{C}{4} and \ion{Al}{3}. There is a bright emission spike around
3900 \AA, which could be the onset of the Ly$\alpha$ / \ion{N}{5} emission.
Based on the SDSS photometry this object is not flagged as a QSO.

\subsection{CQ2324$-$0105 ($z=2.25$?)}
The nature of this object has not been securely established. It has a strong
absorption feature at 5108 \AA \ and a broad emission line at the very blue
end of the wavelength range. There is also a broad emission bump around 6200 \AA.
A possible solution is a reddened QSO at $z=2.25$ (i.e. very similar to
CQ2227+0022), where the absorption feature is $z_{abs} > z_{em}$ absorption
from \ion{C}{4} and the broad emission features are from Ly$\alpha$ and
\ion{C}{3}. Based on the SDSS photometry this object is not flagged as a
QSO. 

\subsection{CQ2342+0043 ($z=1.65$)}
This is a reddened QSO at $z=1.65$ with an estimated extinction of
$A_\mathrm{V} = 0.8$.  The spectrum displays broad emission lines from
\ion{Mg}{2} and \ion{C}{3}, but a significantly narrower \ion{C}{4} line. The
source is also observed spectroscopically by SDSS. From their spectrum we infer
a larger amount of extinction ($A_\mathrm{V} = 1.1$), which also leads
to a better match with the $Y$ band photometry. However, we match the 
$JHK$ photometry very badly. In SDSS the object is flagged as
QSO\_REJECT.

\subsection{CQ2344$-$0001 ($z=1.04$)}
This is a reddened ($A_\mathrm{V} = 0.5$) QSO at $z=1.04$.
The redshift is based on the detection of \ion{Mg}{2}
emission and of the \ion{Fe}{2} emission bump redwards of
\ion{Mg}{2}.
In this case we had to invoke the MW extinction curve to
match the shape of the spectrum so most likely the dust
in this system contains the carriers of the 2175 \AA \
extinction feature. In SDSS the object is not flagged as a QSO.

\subsection{CQ2347$-$0109 ($z=1.08$)}
This is a reddened ($A_\mathrm{V} = 0.9$) QSO at $z=1.08$ observed by
SDSS. The redshift is presumably based on the detection of
emission lines from \ion{Mg}{2} and \ion{C}{3}. In SDSS the object is
flagged as QSO\_SKIRT.

\subsection{CQ2355$-$0041 ($z=1.01$)}
This spectrum displays a single emission line, which we interpret as
\ion{Mg}{2} at $z=1.01$. Given that we can get a good match of the shape of the
optical spectrum with a reddened QSO template ($A_\mathrm{V} = 1.2$) at this redshift
we consider the nature of the object fairly secure.  The object is also covered
by SDSS where a redshift of $z=1.00$ is determined. From the SDSS spectrum we
infer a larger amount of extinction ($A_\mathrm{V} = 1.6$).  There is a very
poor match of the near-IR photometry.  In SDSS the object is flagged as
QSO\_HIZ.

\subsection{CQ2355+0007 (unknown redshift)}
This is a faint target and we are not able to establish its nature based on our
spectrum. The spectrum is much fainter than expected based on the SDSS
photometry (by a factor of 6) so either the object is variable or there was a
problem with the observation. In SDSS the object is not flagged as a
QSO.

\subsection{CQ0009$-$0020 ($z=0.387$)}
This is a star-forming galaxy or possibly a narrow-line AGN.  The spectrum
displays strong \ion{O}{2}, \ion{O}{3} and \ion{Ne}{3} emission and a 4000-\AA
\ break in the continuum. There are no Balmer lines in emission, but note that
the H$\beta$ line falls in the atmospheric A-band. There is also a hint of
\ion{Ne}{5} emission at 4750 \AA. The object is not flagged as a QSO in
SDSS.

\subsection{CQ0022+0020 ($z=0.80$?))}
This is a faint target. The object has a single narrow emission line at 6700
\AA \ and a hint of a broad line at 7900 \AA.  This is most likely a dust
reddened QSO at $z=0.80$, where identified narrow line is \ion{O}{2} and the
broad line is H$\gamma$. The object is also observed by SDSSIII where the same
features are detected with better significance (Noterdaeme, private
communication). The object is not flagged as a QSO in SDSS.

\subsection{CQ0027$-$0019 ($z=3.55$)}
This spectrum displays a single broad emission line, which we interpret as
Ly$\alpha$ at $z=3.55$ as there seems to be Ly$\alpha$ forest absorption bluewards
of it including a strong (possibly damped) Ly$\alpha$ absorber at 4780 \AA \
($z=2.93$). However, \ion{C}{4} emission is very weak or absent. The object
appears similar to CQ0327+0006 at $z=3.50$.  There is no sign of reddening and
the object likely entered our sample due to its high redshift.  The object is
also observed spectroscopically by SDSS where a redshift of $z=3.52$ is derived. 
The object is flagged as QSO\_CAP and QSO\_HIZ in SDSS.

\subsection{CQ0043$-$0000 ($z=0$?)}
This object is observed by SDSS. The object is flagged by SDSS as an L5.5
star, but we believe this classification to be wrong as the near-IR colors are
bluer than for late-type stars. The object has possible emission lines at 5900
\AA\ and 8090 \AA\, but we have not been able to establish the nature of these 
features.

\subsection{CQ0046$-$0011 ($z=2.44$)}
This is a reddened QSO at $z=2.44$. The redshift is based on the
detection of \ion{C}{3} emission and absorption at the positions of 
\ion{C}{4} and Ly$\alpha$ at this redshift. The shape of the optical spectrum
is reasonably well fitted with the template QSO spectrum reddened by
an SMC-like extinction curve with $A_\mathrm{V} = 0.6$.
The object is not flagged as a QSO in
SDSS. The object is also included
in SDSS DR9 (released after our paper was submitted) where a redshift of
2.467 is derived.

\subsection{CQ0105+0000 ($z=0.279$)}
This object is a compact galaxy with an old stellar population at $z=0.279$.
Both in the SDSS and UKIDSS imaging the object is consistent with being a point
source. Hence, this is a galaxy that bears some resemblance to the compact
quiescent galaxies seen at redshifts around 2 \citep[e.g.,][]{VanDokkum98}.
The object is not flagged as a QSO in SDSS.

\subsection{CQ0107+0016 ($z=2.47$)}
This is a reddened QSO at $z=2.47$. The spectrum displays \ion{C}{3} in
emission, \ion{C}{4} is nearly absorbed away and Ly$\alpha$ is weak. Narrow
emission from \ion{N}{5} is also detected. The shape of the spectrum is fairly
well fitted with the template QSO spectrum reddened by an SMC-like extinction
curve with $A_\mathrm{V} = 0.6$ except redwards of H$\beta$ in the restframe
where the reddened template spectrum is too red. 
The object is not flagged as a QSO in SDSS.

\subsection{CQ0127+0114 ($z=1.16$)}
This is a bright reddened QSO at $z=1.16$. The redshift is based on the
detection of a broad \ion{Mg}{2} emission line. This source has also been
observed by the SDSS where a slightly lower redshift ($z=1.14$) is derived. This
source is also detected in the FIRST survey \citep{FIRST}. The QSO is reddened
by an amount corresponding to $A_\mathrm{V} = 1.1$ as determined from both our
and the SDSS spectrum. The near-IR photometry for this source is very blue -
even bluer than the unreddened QSO template. In SDSS the source is 
flagged as QSO\_FIRST\_CAP.

\subsection{CQ0129$-$0059 ($z=0.71$)}
This is a strongly reddened ($A_\mathrm{V} = 1.5$) QSO at $z=0.71$.
The spectrum displays broad emission from Balmer-lines
(H$\gamma$ and H$\delta$), \ion{Ne}{3} and \ion{Mg}{2} and narrower
\ion{O}{2} emission. This source is also detected in the FIRST
survey \citep{FIRST}. In SDSS the source is flagged as QSO\_FIRST\_CAP. 
The near-IR photometry for this source is very blue and the reddened
QSO template fails completely in matching this.

\subsection{CQ0130+0013 ($z=1.05$)}
This is a reddened QSO ($A_\mathrm{V} = 0.9$) at $z=1.05$
observed by the SDSS. The redshift is presumably based on the detection
of \ion{Mg}{2} emission. Also for this source the reddened QSO template
does not match the near-IR photometry. In SDSS the object is flagged
as QSO\_REJECT.

\subsection{CQ0202+0010 ($z=1.61$)}
This is a reddened QSO ($A_\mathrm{V} = 0.9$) at $z=1.61$
observed by the SDSS. The redshift is presumably based on the detection 
of \ion{Mg}{2}, \ion{C}{3} and \ion{C}{4} emission. In SDSS the object is
flagged as QSO\_REJECT, QSO\_HIZ.

\subsection{CQ0211+0030 ($z=3.45$)}
This is a reddened ($A_\mathrm{V} = 0.30$) QSO at a relatively high
redshift of $z=3.45$. The redshift is based on the detection of \ion{C}{4} and
Ly$\alpha$ emission. The source is also detected by the ROSAT survey
\citep{ROSAT}. This object is really what we are searching for in terms of
redshift, but there is no indication of a DLA at $z>2$ in the spectrum. The
spectrum shows associated absorption. The only flags for the
object in the SDSS catalog are ROSAT\_D and ROSAT\_E.

\subsection{CQ0212$-$0023 ($z=1.87$?)}
The nature of this source is not secure. There is an indication
of broad emission lines around 4400 and 5400 \AA, which we interpret
as \ion{C}{3} and \ion{C}{4} at $z=1.87$. If correct the inferred extinction
based on the QSO template reddened by an SMC-like extinction curve is
$A_\mathrm{V} = 1.2$. Again, this model is far too red to match the 
near-IR photometry. In SDSS the object is flagged as QSO\_REJECT.

\subsection{CQ0220$-$0107 ($z=3.43$)}
This is a BAL QSO at $z=3.43$. The redshift is based on the identification of
emission and BAL features from \ion{C}{4}, \ion{Si}{4}, Ly$\alpha$ and
\ion{O}{6}/Ly$\beta$. There is no evidence for reddening. In SDSS the object is
not flagged as a QSO.
The object is also included
in SDSS DR9 (released after our paper was submitted) where a redshift of
3.461 is derived.

\subsection{CQ0222$-$0019 ($z=3.95$)}
This is the 2nd most distant source in our survey. It is also 
flagged as a high-$z$ QSO by SDSS (QSO\_HIZ). There is no evidence
for reddening and the source has hence entered our selection
due to its high redshift. The spectrum displays relatively
strong associated absorption seen both in \ion{C}{4},
Ly$\alpha$ and at the Lyman-limit.
The object is also included
in SDSS DR9 (released after our paper was submitted) where a redshift of
3.947 is derived.

\subsection{CQ0229$-$0029 ($z=2.14$)}
This is a BAL QSO observed by the SDSS. In SDSS the object is listed
to have redshift of $z=1.97$. However, the redshift we infer based on 
identification of the \ion{C}{4}, \ion{C}{3} and \ion{Mg}{2} emission 
lines is $z=2.14$. The
inferred amount of extinction is $A_\mathrm{V}=1.40$ (based on the 
QSO template reddened by SMC-like extinction), but this model fails
completely in matching the near-IR photometry. Based on the SDSS
photometry the object is flagged as QSO\_FAINT, QSO\_HIZ.

\subsection{CQ0239+0115 ($z=0.867$)}
The nature of this object could not been established based on 
our spectrum. 
The object has narrow emission lines at
5122 \AA \ and 5231 \AA \ and a P-Cygni-like feature around 4890
\AA.
There is also tentative evidence for an emission line at 6960 \AA.
The photometry shows a red continuum with a break in the $J$ band 
to a blue continuum from $J$ to $K$. Based on the SDSS photometry the
object is flagged as QSO\_REJECT. After submission of our paper DR9
was released. The object is included here and based on this spectrum
a redshift of $z=0.867$ is reported (presumably based on
narrow \ion{O}{2} and \ion{O}{3} emission as well as 
broad H$\beta$ and H$\delta$). The emission lines in the blue end of the 
spectrum are also detected in the DR9 spectrum. Their nature remains
unclear. Based on the shape of the spectrum we infer strong extinction
corresponding to $A_\mathrm{V} = 1.7$, but the assumption that the underlying
spectrum is similar to the QSO template spectrum may be questionable in
this case.

\subsection{CQ0242$-$0000 ($z=2.48$)}
This is a reddened BAL-like QSO at $z=2.48$ ($A_\mathrm{V} = 0.3$).
The redshift is based on the identification of emission and BAL features
from \ion{C}{4}, \ion{Si}{4} and Ly$\alpha$. Broad \ion{C}{3} emission
is also detected.
Based on the SDSS photometry
the source is flagged as STAR\_CARBON, QSO\_HIZ.

\subsection{CQ0247$-$0052 ($z=0.825$)}
This is a strongly reddened QSO at $z=0.825$ with an
estimated amount of extinction of $A_\mathrm{V} = 1.5$.
The spectrum displays emission from \ion{Mg}{2}, \ion{O}{2},
and indications of \ion{Ne}{3} and Balmer lines.
The reddened QSO model is again too red to match the near-IR 
photometry. Based on the SDSS photometry the object is not
flagged as a QSO.

\subsection{CQ0255+0048 ($z=4.01$)}
This object, observed by the SDSS, is the most distant QSO in 
our sample at a redshift of $z=4.01$. The object shows no sign
of reddening. It has a strong $z_{abs} = z_{em}$ absorber
detected both in Ly$\alpha$ and at the Lyman limit.
Based on the SDSS photometry the object is flagged as QSO\_HIZ.

\subsection{CQ0303+0105 ($z=3.45$)}
This is a $z=3.45$ QSO with relatively weak BAL-features and no
evidence for reddening. \ion{C}{4}
has a P-Cygni profile, Ly$\alpha$ is weak and narrow, but
\ion{N}{5} appears very strong. The narrow feature at 4224 \AA
\ is an energetic cosmic ray hit that has not been 
completely removed by the rejection algorithm. The object is 
also observed spectroscopically by SDSS leading to a consistent
redshift measurement. The object is flagged as QSO\_HIZ based
on the SDSS photometry.

\subsection{CQ0310+0055 ($z=3.78$)}
This is a high redshift ($z=3.78$) QSO observed by the SDSS. There is
no evidence for reddening. The object has strong associated absorption
detected both in \ion{C}{4}, \ion{Si}{4} and at the Lyman-limit.
Based on the SDSS photometry the source is flagged as QSO\_FAINT, QSO\_HIZ.

\subsection{CQ0311+0103 (unknown redshift?)}
The nature of this object is not established based on our spectrum. 
The signal-to-noise
ratio is low, but there seems to be a double absorption line at
5400 \AA \ and possible a broad emission line at 6600 \AA.
The object is also observed by SDSS where a redshift of 3.27 is 
determined. At this redshift the broad feature at 6600 \AA \ would be
\ion{C}{4} emission. The photometry is consistent with a reddened
QSO ($A_\mathrm{V} = 0.3$) at this redshift.
However, we still consider this redshift measurement insecure as
there does not seem to be an onset for Ly$\alpha$ forest absorption
(and Ly$\alpha$ emission or absorption at $z=3.27$ is absent). 
The doublet at 5400 \AA \ is consistent with \ion{Ca}{2} at $z=0.36$.
Based on the SDSS photometry the object is flagged as QSO\_FAINT, QSO\_HIZ.

\subsection{CQ0312+0032 ($z=1.25$)}
This is a reddened QSO at $z=1.25$. The redshift is based on the 
detection of \ion{Mg}{2} emission (with a P-Cygni profile). The shape of the
optical spectrum is well matched with the template spectrum
reddened by an $A_\mathrm{V} = 0.8$ extinction curve
except \ion{C}{3} seems to be weak/absent. Based on the SDSS photometry the 
object is not flagged as a QSO.

\subsection{CQ0312+0035 ($z=1.28$)}
This object looks very similar to CQ0312+0032, except here \ion{Mg}{2} is not
absorbed and \ion{C}{3} is well detected. The redshift is slightly larger
($z=1.28$).  Based on the SDSS photometry the object is not flagged as a
QSO. Note that CQ0312+0032 and CQ0312+0035 are only a few arcmin from
each other on the sky.
The redshifts of the two sources seems to be significantly
different that this must be a coincidence.

\subsection{CQ0321$-$0035 ($z=2.40$)}
This is a very bright BAL QSO at $z=2.40$ observed by the SDSS.
The inferred amount of extinction corresponds to $A_\mathrm{V} = 0.70$.
based on the SDSS photometry the source is flagged as QSO\_HIZ.
Again, the reddened QSO template spectrum fails complete to
match the UKIDSS photometry.

\subsection{CQ0326+0106 ($z=0.85$?)}
The nature of this objet is not well established from our
spectrum. There is
an emission line at around 5200 \AA \ and the shape of the
optical spectrum can be well matched assuming that the line is 
\ion{Mg}{2} at $z=0.85$ and assuming SMC-like extinction
with $A_\mathrm{V} = 1.1$.
The object is also
observed by SDSS where a redshift of $z=0.85$ is determined.
Based on the SDSS spectrum we infer an extinction of
$A_\mathrm{V}=1.30$. 
In SDSS the only flag for the object is SERENDIP\_DISTANT.

\subsection{CQ0327+0006 ($z=3.50$?)}
The spectrum shows a broad emission line around 5500 \AA, which most 
likely is Ly$\alpha$ at $z=3.50$. There is evidence for mild 
reddening ($A_\mathrm{V} = 0.15$).
The object is not flagged as a QSO based on the SDSS
photometry.

\subsection{CQ0329$-$0057 ($z=1.31$)}
This object appears very similar to CQ0312+0032 and CQ0312+0035. It is a
reddened QSO at $z=1.31$ with an inferred extinction of $A_\mathrm{V} = 1.1$.
The redshift is based on the detection of the \ion{Mg}{2} emission line.  The
object is also observed by SDSS. Based on the SDSS spectrum we infer a somewhat
larger amount of extinction ($A_\mathrm{V} = 1.4$). The object is not
flagged as a QSO based on the SDSS photometry.

\subsection{CQ0332$-$0013 ($z=0.438$)}
This is the lowest redshift QSO in our sample.  This object is a reddened
($A_\mathrm{V} = 1.0$) QSO at $z=0.438$. The spectrum displays narrow
\ion{O}{2} and \ion{O}{3} emission and broad lines from H$\beta$ and
\ion{Ne}{3}. The object is not flagged as a QSO based on the SDSS
photometry.

\subsection{CQ0336+0112 ($z=0$)}
This is an M-dwarf. Based on the SDSS photometry the source is not 
flagged as a QSO.

\subsection{CQ0338+0004 ($z=1.45$)}
This is a reddened QSO at $z=1.45$ observed by the SDSS.
The redshift is presumably based on the detection of \ion{Mg}{2} and
\ion{C}{3} emission lines.
Based on the SDSS spectrum we infer an amount of extinction corresponding
to $A_\mathrm{V} = 0.6$. In the SDSS catalog the source is flagged
QSO\_CAP.

\subsection{CQ0339+0022 ($z=1.41$)}
This is a reddened QSO at $z=1.41$. The shape of the optical spectrum
is well matched with the template spectrum reddened by 
SMC-like extinction with $A_\mathrm{V} = 0.6$. The spectrum
displays emission lines from \ion{Mg}{2} and \ion{C}{3}. \ion{Mg}{2}
has a P-Cygni profile and there is broad \ion{Al}{3} absorption bluewards
of the \ion{C}{3} emission line, characteristic of lo-BAL QSOs.
The object is not flagged as a QSO based on the SDSS photometry.

\subsection{CQ0350$-$0031 ($z=2.00$)}
This is a reddened QSO at $z=2.00$. A spectrum has also been secured by
the SDSS team. Based on both our and the SDSS spectrum we infer an amount of
extinction corresponding to $A_\mathrm{V} = 0.4$. The spectrum 
displays emission lines from \ion{C}{3}, \ion{C}{4}, and
\ion{Si}{4}. In the SDSS catalog the source is flagged QSO\_CAP.

\subsection{CQ0354$-$0012 ($z=2.45$)}
This is a reddened QSO at $z=2.45$, but in this case the
dust is Galactic ($E(B-V) = 0.37$) and there is no evidence for excess
extinction in the host galaxy or intervening along the line-of-sight. 
Yet, the spectrum is still an unusual QSO spectrum
in the sense that both \ion{C}{4} and Ly$\alpha$ are strongly absorbed.
The object is not flagged as a QSO based on the SDSS photometry.

\subsection{CQ0354$-$0030 ($z=1.00$)}
This is a reddened QSO at $z=1.00$. 
The redshift is based on the detection of the \ion{Mg}{2} emission line.
The objects is significantly
reddened by Galactic extinction ($E(B-V) = 0.45$), but it also has 
intrinsic excess reddening corresponding to $A_\mathrm{V} = 0.3$.
The object is not flagged as a QSO based on the SDSS photometry.

\subsection{CQ0822+0004 ($z=0.378$)}
This spectrum shows strong Balmer absorption at $z=0.378$, but no
emission lines. This is most likely a compact post-starburst galaxy (so-called E+A galaxy).
The object is flagged as QSO\_HIZ based on the SDSS photometry.

\subsection{CQ0822+0435 ($z=0$)}
This is an M-dwarf. Based on the SDSS photometry the source is not flagged as a QSO. 

\subsection{CQ0826+0728 ($z=1.77$)}
This is a reddened QSO at $z=1.77$. The spectrum displays broad emission
lines from \ion{Mg}{2} and \ion{C}{3} and narrower emission from \ion{C}{4}.
The shape of the spectrum is well meatched with the template spectrum reddened
by an SMC-like extinction curve with $A_\mathrm{V} = 0.6$.
The object is not flagged as a QSO based on the SDSS photometry.

\subsection{CQ0831+0930 ($z=1.96$)}
This is a reddened QSO at $z=1.96$. The inferred amount of extinction
corresponds to $A_\mathrm{V} = 0.75$. The spectrum displays
broad emission from \ion{C}{3}, whereas \ion{C}{4} and \ion{Si}{4}
are only detected in absorption.
The object is not flagged as a QSO based on the SDSS photometry.

\subsection{CQ0831$-$0022 ($z=2.53$)}
This is a mildly reddened QSO at $z=2.53$. The inferred amount of extinction
corresponds to $A_\mathrm{V} = 0.23$. The spectrum displays 
broad \ion{C}{3} emission and complex absorption at the positions
of \ion{C}{4}, \ion{Si}{4}, \ion{N}{5} and Ly$\alpha$.
The object is not flagged as a QSO based on the SDSS photometry.

\subsection{CQ0832+0121 ($z=0.166$)}
This object is a compact, old galaxy (similar to CQ0105+0000).
The object is not flagged as a QSO based on the SDSS photometry.

\subsection{CQ0832+0606 ($z=2.57$)}
This spectrum is a reddened QSO at $z=2.57$ --
similar to that of the nearby CQ0831$-$0022. 
The inferred amount of extinction
corresponds to $A_\mathrm{V} = 0.4$. The spectrum displays
broad \ion{C}{3} emission and complex absorption at the positions
of \ion{C}{4}, \ion{Si}{4}, \ion{N}{5} and Ly$\alpha$.
The object is not flagged as a QSO based on the SDSS photometry.
\bibliographystyle{apj}
\bibliography{thebib}

\begin{thebibliography}{67}
\expandafter\ifx\csname natexlab\endcsname\relax\def\natexlab#1{#1}\fi

\bibitem[{{Arnouts} {et~al.}(1999){Arnouts}, {Cristiani}, {Moscardini},
  {Matarrese}, {Lucchin}, {Fontana}, \& {Giallongo}}]{Arnouts99}
{Arnouts}, S., {Cristiani}, S., {Moscardini}, L., {Matarrese}, S., {Lucchin},
  F., {Fontana}, A., \& {Giallongo}, E. 1999, \mnras, 310, 540

\bibitem[{{Banerji} {et~al.}(2012){Banerji}, {McMahon}, {Hewett},
  {Alaghband-Zadeh}, {Gonzalez-Solares}, \& {Venemans}}]{Banerji12}
{Banerji}, M., {McMahon}, R.~G., {Hewett}, P.~C., {Alaghband-Zadeh}, S.,
  {Gonzalez-Solares}, E., \& {Venemans}, B.~P. 2012, ArXiv e-prints

\bibitem[{{Becker} {et~al.}(1995){Becker}, {White}, \& {Helfand}}]{FIRST}
{Becker}, R.~H., {White}, R.~L., \& {Helfand}, D.~J. 1995, \apj, 450, 559

\bibitem[{{Benn} {et~al.}(1998){Benn}, {Vigotti}, {Carballo},
  {Gonzalez-Serrano}, \& {S{\'a}nchez}}]{Benn98}
{Benn}, C.~R., {Vigotti}, M., {Carballo}, R., {Gonzalez-Serrano}, J.~I., \&
  {S{\'a}nchez}, S.~F. 1998, \mnras, 295, 451

\bibitem[{{Bongiorno} {et~al.}(2007){Bongiorno}, {Zamorani}, {Gavignaud},
  {Marano}, {Paltani}, {Mathez}, {M{\o}ller}, {Picat}, {Cirasuolo},
  {Lamareille}, {Bottini}, {Garilli}, {Le Brun}, {Le F{\`e}vre}, {Maccagni},
  {Scaramella}, {Scodeggio}, {Tresse}, {Vettolani}, {Zanichelli}, {Adami},
  {Arnouts}, {Bardelli}, {Bolzonella}, {Cappi}, {Charlot}, {Ciliegi},
  {Contini}, {Foucaud}, {Franzetti}, {Guzzo}, {Ilbert}, {Iovino}, {McCracken},
  {Marinoni}, {Mazure}, {Meneux}, {Merighi}, {Pell{\`o}}, {Pollo}, {Pozzetti},
  {Radovich}, {Zucca}, {Hatziminaoglou}, {Polletta}, {Bondi}, {Brinchmann},
  {Cucciati}, {de la Torre}, {Gregorini}, {Mellier}, {Merluzzi}, {Temporin},
  {Vergani}, \& {Walcher}}]{Bongiorno2007}
{Bongiorno}, A., {et~al.} 2007, \aap, 472, 443

\bibitem[{{Ellison} {et~al.}(2004){Ellison}, {Churchill}, {Rix}, \&
  {Pettini}}]{Ellison04}
{Ellison}, S.~L., {Churchill}, C.~W., {Rix}, S.~A., \& {Pettini}, M. 2004,
  \apj, 615, 118

\bibitem[{{Ellison} {et~al.}(2001){Ellison}, {Yan}, {Hook}, {Pettini}, {Wall},
  \& {Shaver}}]{Ellison01}
{Ellison}, S.~L., {Yan}, L., {Hook}, I.~M., {Pettini}, M., {Wall}, J.~V., \&
  {Shaver}, P. 2001, \aap, 379, 393

\bibitem[{{Erkal} {et~al.}(2012){Erkal}, {Gnedin}, \& {Kravtsov}}]{Erkal2012}
{Erkal}, D., {Gnedin}, N.~Y., \& {Kravtsov}, A.~V. 2012, ArXiv e-prints

\bibitem[{{Fleuren} {et~al.}(2012){Fleuren}, {Sutherland}, {Dunne}, {Smith},
  {Maddox}, {Gonz{\'a}lez-Nuevo}, {Findlay}, {Auld}, {Baes}, {Bond},
  {Bonfield}, {Bourne}, {Cooray}, {Buttiglione}, {Cava}, {Dariush}, {De Zotti},
  {Driver}, {Dye}, {Eales}, {Fritz}, {Gunawardhana}, {Hopwood}, {Ibar},
  {Ivison}, {Jarvis}, {Kelvin}, {Lapi}, {Liske}, {Micha{\l}owski}, {Negrello},
  {Pascale}, {Pohlen}, {Prescott}, {Rigby}, {Robotham}, {Scott}, {Temi},
  {Thompson}, {Valiante}, \& {Werf}}]{Viking}
{Fleuren}, S., {et~al.} 2012, \mnras, 423, 2407

\bibitem[{{Frank} \& {P{\'e}roux}(2010)}]{Frank10}
{Frank}, S., \& {P{\'e}roux}, C. 2010, \mnras, 406, 2235

\bibitem[{{Franx} {et~al.}(2008){Franx}, {van Dokkum}, {Schreiber}, {Wuyts},
  {Labb{\'e}}, \& {Toft}}]{Franx08}
{Franx}, M., {van Dokkum}, P.~G., {Schreiber}, N.~M.~F., {Wuyts}, S.,
  {Labb{\'e}}, I., \& {Toft}, S. 2008, \apj, 688, 770

\bibitem[{{Fynbo} {et~al.}(2011){Fynbo}, {Ledoux}, {Noterdaeme}, {Christensen},
  {M{\o}ller}, {Durgapal}, {Goldoni}, {Kaper}, {Krogager}, {Laursen}, {Maund},
  {Milvang-Jensen}, {Okoshi}, {Rasmussen}, {Thorsen}, {Toft}, \&
  {Zafar}}]{Fynbo11}
{Fynbo}, J.~P.~U., {et~al.} 2011, \mnras, 413, 2481

\bibitem[{{Gavignaud} {et~al.}(2006){Gavignaud}, {Bongiorno}, {Paltani},
  {Mathez}, {Zamorani}, {M{\o}ller}, {Picat}, {Le Brun}, {Marano}, {Le
  F{\`e}vre}, {Bottini}, {Garilli}, {Maccagni}, {Scaramella}, {Scodeggio},
  {Tresse}, {Vettolani}, {Zanichelli}, {Adami}, {Arnaboldi}, {Arnouts},
  {Bardelli}, {Bolzonella}, {Cappi}, {Charlot}, {Ciliegi}, {Contini},
  {Foucaud}, {Franzetti}, {Guzzo}, {Ilbert}, {Iovino}, {McCracken}, {Marinoni},
  {Mazure}, {Meneux}, {Merighi}, {Pell{\`o}}, {Pollo}, {Pozzetti}, {Radovich},
  {Zucca}, {Bondi}, {Busarello}, {Cucciati}, {de la Torre}, {Gregorini},
  {Lamareille}, {Mellier}, {Merluzzi}, {Ripepi}, {Rizzo}, \&
  {Vergani}}]{Gavignaud2006}
{Gavignaud}, I., {et~al.} 2006, \aap, 457, 79

\bibitem[{{Glikman} {et~al.}(2004){Glikman}, {Gregg}, {Lacy}, {Helfand},
  {Becker}, \& {White}}]{Glikman04}
{Glikman}, E., {Gregg}, M.~D., {Lacy}, M., {Helfand}, D.~J., {Becker}, R.~H.,
  \& {White}, R.~L. 2004, \apj, 607, 60

\bibitem[{{Glikman} {et~al.}(2006){Glikman}, {Helfand}, \& {White}}]{Glikman06}
{Glikman}, E., {Helfand}, D.~J., \& {White}, R.~L. 2006, \apj, 640, 579

\bibitem[{{Glikman} {et~al.}(2007){Glikman}, {Helfand}, {White}, {Becker},
  {Gregg}, \& {Lacy}}]{Glikman07}
{Glikman}, E., {Helfand}, D.~J., {White}, R.~L., {Becker}, R.~H., {Gregg},
  M.~D., \& {Lacy}, M. 2007, \apj, 667, 673

\bibitem[{{Glikman} {et~al.}(2012){Glikman}, {Urrutia}, {Lacy}, {Djorgovski},
  {Mahabal}, {Myers}, {Ross}, {Petitjean}, {Ge}, {Schneider}, \&
  {York}}]{Glikman12}
{Glikman}, E., {et~al.} 2012, \apj, 757, 51

\bibitem[{{Gosling} {et~al.}(2009){Gosling}, {Bandyopadhyay}, \&
  {Blundell}}]{Gosling2009}
{Gosling}, A.~J., {Bandyopadhyay}, R.~M., \& {Blundell}, K.~M. 2009, \mnras,
  394, 2247

\bibitem[{{Gregg} {et~al.}(2002){Gregg}, {Lacy}, {White}, {Glikman}, {Helfand},
  {Becker}, \& {Brotherton}}]{Gregg02}
{Gregg}, M.~D., {Lacy}, M., {White}, R.~L., {Glikman}, E., {Helfand}, D.,
  {Becker}, R.~H., \& {Brotherton}, M.~S. 2002, \apj, 564, 133

\bibitem[{{Hall} {et~al.}(2002){Hall}, {Gunn}, {Knapp}, {Narayanan}, {Strauss},
  {Anderson}, {vanden Berk}, {Heckman}, {Krolik}, {Tsvetanov}, {Zheng},
  {Richards}, {Schneider}, {Fan}, {York}, {Geballe}, {Davis}, {Becker}, \&
  {Brunner}}]{Hall02}
{Hall}, P.~B., {et~al.} 2002, in Astronomical Society of the Pacific Conference
  Series, Vol. 255, Mass Outflow in Active Galactic Nuclei: New Perspectives,
  ed. {D.~M.~Crenshaw, S.~B.~Kraemer, \& I.~M.~George}, 161

\bibitem[{{Hewett} {et~al.}(2006){Hewett}, {Warren}, {Leggett}, \&
  {Hodgkin}}]{Hewett}
{Hewett}, P.~C., {Warren}, S.~J., {Leggett}, S.~K., \& {Hodgkin}, S.~T. 2006,
  \mnras, 367, 454

\bibitem[{{Hopkins} {et~al.}(2004){Hopkins}, {Strauss}, {Hall}, {Richards},
  {Cooper}, {Schneider}, {Vanden Berk}, {Jester}, {Brinkmann}, \&
  {Szokoly}}]{Hopkins04}
{Hopkins}, P.~F., {et~al.} 2004, \aj, 128, 1112

\bibitem[{{Horne}(1986)}]{Horne}
{Horne}, K. 1986, \pasp, 98, 609

\bibitem[{{Hunt} {et~al.}(1998){Hunt}, {Mannucci}, {Testi}, {Migliorini},
  {Stanga}, {Baffa}, {Lisi}, \& {Vanzi}}]{Hunt98}
{Hunt}, L.~K., {Mannucci}, F., {Testi}, L., {Migliorini}, S., {Stanga}, R.~M.,
  {Baffa}, C., {Lisi}, F., \& {Vanzi}, L. 1998, \aj, 115, 2594

\bibitem[{{Ilbert} {et~al.}(2006){Ilbert}, {Arnouts}, {McCracken},
  {Bolzonella}, {Bertin}, {Le F{\`e}vre}, {Mellier}, {Zamorani}, {Pell{\`o}},
  {Iovino}, {Tresse}, {Le Brun}, {Bottini}, {Garilli}, {Maccagni}, {Picat},
  {Scaramella}, {Scodeggio}, {Vettolani}, {Zanichelli}, {Adami}, {Bardelli},
  {Cappi}, {Charlot}, {Ciliegi}, {Contini}, {Cucciati}, {Foucaud}, {Franzetti},
  {Gavignaud}, {Guzzo}, {Marano}, {Marinoni}, {Mazure}, {Meneux}, {Merighi},
  {Paltani}, {Pollo}, {Pozzetti}, {Radovich}, {Zucca}, {Bondi}, {Bongiorno},
  {Busarello}, {de La Torre}, {Gregorini}, {Lamareille}, {Mathez}, {Merluzzi},
  {Ripepi}, {Rizzo}, \& {Vergani}}]{Ilbert06}
{Ilbert}, O., {et~al.} 2006, \aap, 457, 841

\bibitem[{{Jarvis} {et~al.}(2012){Jarvis}, {Bonfield}, {Bruce}, {Geach},
  {McAlpine}, {McLure}, {Gonzalez-Solares}, {Irwin}, {Lewis}, {Kupcu Yoldas},
  {Andreon}, {Cross}, {Emerson}, {Dalton}, {Dunlop}, {Hodgkin}, {Le Fevre},
  {Karouzos}, {Meisenheimer}, {Oliver}, {Rawlings}, {Simpson}, {Smail},
  {Smith}, {Sullivan}, {Sutherland}, {White}, \& {Zwart}}]{Jarvis2012}
{Jarvis}, M.~J., {et~al.} 2012, ArXiv e-prints

\bibitem[{{Jian-Guo} {et~al.}(2012){Jian-Guo}, {Hong-Yan}, {Jian}, {Peng},
  {Hong-Lin}, {Xavier}, {Fred}, {Hui-Yuan}, {Ting-Gui}, \&
  {WeiMin}}]{JianGuo2012}
{Jian-Guo}, W., {et~al.} 2012, ArXiv e-prints

\bibitem[{{Jorgenson} {et~al.}(2006){Jorgenson}, {Wolfe}, {Prochaska}, {Lu},
  {Howk}, {Cooke}, {Gawiser}, \& {Gelino}}]{Jorgenson06}
{Jorgenson}, R.~A., {Wolfe}, A.~M., {Prochaska}, J.~X., {Lu}, L., {Howk},
  J.~C., {Cooke}, J., {Gawiser}, E., \& {Gelino}, D.~M. 2006, \apj, 646, 730

\bibitem[{{Kaplan} {et~al.}(2010){Kaplan}, {Prochaska}, {Herbert-Fort},
  {Ellison}, \& {Dessauges-Zavadsky}}]{Kaplan10}
{Kaplan}, K.~F., {Prochaska}, J.~X., {Herbert-Fort}, S., {Ellison}, S.~L., \&
  {Dessauges-Zavadsky}, M. 2010, \pasp, 122, 619

\bibitem[{{Khare} {et~al.}(2012){Khare}, {vanden Berk}, {York}, {Lundgren}, \&
  {Kulkarni}}]{Khare12}
{Khare}, P., {vanden Berk}, D., {York}, D.~G., {Lundgren}, B., \& {Kulkarni},
  V.~P. 2012, \mnras, 419, 1028

\bibitem[{{Laureijs} {et~al.}(2011){Laureijs}, {Amiaux}, {Arduini},
  {Augu{\`e}res}, {Brinchmann}, {Cole}, {Cropper}, {Dabin}, {Duvet}, {Ealet},
  \& et~al.}]{Euclid}
{Laureijs}, R., {et~al.} 2011, ArXiv e-prints

\bibitem[{{Maddox} {et~al.}(2008){Maddox}, {Hewett}, {Warren}, \&
  {Croom}}]{Maddox08}
{Maddox}, N., {Hewett}, P.~C., {Warren}, S.~J., \& {Croom}, S.~M. 2008, \mnras,
  386, 1605

\bibitem[{{McCracken} {et~al.}(2012){McCracken}, {Milvang-Jensen}, {Dunlop},
  {Franx}, {Fynbo}, {Le F{\`e}vre}, {Holt}, {Caputi}, {Goranova}, {Buitrago},
  {Emerson}, {Freudling}, {Hudelot}, {L{\'o}pez-Sanjuan}, {Magnard}, {Mellier},
  {M{\o}ller}, {Nilsson}, {Sutherland}, {Tasca}, \& {Zabl}}]{Henry2012}
{McCracken}, H.~J., {et~al.} 2012, \aap, 544, A156

\bibitem[{{Murphy} \& {Liske}(2004)}]{Murphy04}
{Murphy}, M.~T., \& {Liske}, J. 2004, \mnras, 354, L31

\bibitem[{{Nilsson} \& {M{\o}ller}(2009)}]{NilssonMoller}
{Nilsson}, K.~K., \& {M{\o}ller}, P. 2009, \aap, 508, L21

\bibitem[{{Nilsson} {et~al.}(2009){Nilsson}, {Tapken}, {M{\o}ller},
  {Freudling}, {Fynbo}, {Meisenheimer}, {Laursen}, \&
  {{\"O}stlin}}]{Nilsson2009}
{Nilsson}, K.~K., {Tapken}, C., {M{\o}ller}, P., {Freudling}, W., {Fynbo},
  J.~P.~U., {Meisenheimer}, K., {Laursen}, P., \& {{\"O}stlin}, G. 2009, \aap,
  498, 13

\bibitem[{{Nishiyama} {et~al.}(2008){Nishiyama}, {Nagata}, {Tamura}, {Kandori},
  {Hatano}, {Sato}, \& {Sugitani}}]{Nishiyama2008}
{Nishiyama}, S., {Nagata}, T., {Tamura}, M., {Kandori}, R., {Hatano}, H.,
  {Sato}, S., \& {Sugitani}, K. 2008, \apj, 680, 1174

\bibitem[{{Nishiyama} {et~al.}(2009){Nishiyama}, {Tamura}, {Hatano}, {Kato},
  {Tanab{\'e}}, {Sugitani}, \& {Nagata}}]{Nishiyama2009}
{Nishiyama}, S., {Tamura}, M., {Hatano}, H., {Kato}, D., {Tanab{\'e}}, T.,
  {Sugitani}, K., \& {Nagata}, T. 2009, \apj, 696, 1407

\bibitem[{{Noterdaeme} {et~al.}(2012){Noterdaeme}, {Laursen}, {Petitjean},
  {Vergani}, {Maureira}, {Ledoux}, {Fynbo}, {L{\'o}pez}, \&
  {Srianand}}]{Noterdaeme12}
{Noterdaeme}, P., {et~al.} 2012, \aap, 540, A63

\bibitem[{{Noterdaeme} {et~al.}(2009){Noterdaeme}, {Ledoux}, {Srianand},
  {Petitjean}, \& {Lopez}}]{Noterdaeme09b}
{Noterdaeme}, P., {Ledoux}, C., {Srianand}, R., {Petitjean}, P., \& {Lopez}, S.
  2009, \aap, 503, 765

\bibitem[{{Noterdaeme} {et~al.}(2010){Noterdaeme}, {Petitjean}, {Ledoux},
  {L{\'o}pez}, {Srianand}, \& {Vergani}}]{Noterdaeme10}
{Noterdaeme}, P., {Petitjean}, P., {Ledoux}, C., {L{\'o}pez}, S., {Srianand},
  R., \& {Vergani}, S.~D. 2010, \aap, 523, A80+

\bibitem[{{Pei}(1992)}]{Pei92}
{Pei}, Y.~C. 1992, \apj, 395, 130

\bibitem[{{Pei} {et~al.}(1991){Pei}, {Fall}, \& {Bechtold}}]{Pei1991}
{Pei}, Y.~C., {Fall}, S.~M., \& {Bechtold}, J. 1991, \apj, 378, 6

\bibitem[{{Pei} {et~al.}(1999){Pei}, {Fall}, \& {Hauser}}]{Pei99}
{Pei}, Y.~C., {Fall}, S.~M., \& {Hauser}, M.~G. 1999, \apj, 522, 604

\bibitem[{{Peth} {et~al.}(2011){Peth}, {Ross}, \& {Schneider}}]{Peth11}
{Peth}, M.~A., {Ross}, N.~P., \& {Schneider}, D.~P. 2011, \aj, 141, 105

\bibitem[{{Pontzen} \& {Pettini}(2009)}]{Pontzen09}
{Pontzen}, A., \& {Pettini}, M. 2009, \mnras, 393, 557

\bibitem[{{Prochaska} {et~al.}(2003){Prochaska}, {Gawiser}, {Wolfe}, {Castro},
  \& {Djorgovski}}]{Prochaska03}
{Prochaska}, J.~X., {Gawiser}, E., {Wolfe}, A.~M., {Castro}, S., \&
  {Djorgovski}, S.~G. 2003, \apjl, 595, L9

\bibitem[{{Rauch}(1998)}]{Rauch98}
{Rauch}, M. 1998, \araa, 36, 267

\bibitem[{{Richards} {et~al.}(2002){Richards}, {Fan}, {Newberg}, {Strauss},
  {Vanden Berk}, {Schneider}, {Yanny}, {Boucher}, {Burles}, {Frieman}, {Gunn},
  {Hall}, {Ivezi{\'c}}, {Kent}, {Loveday}, {Lupton}, {Rockosi}, {Schlegel},
  {Stoughton}, {SubbaRao}, \& {York}}]{Richards02}
{Richards}, G.~T., {et~al.} 2002, \aj, 123, 2945

\bibitem[{{Richards} {et~al.}(2006){Richards}, {Lacy}, {Storrie-Lombardi},
  {Hall}, {Gallagher}, {Hines}, {Fan}, {Papovich}, {Vanden Berk}, {Trammell},
  {Schneider}, {Vestergaard}, {York}, {Jester}, {Anderson}, {Budav{\'a}ri}, \&
  {Szalay}}]{Richards06}
---. 2006, \apjs, 166, 470

\bibitem[{{Schlegel} {et~al.}(1998){Schlegel}, {Finkbeiner}, \&
  {Davis}}]{Schlegel98}
{Schlegel}, D.~J., {Finkbeiner}, D.~P., \& {Davis}, M. 1998, \apj, 500, 525

\bibitem[{{Sumi}(2004)}]{Sumi2004}
{Sumi}, T. 2004, \mnras, 349, 193

\bibitem[{{Taylor} {et~al.}(2010){Taylor}, {Franx}, {Glazebrook}, {Brinchmann},
  {van der Wel}, \& {van Dokkum}}]{Taylor10}
{Taylor}, E.~N., {Franx}, M., {Glazebrook}, K., {Brinchmann}, J., {van der
  Wel}, A., \& {van Dokkum}, P.~G. 2010, \apj, 720, 723

\bibitem[{{Telfer} {et~al.}(2002){Telfer}, {Zheng}, {Kriss}, \&
  {Davidsen}}]{Telfer02}
{Telfer}, R.~C., {Zheng}, W., {Kriss}, G.~A., \& {Davidsen}, A.~F. 2002, \apj,
  565, 773

\bibitem[{{Truemper}(1982)}]{ROSAT}
{Truemper}, J. 1982, Advances in Space Research, 2, 241

\bibitem[{{Urrutia} {et~al.}(2009){Urrutia}, {Becker}, {White}, {Glikman},
  {Lacy}, {Hodge}, \& {Gregg}}]{Urrutia09}
{Urrutia}, T., {Becker}, R.~H., {White}, R.~L., {Glikman}, E., {Lacy}, M.,
  {Hodge}, J., \& {Gregg}, M.~D. 2009, \apj, 698, 1095

\bibitem[{{van Dokkum}(2001)}]{LACosmic}
{van Dokkum}, P.~G. 2001, \pasp, 113, 1420

\bibitem[{{van Dokkum} {et~al.}(2008){van Dokkum}, {Franx}, {Kriek}, {Holden},
  {Illingworth}, {Magee}, {Bouwens}, {Marchesini}, {Quadri}, {Rudnick},
  {Taylor}, \& {Toft}}]{VanDokkum98}
{van Dokkum}, P.~G., {et~al.} 2008, \apjl, 677, L5

\bibitem[{{Vanden Berk} {et~al.}(2001){Vanden Berk}, {Richards}, {Bauer},
  {Strauss}, {Schneider}, {Heckman}, {York}, {Hall}, {Fan}, {Knapp},
  {Anderson}, {Annis}, {Bahcall}, {Bernardi}, {Briggs}, {Brinkmann}, {Brunner},
  {Burles}, {Carey}, {Castander}, {Connolly}, {Crocker}, {Csabai}, {Doi},
  {Finkbeiner}, {Friedman}, {Frieman}, {Fukugita}, {Gunn}, {Hennessy},
  {Ivezi{\'c}}, {Kent}, {Kunszt}, {Lamb}, {Leger}, {Long}, {Loveday}, {Lupton},
  {Meiksin}, {Merelli}, {Munn}, {Newberg}, {Newcomb}, {Nichol}, {Owen}, {Pier},
  {Pope}, {Rockosi}, {Schlegel}, {Siegmund}, {Smee}, {Snir}, {Stoughton},
  {Stubbs}, {SubbaRao}, {Szalay}, {Szokoly}, {Tremonti}, {Uomoto}, {Waddell},
  {Yanny}, \& {Zheng}}]{Vandenberk01}
{Vanden Berk}, D.~E., {et~al.} 2001, \aj, 122, 549

\bibitem[{{Vladilo} \& {P{\'e}roux}(2005)}]{Vladilo05}
{Vladilo}, G., \& {P{\'e}roux}, C. 2005, \aap, 444, 461

\bibitem[{{Warren} {et~al.}(2007){Warren}, {Hambly}, {Dye}, {Almaini}, {Cross},
  \& {Edge}}]{Warren07}
{Warren}, S.~J., {Hambly}, N.~C., {Dye}, S., {Almaini}, O., {Cross}, N.~J.~G.,
  \& {Edge}, A.~C. e.~a. 2007, \mnras, 375, 213

\bibitem[{{Warren} {et~al.}(2000){Warren}, {Hewett}, \& {Foltz}}]{Warren00}
{Warren}, S.~J., {Hewett}, P.~C., \& {Foltz}, C.~B. 2000, \mnras, 312, 827

\bibitem[{{Webster} {et~al.}(1995){Webster}, {Francis}, {Petersont},
  {Drinkwater}, \& {Masci}}]{Webster95}
{Webster}, R.~L., {Francis}, P.~J., {Petersont}, B.~A., {Drinkwater}, M.~J., \&
  {Masci}, F.~J. 1995, \nat, 375, 469

\bibitem[{{Wolfe} {et~al.}(2005){Wolfe}, {Gawiser}, \& {Prochaska}}]{Wolfe05}
{Wolfe}, A.~M., {Gawiser}, E., \& {Prochaska}, J.~X. 2005, \araa, 43, 861

\bibitem[{{Wright}(1981)}]{Wright81}
{Wright}, E.~L. 1981, \apj, 250, 1

\bibitem[{{Wu} {et~al.}(2012){Wu}, {Zuo}, {Yang}, {Yang}, \& {Wang}}]{Wu2012}
{Wu}, X.-B., {Zuo}, W., {Yang}, J., {Yang}, Q., \& {Wang}, F. 2012, ArXiv
  e-prints

\bibitem[{{Zafar} {et~al.}(2012){Zafar}, {Watson}, {El{\'{\i}}asd{\'o}ttir},
  {Fynbo}, {Kr{\"u}hler}, {Schady}, {Leloudas}, {Jakobsson}, {Th{\"o}ne},
  {Perley}, {Morgan}, {Bloom}, \& {Greiner}}]{Tayyaba12}
{Zafar}, T., {et~al.} 2012, \apj, 753, 82

\end{thebibliography}

\end{document}